\documentclass[12pt, letterpaper]{article}
\usepackage[margin=1in]{geometry}

\usepackage[utf8]{inputenc}
\usepackage[english]{babel}

\usepackage{outlines}
\usepackage[colorlinks=true,allcolors=blue,urlcolor=black]{hyperref}
\usepackage{csquotes}
\usepackage{amsmath}
\usepackage{amssymb}
\usepackage{graphicx}
\usepackage{xcolor}
\usepackage{soul}
\usepackage{chngpage}
\usepackage{float}
\usepackage{amsthm}
\usepackage{mathtools}
\usepackage{bbm}
\usepackage{enumitem}
\usepackage{algorithm}
\usepackage{algpseudocode}
\usepackage{multirow}
\usepackage{setspace}
\usepackage{times}
\usepackage{authblk}
\usepackage{subcaption}

\usepackage{caption} 
\DeclareMathOperator{\CADT}{CADT}
\DeclareMathOperator{\CALD}{CALD}
\DeclareMathOperator{\ASDT}{ASDT}
\DeclareMathOperator{\ASLD}{ASLD}

\DeclareMathOperator{\E}{E}
\DeclareMathOperator{\PP}{P}
\DeclareMathOperator{\Cov}{Cov}
\DeclareMathOperator{\Var}{Var}

\newcommand{\boldX}{\mathbf{X}}
\newcommand{\boldx}{\mathbf{x}}
\newcommand{\boldO}{\mathbf{O}}
\newcommand{\boldo}{\mathbf{o}}

\newcommand{\meanin}{n^{-1}\sum_{i=1}^{n}}

\usepackage{natbib}
\setcitestyle{round}
\newtheorem{theorem}{Theorem}

\newtheorem{assumption}{Assumption}
\newtheorem{corollary}{Corollary}[theorem]

\allowdisplaybreaks

\title{Difference-in-differences with stochastic policy shifts of a continuous treatment}
\author{Michael Jetsupphasuk\thanks{Corresponding author. Email: jetsupphasuk@unc.edu}, Chenwei Fang, Didong Li, Michael G. Hudgens}
\date{}

\affil{Department of Biostatistics, University of North Carolina at Chapel Hill}

\begin{document}
	
\maketitle

\vspace{-2em}

\begin{abstract}
    Treatment effects of stochastic policy shifts quantify differences in outcomes across counterfactual scenarios with varying treatment distributions. Stochastic policy shifts may be of interest in settings where it is unrealistic or infeasible to deterministically manipulate treatments. In this paper, methods are developed to draw inference about stochastic policy effects under difference-in-differences (DiD) designs with a continuous treatment. The proposed causal estimand is the expected effect of modifying the continuous dose distribution among the treated, i.e., those that received a non-zero dose. Several possible stochastic policies are discussed and a general framework for identification and estimation is proposed. One stochastic policy applicable to many settings is the exponential tilt, which increments the conditional density function of the continuous dose. For the exponential tilt policy, a double/debiased machine learning estimator is proposed that allows for data-adaptive, nonparametric nuisance function estimation. Under mild convergence rate conditions, the estimator is shown to be root-$n$ consistent and asymptotically normal with variance attaining the nonparametric efficiency bound. The proposed method is used to study the effect of hydraulic fracturing activity on employment and income.
\end{abstract}

\section{Introduction}

Difference-in-differences (DiD) is a causal inference method that has risen in prominence in recent years, both in applications and in methodological development. \citet{goldsmith-pinkham_tracking_2024} estimated that within the applied microeconomics working paper series of the National Bureau of Economic Research, DiD methods were used in about 10\% of papers in 2002 but over 30\% in 2024. In the canonical DiD design, there are two time periods with all units untreated in the first period and some (but not all) units treated in the second period. The key assumption in the canonical DiD design stipulates that the expected potential outcome trends for both treated and untreated groups are parallel in the counterfactual scenario where neither group received treatment. Under this parallel trends assumption, the average treatment effect on the treated (ATT) is identified. DiD has been extended in recent years to allow for parallel trends conditional on observed covariates \citep{abadie_semiparametric_2005}, semi- or nonparametric estimation \citep{santanna_doubly_2020, chang_doubledebiased_2020}, interference \citep{xu_difference--differences_2025, jetsupphasuk_difference--differences_2025}, staggered treatment rollouts \citep{callaway_difference--differences_2021}, among many other advances \citep{roth_whats_2023}. Of particular relevance to this paper, recent studies have proposed nonparametric estimators when treatment is continuous \citep{callaway_difference--differences_2025, hettinger_multiply_2025, zhang_continuous_2025}. 

The causal estimands considered in DiD designs typically focus on contrasts of potential outcomes under deterministic interventions, e.g., the ATT, where treated individuals' outcomes are compared to the counterfactual scenario where no one is treated. Deterministic interventions may be relevant in many settings. For example, if there is interest in the effects of prescribing a drug, examining causal contrasts for prescribing or not prescribing the drug is sensible since prescriptions are in control of the physician. However, there may be settings where treatment levels cannot be deterministically set but the likelihood of a treatment may be manipulable. In these settings, stochastic policies (or stochastic interventions) may represent more realistic interventions. For instance, consider the evaluation of health effects due to wildfire-derived air pollution. There are no reasonable interventions that can set individuals' exposure to wildfire smoke to some exact pre-specified level, but there are interventions that can reduce the probability of large wildfires, such as prescribed burns. Another example is the evaluation of public health campaigns to influence healthy behavior. Public health campaigns may influence the probability that individuals, for instance, adopt healthy eating habits, but such campaigns do not deterministically impose healthy eating habits upon individuals. 

Causal estimands based on stochastic policies compare potential outcomes under (possible) counterfactuals where the distribution of treatments may be different than observed. For example, with a binary treatment, there may be interest in an effect measure that compares the scenario where 70\% of individuals are treated with the scenario where 20\% of individuals are treated. When treatment is continuous there may be interest in comparing counterfactual scenarios where the probability of being exposed to large doses is high compared to when smaller doses are more likely. 

The methods proposed in this paper are used to estimate the economic effects of hydraulic fracturing (``fracking") in the United States from 2008 to 2012. Governments may decide to allow or disallow fracking based on estimated effects of potential fracking productivity, where potential fracking productivity is measured by a continuous ``prospectivity score." Governments may be interested in effects for a range of possible prospectivity scores, where some values are more likely than others, motivating causal estimands based on different distributions of the prospectivity score. Additionally, there may be interest in effects of fracking productivity shifts in the future, e.g., due to technological advances. The analysis in this paper focuses on the employment and income effects associated with shifts in the prospectivity score distribution toward both the minimal and maximal values. The application builds on prior work by \citet{bartik_local_2019}, who estimated deterministic effects of dichotomized prospectivity scores on various socioeconomic outcomes.

In this paper, stochastic policies are considered in the DiD design with a continuous treatment. Section \ref{sec:estimands} discusses the causal estimand of interest under the canonical two time period design. Section \ref{sec:estimation} presents the proposed estimators and their large sample properties; in particular, efficient influence function-based estimators are proposed and shown to be asymptotically normal and nonparametric efficient. Extensions to the multiple time periods setting are developed in Section \ref{sec:mtp}. The proposed methods are evaluated in simulation studies presented in Section \ref{sec:sim} and used to study the economic effects of fracking in Section \ref{sec:application}. Finally, Section \ref{sec:discussion} concludes.

\section{Causal estimands with stochastic policies} \label{sec:estimands}

\subsection{Data structure}

Consider a treatment (or exposure) $A \in [0,1]$ that follows a mixture distribution with a point mass at $0$ and a continuous dose $D = A | A>0$ with support $\mathcal{D} \subseteq (0,1]$. Throughout, $A=0$ is referred to as ``untreated" and $A>0$ as ``treated" since it is often the case that the reference level $0$ denotes lack of treatment. In general, however, $A=0$ may refer to the minimum level of treatment or some background level of treatment. 

Consider the two period setting with time denoted $t \in \{0,1\}$ and units indexed by $i=1,\dots,n$. All units are untreated at $t=0$ and some units are treated at $t=1$. Let $A_i$ refer to treatment for unit $i$ at time $t=1$. Let $Y_{it}$ denote the outcome observed at time $t$ for unit $i$, and let $\Delta Y_i = Y_{i1} - Y_{i0}$. The pre-treatment ($t=0$) covariate vector for unit $i$ is denoted $\boldX_i$ where the support is denoted $\mathcal{X}$. The observed data are $\boldO_i = (Y_{i0}, Y_{i1}, A_i, \boldX_i) \sim \mathbbm{P}$ which are assumed to be independent and identically distributed (iid) for $i=1,\dots,n$. Potential outcomes under (possible) counterfactual treatment $a$ are denoted $Y_{it}(a)$. Potential outcomes are related to observed outcomes by Assumption \ref{assump:consistency}, the standard causal consistency assumption. 

\begin{assumption}[Causal consistency]
    \label{assump:consistency}
    If $A_i=a$, then $Y_{it} = Y_{it}(a)$.
\end{assumption}

\subsection{Causal estimands and identification}

Define the average stochastic dose effect among the treated (ASDT) to be
\begin{align*}
     \ASDT(Q) &= \E \left[ \int_{\mathcal{D}} \{Y_{i1}(d) - Y_{i1}(0) \}\mathrm{d}Q(d | \boldX_i, A_i>0) \bigg| A_i>0 \right],
\end{align*}
where $Q$ is the (possibly) counterfactual stochastic policy for the dose distribution conditional on covariates $\boldX=\boldx$, and $Q(d | \boldx, A_i>0) = \PP(D<d | \boldX_i=\boldx, A_i>0)$ is the corresponding conditional distribution function of $D$ given $\boldX_i=\boldx$ and $A_i>0$. Define $\CADT(d, \boldx) = \E[Y_{i1}(d) - Y_{i1}(0) | \boldX_i=\boldx, A_i > 0 ]$ as the conditional average dose effect among the treated (CADT), which describes the average effect of dose $d$ relative to no treatment, conditional on covariates and being treated. Then, the ASDT can be expressed as the average CADT of all doses $d \in \mathcal{D}$ over the counterfactual distribution $Q$ and the distribution of covariates, i.e., $\ASDT(Q) = \E \left[ \int_{\mathcal{D}} \CADT(d,\boldX_i) \mathrm{d}Q(d | \boldX_i, A_i>0) \bigg| A_i > 0 \right]$.

The ASDT includes some common estimands as special cases. For instance, if $Q$ is a degenerate distribution with point mass at $d' \in \mathcal{D}$ for all $\boldx$, then the ASDT reduces to the average dose effect among the treated (ADT) at level $d'$, i.e., $\E[Y_{i1}(d') - Y_{i1}(0) | A_i > 0]$. In this sense, the ASDT parameter generalizes the ADT parameter which has been previously studied \citep{callaway_difference--differences_2025, hettinger_multiply_2025}. Additionally, consider specifying a generic stochastic policy $Q$ that does not depend on covariates $\boldx$. Then, the ASDT reduces to  $\int_{\mathcal{D}} \E[Y_{i1}(d) - Y_{i1}(0) | A_i > 0 ]\mathrm{d}Q(d | A_i>0)$, the average treatment effect of a driver-unconfounded treatment on the treated (ADUTT), introduced by \citet{hettinger_causal_2026}. The ADUTT averages the ADT of all doses $d \in \mathcal{D}$ over the counterfactual distribution $Q$. In general, $Q$ is user-specified and should be chosen to fit the context of a study. Several stochastic policies that may be of general interest, aside from the special cases presented above, are discussed in Section \ref{sec:stochastic}. 

Define another causal estimand of interest to be the average stochastic local dose effect (ASLD),
\begin{align*}
    \ASLD(Q) = \E \left[ \int_{\mathcal{D}} \E[Y_{i1}(d) - Y_{i1}(0) | \boldX_i, D=d, A_i > 0 ]\mathrm{d}Q(d | \boldX_i, A_i>0) \bigg| A_i > 0 \right]
\end{align*}
which is similar to the ASDT except the CADT is replaced with the conditional average local dose effect (CALD), $\CALD(d, \boldx) = \E[Y_{i1}(d) - Y_{i1}(0) | \boldX_i=\boldx, D=d, A_i > 0 ]$, where the term ``local" refers to conditioning on the dose group that corresponds to the counterfactual dose in the causal contrast \citep{callaway_difference--differences_2025}. The ASDT is arguably more interpretable than the ASLD since the ASDT averages over the CADT where the conditioning group does not depend on the dose $d$. In contrast, the ASLD averages over the CALD where both the dose in the causal contrast and the conditioning group vary simultaneously. In other words, variation between $\CADT(d,\boldx)$ and $\CADT(d',\boldx)$ for $d \neq d'$ is due to differences in dose effects while variation between $\CALD(d,\boldx)$ and $\CALD(d',\boldx)$ is due to differences in both dose and group effects. 

In order to express the proposed causal estimands in terms of the observed data distribution, identification assumptions are needed. Assumption \ref{assump:no_anticip} is standard in DiD designs and states that treatments at time $t=1$ do not affect potential outcomes at time $t=0$. 

\begin{assumption}[No anticipation]
\label{assump:no_anticip}
        $Y_{i0}(0) = Y_{i0}(a)$ for all $a$. 
\end{assumption}

Assumptions \ref{assump:parallel-cond-a0} and \ref{assump:parallel-cond-d} are parallel trends assumptions that are conditional on covariates. Assumption \ref{assump:parallel-cond-a0} reduces to the conditional parallel trends assumption in the classic DiD setting when the treatment $A$ is dichotomized into treated and untreated groups. Assumption \ref{assump:parallel-cond-a0} states that, under the no treatment counterfactual, the expected potential outcome trends of the treated and untreated groups are parallel conditional on covariates. Assumption \ref{assump:parallel-cond-d} differs from the typical parallel trends assumption by instead making a supposition about the potential outcomes for doses $d \in \mathcal{D}$. Let $\E[Y_{i1}(d) - Y_{i0}(d) | \boldX, A_i=d']$ be the expected potential outcome trend under counterfactual dose $d$, where, borrowing terminology from \citet{callaway_difference--differences_2025}, $d'=d$ defines the ``local" group and $d' \in \mathcal{D} \setminus d$ defines ``non-local" groups. Then, Assumption \ref{assump:parallel-cond-d} states that, for any dose, the local trend is parallel (conditional on covariates) to the average trend over all local and non-local trajectories for that counterfactual dose. 

\begin{assumption}[Conditional untreated parallel trends between treated and untreated groups]
    \label{assump:parallel-cond-a0}
    ~\\
    $\E[Y_{i1}(0) - Y_{i0}(0) | \boldX_i, A_i>0] = \E[Y_{i1}(0) - Y_{i0}(0) | \boldX_i, A_i=0]$.
\end{assumption}

\begin{assumption}[Conditional dose-specific parallel trends between local and treated dose groups]
    \label{assump:parallel-cond-d}
    For all doses $d \in \mathcal{D}$, $\E[Y_{i1}(d) - Y_{i0}(d) | \boldX_i, A_i=d] = \E[Y_{i1}(d) - Y_{i0}(d) | \boldX_i, A_i > 0]$.
\end{assumption}

\begin{figure}[!h]
    \centering
    \begin{subfigure}{0.45\textwidth}
        \centering
        \includegraphics[width=1\textwidth]{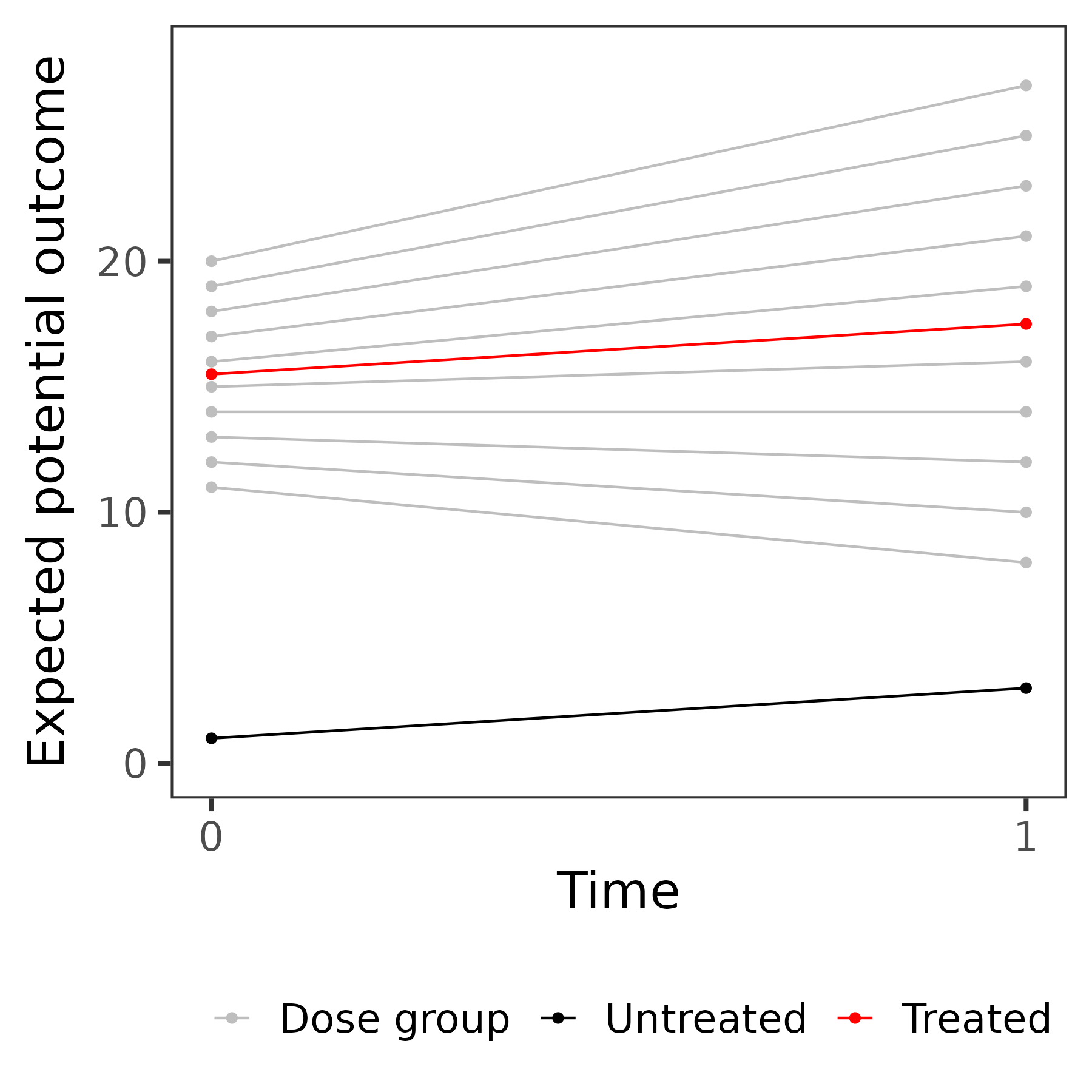}
        \caption{}
        \label{fig:parallel-trends-a}
    \end{subfigure}
    \hfill
    \begin{subfigure}{0.45\textwidth}
        \centering
        \includegraphics[width=1\textwidth]{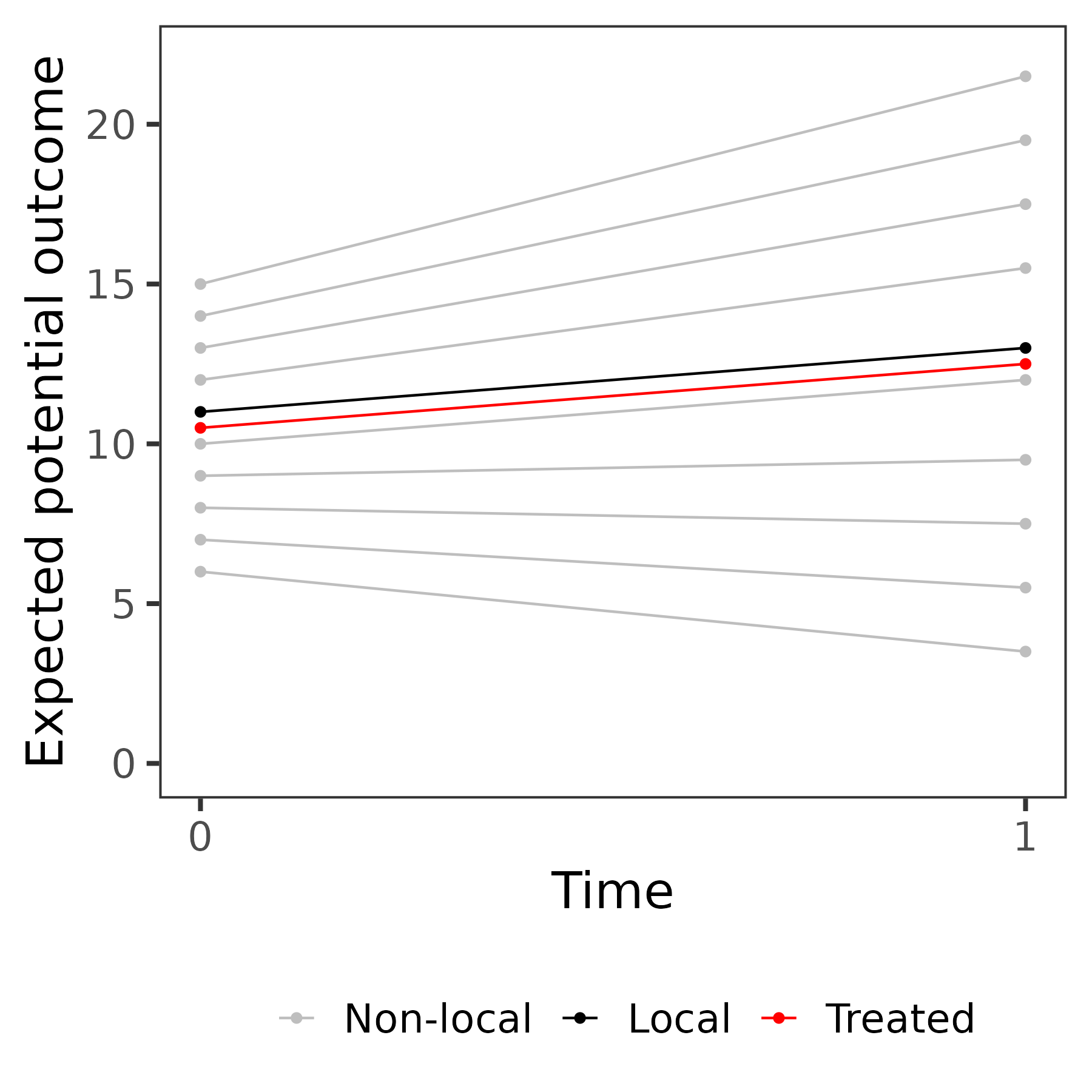}
        \caption{}
    \end{subfigure}
    \caption{Illustration of unconditional parallel trends assumptions: (a) Assumption \ref{assump:parallel-cond-a0} is illustrated, where $\E[Y_t(0)|A=d]$ for 10 non-zero dose levels $d$ (grey), $\E[Y_t(0)|A=0]$ (black), and $\E[Y_t(0)|A>0]$ (red) are shown; (b) Assumption \ref{assump:parallel-cond-d} is illustrated for counterfactual doses $d > 0$, where $\E[Y_t(d)|A=a]$ for 9 dose levels $a \neq d$ (grey), $\E[Y_t(d)|A=d]$ (black), and $\E[Y_t(d)|A>0]$ (red) are shown.}
    \label{fig:parallel-trends}
\end{figure}

\noindent
Figure \ref{fig:parallel-trends} illustrates versions of Assumptions \ref{assump:parallel-cond-a0} and \ref{assump:parallel-cond-d} that are unconditional on covariates for 10 dose levels $d \in \mathcal{D}$. These parallel trends assumptions are common in DiD designs with continuous treatments and are often implicitly made when two-way fixed effects models are imposed and causally interpreted; see \citet{callaway_difference--differences_2025} for further discussion.

Assumptions \ref{assump:parallel-cond-a0} and \ref{assump:parallel-cond-d} are part of a sufficient set of conditions to identify the ASDT. In contrast to the parallel trends assumption utilized in the canonical DiD setup which supposes parallel expected potential outcome trends under only the \textit{no treatment} counterfactual, Assumption \ref{assump:parallel-cond-d} places a restriction on trends under \textit{dose} counterfactuals. On the other hand, to identify the ASLD, it is not necessary to make any assumption on trends under \textit{dose} counterfactuals, i.e., Assumption \ref{assump:parallel-cond-d} need not hold. Instead, the ASLD is shown below to be identified under a set of conditions that does not include Assumptions \ref{assump:parallel-cond-a0} and \ref{assump:parallel-cond-d} but does include Assumption \ref{assump:parallel-cond-alt}. Assumption \ref{assump:parallel-cond-alt} is stronger than Assumption \ref{assump:parallel-cond-a0} in the sense that Assumption \ref{assump:parallel-cond-alt} implies Assumption \ref{assump:parallel-cond-a0}. Assumption \ref{assump:parallel-cond-alt} states that, under the no treatment counterfactual, the expected potential outcome trends for all treatment groups are parallel, conditional on covariates. This assumption (unconditional on covariates) would suppose that in the left-hand plot in Figure \ref{fig:parallel-trends}, all the grey lines are parallel to the black line. However, no assumption on potential outcome trends for dose counterfactuals (right-hand plot in Figure \ref{fig:parallel-trends}) is needed to identify the ASLD.

\begin{assumption}[Conditional untreated parallel trends between all treatment groups]
    \label{assump:parallel-cond-alt}
    For all doses $d \in \mathcal{D}$, $\E[Y_{i1}(0) - Y_{i0}(0) | \boldX_i, A_i=d] = \E[Y_{i1}(0) - Y_{i0}(0) | \boldX_i, A_i = 0]$.
\end{assumption}

In some settings, parallel trends unconditional on covariates may be justified. Under unconditional parallel trends (and not assuming conditional parallel trends), there may be interest in stochastic intervention effects that shift the marginal dose density function, i.e., one may consider the special case where the causal estimands and parallel trends assumptions introduced above are unconditional on covariates. For brevity, the formal results for the unconditional parallel trends assumptions are discussed in the Appendix. 

For identification, we consider a weaker version of positivity than is typically assumed in causal inference settings with deterministic interventions. Part (i) of Assumption \ref{assump:positivity-ident-c} states standard positivity for dichotomized treatment where $\pi_{A>0}(\boldx) = \PP(A>0 | \boldX=\boldx)$ is the probability of being treated (with any dose), conditional on covariates. Let $\pi_D(d|\boldx)$ denote the density function of the observed dose $D$ conditional on $\boldX=\boldx$, also known as the generalized propensity score, and let $q(d|\boldX=\boldx)$ denote the conditional density function of the stochastic policy $Q$. Part (ii) of Assumption \ref{assump:positivity-ident-c} restricts choice of $Q$ to distributions that have zero density in regions of $\mathcal{D}$ where the distribution of the observed dose has zero density.

\begin{assumption}[Positivity conditional on covariates]
    \label{assump:positivity-ident-c}
    (i) For some $0 < \epsilon < 1/2$, $\epsilon < \pi_{A>0}(\boldX_i) < 1-\epsilon$. (ii) For any $d \in \mathcal{D}$, if $\pi_D(d|\boldX_i) = 0$, then $q(d|\boldX_i) = 0$.
\end{assumption}

Denote the dose-specific expected outcome change as $\mu_d(\boldx) = \E[\Delta Y | \boldX=\boldx, D=d, A>0]$. Further, let $\mu_{A=0}(\boldx) = \E[\Delta Y | \boldX=\boldx, A=0]$. Theorem \ref{thm:identification} shows that the stochastic intervention estimand $\ASDT(Q)$ is identifiable from the observable random variables. Under conditional parallel trends Assumptions \ref{assump:parallel-cond-a0} and \ref{assump:parallel-cond-d}, the identified estimand takes on a familiar form as in the classic DiD setting with a binary treatment. The first component, $\Psi^{(1)}(\mathbb{P})$, averages the dose-specific conditional outcome means $\mu_d(\boldx)$ over the dose distribution $Q$, then marginalizes over the covariate distribution $\mathcal{X}$, all conditional on being treated. The second component, $\Psi^{(2)}(\mathbb{P})$, appears in the classic DiD identification result when treatment is dichotomous and represents the identified potential outcome trend $\E[Y_{i1}(0)-Y_{i0}(0)|A_i>0]$. 

\begin{theorem}[Identification]
    \label{thm:identification}
    Under Assumptions \ref{assump:consistency}, \ref{assump:no_anticip}, \ref{assump:parallel-cond-a0}, \ref{assump:parallel-cond-d}, and \ref{assump:positivity-ident-c},
    \begin{align*}
        \ASDT(Q) &= \Psi(\mathbb{P}) \equiv \Psi^{(1)}(\mathbb{P}) - \Psi^{(2)}(\mathbb{P}),
    \end{align*}
    where $\Psi^{(1)}(\mathbb{P}) = \E \left[ \int_{\mathcal{D}} \mu_{d}(\boldX) \mathrm{d}Q(d|\boldX, A>0) \big| A>0 \right]$ and $\Psi^{(2)}(\mathbb{P}) = \E[ \mu_{A=0}(\boldX) | A>0]$.

    Under Assumptions \ref{assump:consistency}, \ref{assump:no_anticip}, \ref{assump:parallel-cond-alt}, and \ref{assump:positivity-ident-c},
    \begin{align*}
        \ASLD(Q) &= \Psi(\mathbb{P}).
    \end{align*}
\end{theorem}

\noindent
Theorem \ref{thm:identification} also states that the causal estimand $\ASLD(Q)$ is identified by the same statistical functional $\Psi(\mathbb{P})$ if Assumption \ref{assump:parallel-cond-alt} is supposed in lieu of Assumptions \ref{assump:parallel-cond-a0} and \ref{assump:parallel-cond-d}. To ease exposition, only the causal estimand $\ASDT(Q)$ will be referred to for the remainder of the text. 

\subsection{Stochastic policies} \label{sec:stochastic}

Thus far, causal effects with generic stochastic policies $Q$ have been defined. In this subsection, various possible stochastic policies that may be of interest in different settings are discussed. Several of the stochastic interventions discussed here entail shifting the generalized dose propensity score $\pi_D(d|\boldX)$ where $\boldX$ is a set of covariates that satisfy conditional positivity and parallel trends. 

\textbf{\textit{Exponential tilt}}. The exponential tilt \citep{diaz_causal_2020, schindl_incremental_2026} shifts the observed treatment distribution according to an increment parameter $\delta$; in particular, the exponential tilt has density function $$q_{\delta}(d|\boldx) = \frac{\exp(\delta d) \pi_D(d|\boldx)}{\int \exp(\delta b) \pi_D(b | \boldx) \mathrm{d}b}.$$
\noindent
The increment $\delta$ can be expressed as the rate of change of the log ratio comparing the stochastic intervention and the observed distribution with respect to $d$, i.e., \\
$\delta = \frac{\partial}{\partial d} \left\{ \log(q_{\delta}(d|\boldx) / \pi_D(d|\boldx)) \right\}$ \citep{schindl_incremental_2026}. As $\delta \rightarrow - \infty$, the exponential tilt places increasing density towards the left-hand side of the dose support. As $\delta \rightarrow \infty$ the exponential tilt moves density towards the right-hand side of the support. For example, if the conditional distribution $D|X,A>0$ has support $(0,1]$, then the ASDT tends towards $\E[Y_{it}(0) - Y_{it}(0)|A_i>0] = 0$ as $\delta \rightarrow \infty$. Similarly, the ASDT tends towards $\E[Y_{it}(1) - Y_{it}(0)|A_i>0]$ when $\delta \rightarrow -\infty$, i.e., the effect of the maximum dose among the treated. 
\begin{figure}[!h]
    \centering
    \includegraphics[width=0.8\linewidth]{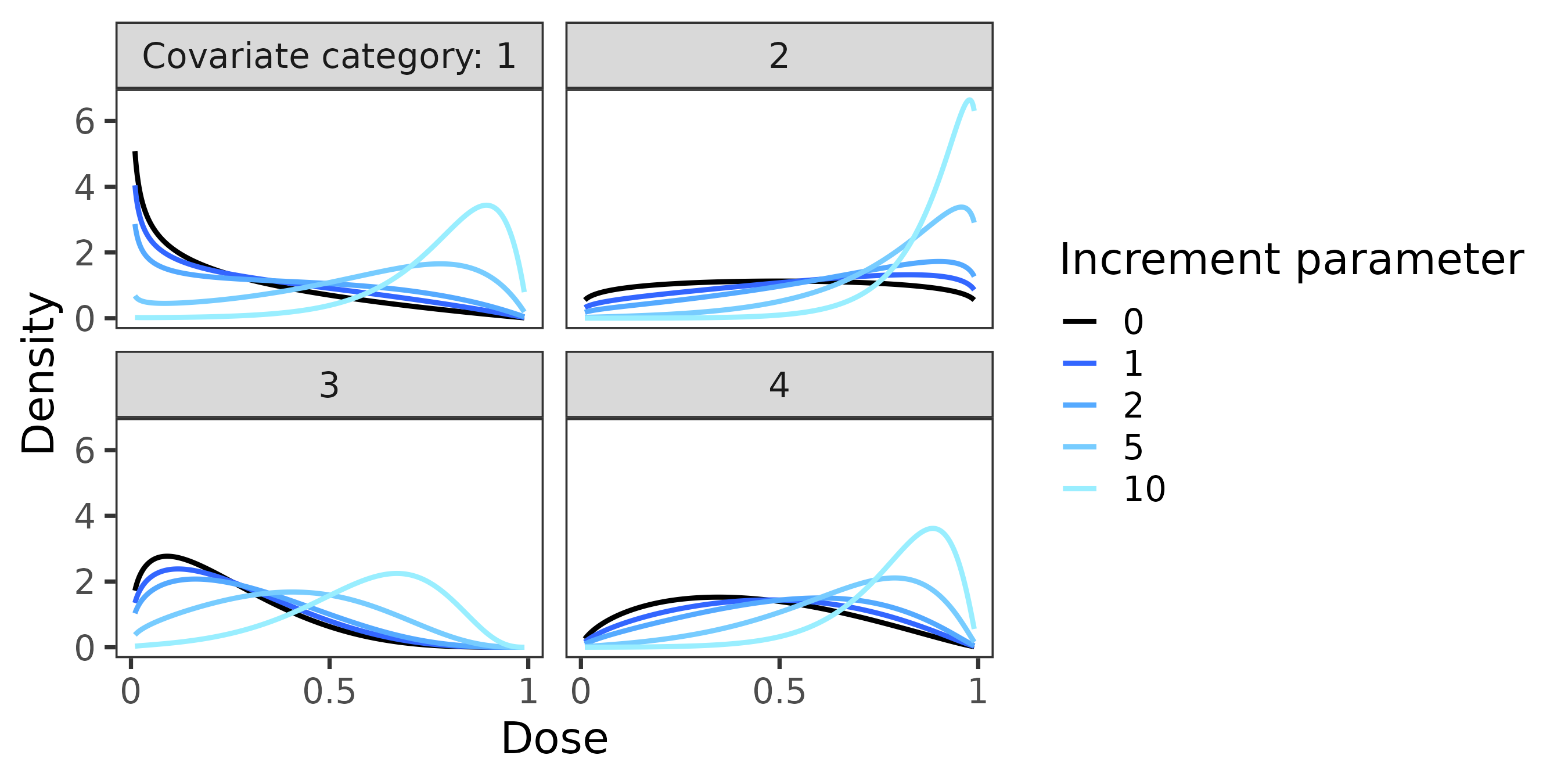}
    \caption{Exponential tilt with a four-level categorical covariate $X$ for increment parameter $\delta$.}
    \label{fig:exp-tilt}
\end{figure}
\noindent
The exponential tilt automatically satisfies Assumption \ref{assump:positivity-ident-c}(ii) and is illustrated in Figure \ref{fig:exp-tilt} for a conditional density function with a four-level categorical covariate, where the black line is the observed density function (i.e., $\delta=0$) and the blue lines vary by increment parameter $\delta$.

\begin{figure}[!h]
    \centering
    \includegraphics[width=0.8\linewidth]{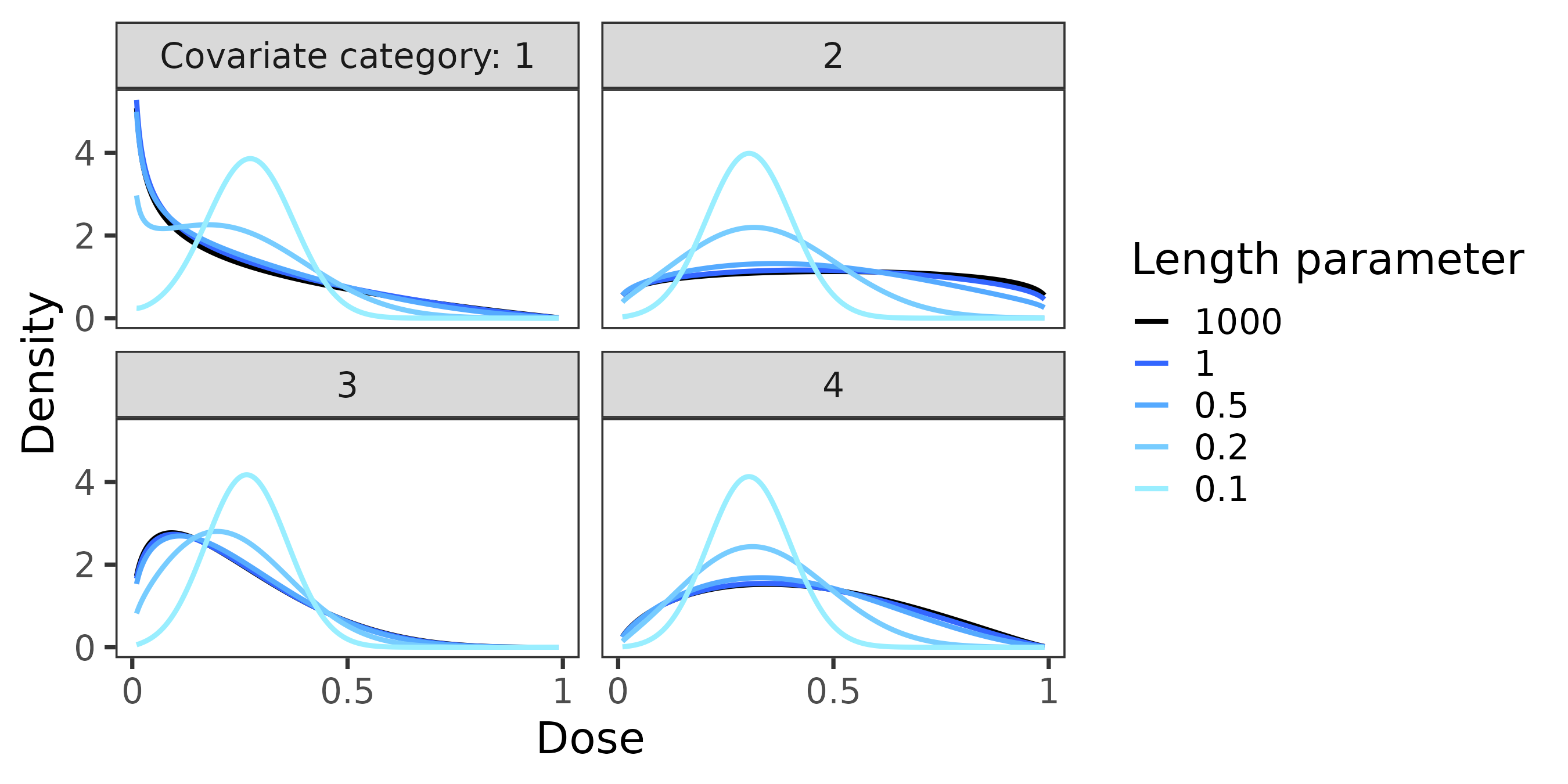}
    \caption{Gaussian concentration centered around specified $d' = 0.3$ with a four-level categorical covariate $X$ for length parameter $l$.}
    \label{fig:gauss-kernel}
\end{figure}

\textbf{\textit{Gaussian concentration around specified $d'$}}. Rather than shifting the observed dose distribution towards $0$ or $1$, there may be interest in shifting the distribution to some point $d'$ in the middle of the support of $\mathcal{D}$. Then, one can define the Gaussian concentration around a specified $d'$ with length parameter $l \in (0, \infty)$, 
\begin{align*}
    q_{l, d'}(d|\boldx) &= \frac{\exp\{- (d-d')^2 / 2l ^2\} \pi_D(d|\boldx)}{\int \exp\{-(b-d')^2 / 2l^2\} \pi_D(b | \boldx) \mathrm{d}b}.
\end{align*}
This stochastic policy resembles a Gaussian kernel (also known as the squared exponential kernel or radial basis function kernel) with weights $\pi_D(d|\boldx)$. Assuming $d'$ is in the support of $D | X, A>0$, this stochastic policy shifts $\pi_D(d|\boldx)$ to concentrate density around the specified $d'$ where smaller length $l$ increases the amount of concentration. When $d'$ is not in the support of $D | X, A>0$, this stochastic policy concentrates density in the neighborhood around $d'$. For example, suppose that the support of $D | X, A>0$ is $(0, 0.4) \cup (0.6, 1)$ and $d' = 0.5$. Then, the Gaussian concentration policy would have increasing density around the points $0.4$ and $0.6$ as the length parameter $l$ becomes smaller. This stochastic policy may also be parameterized with a concentration parameter $c = l^{-2}$. Then, $c$ has a similar interpretation as $\delta$ in the exponential tilt policy since $c = \frac{\partial}{\partial d} \left\{ \log(q_{c,d'}(d|\boldx) / \pi_D(d|\boldx)) \right\} / (d' - d)$ for $d \neq d'$. Thus, the concentration parameter describes how the rate of change in the log density ratios is positive when $d$ increases towards $d'$ and negative thereafter. The conditional distribution of the observed dose is recovered as a special case, i.e., $q_{l, d'}(d|\boldx) \rightarrow \pi_D(d|\boldx)$ as $l \rightarrow \infty$, and the Gaussian concentration policy satisfies Assumption \ref{assump:positivity-ident-c}(ii) by definition. Figure \ref{fig:gauss-kernel} illustrates the Gaussian concentration stochastic policy around the dose $d' = 0.3$ with varying lengths $l$.

\begin{figure}[!h]
    \centering
    \includegraphics[width=0.8\linewidth]{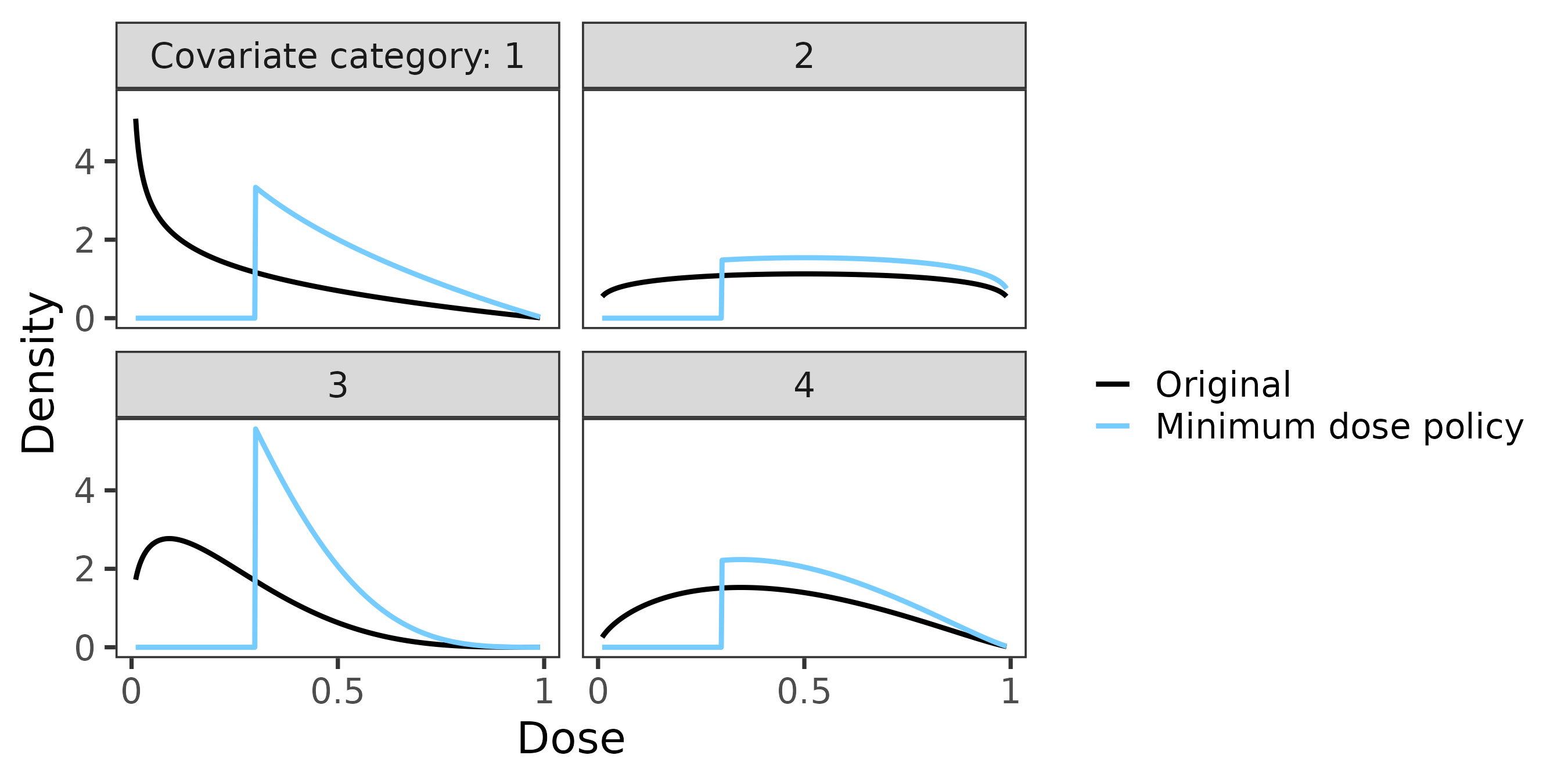}
    \caption{Minimum dose policy at specified $d^* = 0.3$ with a four-level categorical covariate $X$.}
    \label{fig:min-dose}
\end{figure}

\textbf{\textit{Minimum dose policy}}. Suppose there is interest in considering the counterfactual where all units receive at minimum some pre-specified dose $d^*$ \citep{taubman_intervening_2009}. Further suppose that the conditional dose distribution has support in the region $[d^*,1]$, i.e., $\int_{d^*}^{1} \pi_D(b|\boldX) db > 0$. Then, the minimum dose policy may be of interest, where the counterfactual dose density function is $$q_{d^*}(d|\boldx) = \frac{\pi_D(d|\boldx) \mathbbm{1}(d > d^*)}{\int_b \pi_D(b|\boldx) \mathbbm{1}(b > d^*) \mathrm{d}b}.$$ 
\noindent
Similar to the other nonparametric stochastic policies discussed in this section, Assumption \ref{assump:positivity-ident-c}(ii) is satisfied by definition. An example of the minimum dose policy is shown in Figure \ref{fig:min-dose}.

\textbf{\textit{Parametric shift policy}}. If the observed dose distribution is assumed to follow a parametric model, then stochastic policy shifts may be defined as changes to the parameters of the dose distribution. For instance, if dose is modeled as a truncated normal distribution \\
$D \sim \mathrm{TruncNorm}(\mathrm{logit}^{-1}(f(\boldX)), \sigma^2, 0, 1)$, then one may consider the stochastic policy $Q \sim \mathrm{TruncNorm}(\mathrm{logit}^{-1}(f(\boldX) + \eta), \sigma^2, 0, 1)$ where $\eta$ is the increment parameter, $\mathrm{logit}^{-1}$ is the inverse logit function, and the parameters of $\mathrm{TruncNorm}$ are the mean, variance, lower limit, and upper limit, respectively. 

\textbf{\textit{Parametric policy}}. One may also specify any parametric distribution with support on $\mathcal{D}$ that does not depend on the data. For instance, there may be interest in investigating effects when all dose levels are equally likely, i.e., under a uniform distribution on the interval $(0,1)$, provided that $\mathcal{D} = (0,1)$.

\section{Estimation and inference} \label{sec:estimation}

One simple type of estimator of the statistical estimand $\Psi(\mathbb{P})$ is the plug-in estimator, i.e., 
\begin{align*}
    \hat \tau^{\mathrm{plug-in}} &= \meanin \frac{\mathbbm{1}(A_i>0)}{\E_n \{\mathbbm{1}(A>0) \}} \left\{ \int_{\mathcal{D}} \hat \mu_d(\boldX_i) \mathrm{d}\hat Q(d|\boldX_i, A_i>0) - \hat \mu_{A=0}(\boldX_i) \right\},
\end{align*}

\noindent
where $\hat \mu_d$ and $\hat \mu_{A=0}$ are estimators of $\mu_d$ and $\mu_{A=0}$, respectively, $\E_n R = \meanin R_i$ for generic random variable $R$, and $\hat Q$ is an estimator of $Q$ in settings where $Q$ depends on the observed data distribution. Throughout this paper, integrals are left in estimators, e.g., $\int_{\mathcal{D}} \hat \mu_d(\boldX_i) \mathrm{d}\hat Q(d|\boldX_i, A_i>0)$. These integrals may not analytically tractable but can be approximated; for instance, Monte Carlo integration with uniform sampling was used to approximate these integrals in the simulations and application presented in Sections \ref{sec:sim} and \ref{sec:application}, respectively. If correctly specified parametric models are assumed for all the nuisance functions, then it follows from standard estimating equaion theory that $\hat \tau^{\mathrm{plug-in}}$ is $\sqrt{n}$ consistent and asymptotically normal (CAN) under mild regularity conditions. However, $\hat \tau^{\mathrm{plug-in}}$ may be biased if the parametric models are mis-specified. In contrast, nonparametric, machine learning methods can more flexibly estimate nuisance functions and mitigate model mis-specification concerns. 

When machine learning is used to estimate nuisance functions, plug-in estimators like $\hat \tau^{\mathrm{plug-in}}$ may not be $\sqrt{n}$-CAN since plug-in estimators typically inherit the convergence rates of the nuisance component estimators and nonparametric estimators often do not converge at the parametric $\sqrt{n}$ rate. Below, an estimator is proposed that allows for machine learning estimators of nuisance functions, is $\sqrt{n}$-CAN, and achieves the nonparametric efficiency bound asymptotically \citep{kennedy_semiparametric_2023}. 

The efficient influence function (EIF) characterizes the nonparametric efficiency bound and can be used to construct an estimator with the properties mentioned above. Consider stochastic interventions that are smooth functions of the observed propensity score. In particular, let $q_{\nu}(d|\boldx) = f_{\nu}( \pi_D(d|\boldx)) [\bar q(\boldx; \pi)]^{-1}$ where $\nu$ is a finite dimensional parameter, $f_{\nu}: [0,\infty) \mapsto [0,\infty)$, $\bar q(\boldx; \pi) = \int_b f_{\nu}(\pi_D(b|\boldx))\mathrm{d}b$, and assume the derivative $f_{\nu}'(c) = \{ \partial / \partial c \} f_{\nu}(c)$ exists. Theorem \ref{thm:eif} provides the EIF for the identified causal estimand $\Psi(\mathbb{P})$, where $0/0$ is defined to be equal to $0$ for ease of notation. 

\begin{theorem}[Efficient influence function]
    \label{thm:eif}
    For a generic stochastic policy based on the observed propensity score $f_{\nu}( \pi_D(d|\boldX)) [\bar q(\boldX; \pi)]^{-1}$, the EIF of $\Psi(\mathbb{P})$ is $\varphi(\boldO; \mathbb{P}) = \varphi^{(1)}(\boldO; \mathbb{P}) - \varphi^{(2)}(\boldO; \mathbb{P})$, where 
    \begin{align*}
        \varphi^{(1)}(\boldO; \mathbb{P}) &= \frac{\mathbbm{1}(A>0)}{\PP(A>0)} \bigg \{ \frac{q_{\nu}(D|\boldX)}{\pi_D(D|\boldX)} (\Delta Y - \mu_D(\boldX)) \\
        &\hspace{4em} + \frac{f'_{\nu}(\pi_D(D|\boldX)) \mu_D(\boldX)}{\bar q(\boldX; \pi)} - \frac{\int_b f'_{\nu}(\pi_D(b|\boldX)) \pi_D(b|\boldX) \mu_b(\boldX) \mathrm{d}b}{\bar q(\boldX; \pi)}  \\
        &\hspace{4em} - \frac{\int_b \mu_b(\boldX) q_{\nu}(b|\boldX) \mathrm{d}b}{\bar q(\boldX; \pi)}\left(f'_{\nu}(\pi_D(D|\boldX)) - \int_b f'_{\nu}(\pi_D(b|\boldX)) \pi_D(b|\boldX) \mathrm{d}b \right) \\
        &\hspace{4em} + \int_b \mu_b(\boldX) q_{\nu}(b|\boldX) \mathrm{d}b - \Psi^{(1)}(\mathbb{P}) \bigg \}, \\
        \varphi^{(2)}(\boldO; \mathbb{P}) &= \frac{\mathbbm{1}(A=0)}{\PP(A>0)} \left\{ \frac{\pi_{A>0}(\boldX)}{1-\pi_{A>0}(\boldX)} \left(\Delta Y - \mu_{A=0}(\boldX) \right) \right\} + \frac{\mathbbm{1}(A>0)}{\PP(A>0)} \left \{ \mu_{A=0}(\boldX) - \Psi^{(2)}(\mathbb{P}) \right \}.
    \end{align*}
    For stochastic policies that do not depend on the observed data distribution (e.g., density function $q_{\alpha}(d)$ where $\alpha$ is a finite-dimensional parameter), $\varphi^{(1)}(\boldO; \mathbb{P})$ simplifies to
    \begin{align*}
        \varphi^{(1)}(\boldO; \mathbb{P}) &= \frac{\mathbbm{1}(A>0)}{\PP(A>0)} \bigg \{ \frac{q_{\alpha}(D)}{\pi_D(D|\boldX)} (\Delta Y - \mu_D(\boldX)) +  \mu_D(\boldX)q_{\alpha}(D) - \Psi^{(1)}(\mathbb{P}) \bigg \}.
    \end{align*}
\end{theorem}

The EIF for stochastic interventions that do not depend on the observed data distribution, such as the parametric policy discussed in Section \ref{sec:stochastic}, has a similar form to EIFs for ATT-style statistical estimands (e.g., \citet{renson_pulling_2025}) where treatment indicators are replaced with the density function of the stochastic intervention. However, when the stochastic intervention depends on nuisance functions such as in the first part of Theorem \ref{thm:eif}, additional terms are needed to adjust for estimating those nuisance parameters. 

The exponential tilt, which we focus on as the main stochastic intervention of interest for the remainder of this paper, leads to convenient cancellations and simplifies the EIF and subsequent estimator. The EIF under the exponential tilt intervention is provided in Corollary \ref{clly:eif-exptilt}. In the Appendix, the EIF and corresponding estimator for the Gaussian concentration policy are discussed.

\begin{corollary}[EIF for exponential tilt policy]
    \label{clly:eif-exptilt}
    Let $\Psi^{\mathrm{tilt}}(\mathbb{P})$ and $\Psi^{\mathrm{tilt}, 1}(\mathbb{P})$  be the estimands $\Psi(\mathbb{P})$ and $\Psi^{(1)}(\mathbb{P})$ under the exponential tilt policy, respectively. Then the EIF of $\Psi^{\mathrm{tilt}}(\mathbb{P})$ is $\varphi^{\mathrm{tilt}}(\boldO; \mathbb{P}) = \varphi^{\mathrm{tilt}, 1}(\boldO; \mathbb{P}) - \varphi^{(2)}(\boldO; \mathbb{P})$, where $\varphi^{(2)}(\boldO; \mathbb{P})$ is given in Theorem \ref{thm:eif} and
    \begin{align*}
        \varphi^{\mathrm{tilt}, 1}(\boldO; \mathbb{P}) &= \frac{\mathbbm{1}(A>0)}{\PP(A>0)} \bigg \{ \frac{q_{\delta}(D|\boldX)}{\pi_D(D|\boldX)}\left(\Delta Y - \int_{\mathcal{D}} \mu_b(\boldX) q_{\delta}(b|\boldX)\mathrm{d}b  \right) \\
        &\hspace{1em} + \int_{\mathcal{D}} \mu_b(\boldX) q_{\delta}(b|\boldX)\mathrm{d}b - \Psi^{\mathrm{tilt}, 1}(\mathbb{P}) \bigg \}.
    \end{align*}
\end{corollary}

Define the estimator $\hat \psi$ of $\Psi^{\mathrm{tilt}}(\mathbb{P})$ to be the so-called ``one-step" estimator that adjusts the plug-in estimator with an estimator of the EIF \citep{kennedy_semiparametric_2023}, i.e., 
\begin{align*}
    \hat \psi &= \hat \tau^{\mathrm{plug-in}} + \meanin \left\{ \frac{\mathbbm{1}(A_i>0)}{\E_n \{\mathbbm{1}(A>0)\}} \frac{\hat q_{\delta}(D_i|\boldX_i)}{\hat \pi_D(D_i|\boldX_i)}\left(\Delta Y_i - \int_{\mathcal{D}} \hat \mu_b(\boldX_i) \hat q_{\delta}(b|\boldX_i)\mathrm{d}b  \right)  \right\}, \\
    &\hspace{1em} - \meanin \frac{\mathbbm{1}(A_i=0)}{\E_n \{\mathbbm{1}(A=0) \hat \pi_{A>0}(\boldX) / [1 - \hat \pi_{A>0}(\boldX)] \}} \left\{ \frac{\hat \pi_{A>0}(\boldX_i)}{1 - \hat \pi_{A>0}(\boldX_i)} \left(\Delta Y_i - \hat \mu_{A=0}(\boldX_i) \right) \right\}.
\end{align*}

\noindent
There is a wide suite of nonparametric estimators for $\mu_D(\boldX)$ and $\mu_{A=0}(\boldX)$ such as random forests \citep{breiman_random_2001}, support vector machines \citep{steinwart_support_2008}, neural networks \citep{goodfellow_deep_2016}, Bayesian additive regression trees (BART) \citep{chipman_bart_2010}, and highly adaptive LASSO (HAL) \citep{benkeser_highly_2016}. However, there is relatively less work on machine learning estimators for conditional densities like the generalized propensity score $\pi_D(d|\boldX)$. One method, used in the simulations and data application below, is to approximate the conditional density function with a kernel-transformed conditional mean $\E[b^{-1} K(b^{-1}(D-d)) | \boldX=\boldx]$ where $K(\cdot)$ is a kernel function and $b$ is a user-specified bandwidth \citep{schindl_incremental_2026}. Another flexible class of methods entails transforming the problem into a discrete hazard estimation problem where estimators like HAL may be used in a pooled logistic regression \citep{hejazi_haldensify_2022}. For estimating any nuisance function, an ensemble method such as the Superlearner can be used, which outputs a weighted average of predicted values from a user-specified set of candidate estimators \citep{van_der_laan_super_2007}. 

Many machine learning estimators offer robustness to model mis-specification but may result in overfitting. Cross-fitting is often used to allow for nonparametric nuisance function estimators while preserving desirable inference properties \citep{chernozhukov_doubledebiased_2018}. Algorithm \ref{alg:cf} describes an implementation of cross-fitting when using nonparametric nuisance function estimators. The resulting proposed cross-fit estimator is denoted $\hat \psi^{\mathrm{CF}}$. 

\begin{algorithm}
    \caption{Cross-fitting algorithm} \label{alg:cf}
    \begin{algorithmic}
        \State Randomly partition the observation indices $\mathcal{N} = \{1, \dots, n\}$ into $K$ disjoint folds, $\mathcal{I}_1, \dots, \mathcal{I}_K$. 
        \State $k \gets 1$.
        \While{$k \leq K$}
            \State Estimate the nuisance functions using data indexed by $\mathcal{N} \setminus \mathcal{I}_k$. 
            \State Compute $\hat \psi^{(k)}$ using data indexed by $\mathcal{I}_k$ and nuisance function estimates from previous step.
            \State $k \gets k + 1$.
        \EndWhile
        \State $\hat \psi^{\mathrm{CF}} = \sum_{k=1}^{K} (|\mathcal{I}_k| / n) \hat \psi^{(k)}$.
    \end{algorithmic}
\end{algorithm}

To derive the large sample properties of the proposed estimator $\hat \psi^{\mathrm{CF}}$, several additional assumptions are made. First, Assumption \ref{assump:positivity-inf} asserts that the dose propensity scores $\pi_D(d|\boldx)$ are bounded away from zero in some subset of the support $\mathcal{D}$.

\begin{assumption}[Local positivity of dose]
    \label{assump:positivity-inf}
    There exists $\epsilon_{\pi} > 0$ and an interval $[r_l, r_u] \subseteq \mathcal{D}$ with positive length $r_u - r_l > 0$ such that  $\pi_D(d|\boldx) \geq \epsilon_{\pi}$ for all $d \in [r_l, r_u]$.
\end{assumption}

\noindent
In typical causal inference with nonparametric efficient estimators, positivity (as described in Assumption \ref{assump:positivity-inf}) is usually supposed to hold across the support $\mathcal{D}$ \citep{kennedy_non-parametric_2017, mcclean_comparing_2026}. In contrast, Assumption \ref{assump:positivity-inf} only asserts that the dose propensity score is bounded away from zero in some interval within the support $\mathcal{D}$. Thus, Assumption \ref{assump:positivity-inf} allows for ``holes" in $\mathcal{D}$ (e.g., $\mathcal{D} = (0, 0.2) \cup (0.5, 1)$) as well as positive, but not necessarily bounded away from zero, dose propensity scores (i.e., $\pi_D(d|\boldx) > 0$ for some $d \in \mathcal{D}$). Assumption \ref{assump:positivity-inf} is equivalent to the positivity assumption made in \citet{schindl_incremental_2026} except that \citet{schindl_incremental_2026} assumed that the interval where strong positivity held was on the boundaries of the support, i.e., $[0, r_l]$ and $[r_u, 1]$ where both intervals have positive length. Additionally, Assumption \ref{assump:bounded} assumes that the outcomes and covariates are bounded and that the dose propensity score is bounded. 

\begin{assumption}[Bounded data and propensity scores]
    \label{assump:bounded}
    The outcomes $Y$ and covariates $\boldX$ are bounded random variables. Also, there exists $\epsilon_{\pi}^{\mathrm{max}} \in (0, \infty)$ such that $\pi_D(d|\boldx) \leq \epsilon_{\pi}^{\mathrm{max}}$ for all $d \in \mathcal{D}$ and $\boldx \in \mathcal{X}$.
\end{assumption}

Next, convergence rate conditions for the nuisance function estimators are provided. For some function $f$, let $\| f \|^2$ denote the squared $L_2(\mathbb{P})$ norm, i.e., $\| f \|^2 = \int f(x)^2 \mathrm{d}\mathbb{P}(x)$. Further, define the mixed $L_2(\mathbb{P})$-sup norm as $\| f \|^2_{L^2_x, L^{\infty}_d} = \int (\sup_d | f(x, d) |)^2 \mathrm{d}\mathbb{P}(x) $ \citep{schindl_incremental_2026}. 

\begin{assumption}[Convergence rates of nuisance function estimators]
    \label{assump:conv-rates}
    \begin{align*}
        \| \hat \mu_d - \mu_d \| &= o_{\mathbb{P}}(1) \\
        \| \hat \pi_D - \pi_D \|_{L^2_x, L^{\infty}_d} \times \| \hat \mu_d - \mu_d \|_{L^2_x, L^{\infty}_d} &= o_{\mathbb{P}}(n^{-1/2}) \\
        \| \hat \pi_D - \pi_D \|^2_{L^2_x, L^{\infty}_d} &= o_{\mathbb{P}}(n^{-1/2}) \\
        \| \hat \pi_{A>0} - \pi_{A>0} \| &= o_{\mathbb{P}}(1) \\
        \| \hat \mu_{A=0} - \mu_{A=0} \| &= o_{\mathbb{P}}(1) \\
        \| \hat \pi_{A>0} - \pi_{A>0} \| \times \| \hat \mu_{A=0} - \mu_{A=0} \| &= o_{\mathbb{P}}(n^{-1/2}).
    \end{align*}
\end{assumption}

\noindent
The convergence rate conditions in Assumption \ref{assump:conv-rates} are similar to the standard conditions in the double machine learning literature \citep{kennedy_semiparametric_2023}. All nuisance function estimators are assumed to be consistent and convergence rates are assumed to be fast enough such that squares or products are $o_{\mathbb{P}}(n^{-1/2})$, i.e., individual nuisance function estimators need not converge at the parametric rate. These convergence rate conditions allow for flexible nuisance function estimators since nonparametric and machine learning estimators typically converge at a slower than $\sqrt{n}$ rate. 

Theorem \ref{thm:asymp-norm} provides the main inferential result, stating that the proposed cross-fitted estimator $\hat \psi^{\mathrm{CF}}$ is $\sqrt{n}$-CAN for the target parameter, with asymptotic variance $\sigma^2$ achieving the nonparametric efficiency bound. 

\begin{theorem}[Asymptotic normality]
    \label{thm:asymp-norm}
    Under Assumptions \ref{assump:positivity-inf} -- \ref{assump:conv-rates}, $$\sqrt{n}(\hat \psi^{\mathrm{CF}} - \Psi^{\mathrm{tilt}}(\mathbb{P})) \rightarrow_d N(0, \sigma^2), $$ where $\sigma^2 = \E[\varphi(\boldO; \mathbb{P})^2]$ is the nonparametric efficiency bound.
\end{theorem}

\noindent
Theorem \ref{thm:var-est} states that consistent estimation of $\sigma^2$ is achieved by simply using a plug-in estimator $\hat \sigma^2$. The exact expression of $\hat \sigma^2$ is provided in the Appendix. 

\begin{theorem}[Consistent variance estimation]
    \label{thm:var-est}
    Under the same assumptions as Theorem \ref{thm:asymp-norm}, \\ $\hat \sigma^2 \rightarrow_p \sigma^2$.
\end{theorem}


\section{Multiple time periods} \label{sec:mtp}

In this section, we consider the setting with multiple time periods and treatment sequences that follow a staggered adoption pattern \citep{callaway_difference--differences_2025, callaway_event_2024, callaway_difference--differences_2021}. In particular, let $A_{it}$ and $Y_{it}$ denote, respectively, the treatment and outcome of unit $i$ at time $t \in \{0,1,\dots,\mathcal{T}\}$. As before, $A_{i0} = 0$ for all units, i.e., all units are untreated at time $t=0$. Treatment sequences are assumed to follow a staggered adoption pattern in the sense that once a unit is treated at level $d \in \mathcal{D}$, that unit remains treated at that level for the remainder of the study period, i.e., $A_{is} = d$ implies $A_{it} = d$ for $s \leq t \leq \mathcal{T}$. Thus, treatment sequences can be completely characterized by the dose $D_i = A_{i \mathcal{T}} | A_{i \mathcal{T}}>0$ and timing of treatment initiation $G_i = \mathbbm{1}(A_{i\mathcal{T}} > 0) \arg \min_{t} \{ A_{it} > 0 \} + \mathbbm{1}(A_{i\mathcal{T}} = 0) \infty$, which takes values $g \in \mathcal{G} = \{1,\dots,\mathcal{T},\infty\}$ where $G_i = \infty$ denotes that unit $i$ was untreated for the duration of the study. Potential outcomes are defined to be functions of dose $d$ and treatment initiation cohort $g$, i.e., $Y_{it}(d,g)$ is the potential outcome for unit $i$ at time $t$ if (possibly counter to fact) initially exposed to dose $d$ at time $g \in \mathcal{G} \setminus \infty$, and $Y_{it}(\infty)$ is the potential outcome if (possibly counter to fact) exposed to no treatment within the study period.

The generalization of the two time period $\ASDT(Q)$ to multiple time periods is given by
\begin{align*}
    \ASDT(Q, g, t) = \E \left[ \int_{\mathcal{D}} \E[Y_{it}(d,g) - Y_{it}(\infty) | \boldX_i, G_i=g] \mathrm{d}Q(d|\boldX_i, G_i=g) \bigg| G_i=g \right],
\end{align*}
where  $t \geq g$. In words, the $\ASDT(Q, g, t)$ is the average stochastic dose effect for policy $Q$, among units in treatment initiation cohort $g$, and at calendar time $t$. Let treatment duration $e = t-g$ denote the number of time periods since treatment initiation. In staggered adoption designs, there is often interest in parameters that aggregate over treatment cohorts $g$, treatment duration $e$ or both. Below, several aggregated parameters are defined, which are similar to aggregated ATT-style parameters considered by \citet{callaway_difference--differences_2021} in the staggered adoption setting with binary treatments. 

Event study parameters summarize effects over treatment initiation cohorts and describe variation by treatment duration; let $\ASDT^{\mathrm{es}}(Q,e)$ denote the following event study parameter,
\begin{align*}
    \ASDT^{\mathrm{es}}(Q,e) &= \E[ \ASDT(Q,G,G+e) | G+e \in [1,\mathcal{T}]] \\
    &= \sum_{g=1}^{\mathcal{T}} \PP(G=g | G+e \in [1,\mathcal{T}]) \ASDT(Q,g,g+e).
\end{align*}
As discussed in \citet{callaway_difference--differences_2021}, $\ASDT^{\mathrm{es}}(Q,e_2) - \ASDT^{\mathrm{es}}(Q,e_1)$ for $0 \leq e_1 < e_2 \leq \mathcal{T}-1$ combines comparisons of time dynamic effects $\ASDT(Q,g,g+e_2) - \ASDT(Q,g,g+e_1)$ and composition effects due to differences in which cohorts are included in $\ASDT^{\mathrm{es}}(Q,e_2)$ versus $\ASDT^{\mathrm{es}}(Q,e_1)$. Instead, there may be interest in the ``balanced" event study parameter,
\begin{align*}
    \ASDT^{\mathrm{es, bal}}(Q,e; e') &= \sum_{g=1}^{\mathcal{T}} \PP(G=g | G+e' \in [1,\mathcal{T}]) \ASDT(Q,g,g+e),
\end{align*}
where $0 \leq e \leq e' \leq \mathcal{T}-1$ so that $\ASDT^{\mathrm{es, bal}}(Q,e_2; e') - \ASDT^{\mathrm{es, bal}}(Q,e_1; e')$ is equal to a weighted average of $\ASDT(Q,g,g+e_2) - \ASDT(Q,g,g+e_1)$, highlighting only time dynamic effects and not composition effects. 

The cohort parameter $\ASDT^{\mathrm{cohort}}(Q,g; e_1, e_2)$ aggregates over treatment duration in order to highlight potential heterogeneity by timing of treatment initiation, which may be of interest if, for instance, effects are anticipated to depend on time-varying macroeconomic factors. The cohort parameter is defined as,
\begin{align*}
    \ASDT^{\mathrm{cohort}}(Q,g; e_1, e_2) &= \sum_{t=g+e_1}^{g+e_2} \frac{1}{e_2 - e_1 + 1} \ASDT(Q,g,t),
\end{align*}
where $e_2 \geq e_1$ and $g + e_2 \leq \mathcal{T}$. 

The aggregated parameter $\ASDT^{\mathrm{overall}}(Q)$ may be of interest to highlight variation due to different dose distributions $Q$ (e.g., different increments $\delta$ for the exponential tilt), 
\begin{align*}
    \ASDT^{\mathrm{overall}}(Q) &= \sum_{g=1}^{\mathcal{T}} \sum_{t=1}^{\mathcal{T}} \frac{\mathbbm{1}(t \geq g)}{\mathcal{T} - g + 1} \PP(G=g | G \leq \mathcal{T}) \ASDT(Q, g, t).
\end{align*}
Different aggregations of $\ASDT(Q, g, t)$ across treatment cohorts and time exist but $\ASDT^{\mathrm{overall}}(Q)$ is displayed here since it avoids difficulties in interpretation due to composition effects that may arise when aggregating over parameters such as $\ASDT^{\mathrm{es}}(Q,e)$, for reasons discussed above \citep{callaway_difference--differences_2021}. 

Identification assumptions and results for parameters in the multiple time periods setting are similar to the two time period setting. The observed data with multiple time periods is $\boldO = \{\boldO_i\}_{i=1}^{n}$ where $\boldO_i = (Y_{i0}, \dots, Y_{i\mathcal{T}}, G_i, D_i, \boldX_i)$ is assumed to be an iid draw from the distribution $\mathbb{P}$. Each aggregated parameter can be written as $\sum_{g=1}^{\mathcal{T}} \sum_{t=1}^{\mathcal{T}} \omega(g,t) \ASDT(Q, g, t)$ where $\omega(g,t)$ is an identified quantity that varies by choice of aggregation and may depend on $\boldO$. Thus, for identification of all causal parameters discussed in this section, it suffices to show that $\ASDT(Q, g, t)$ is identifiable. 

Assumption \ref{assump:ident-mtp} states identification assumptions. Assumptions \ref{assump:ident-mtp}(a) and \ref{assump:ident-mtp}(b) are multiple time period analogues to Assumptions \ref{assump:consistency} and \ref{assump:no_anticip}, respectively. Together, Assumptions \ref{assump:ident-mtp}(a) and \ref{assump:ident-mtp}(b) relate observed outcomes to potential outcomes by $Y_{it} = Y_{it}(\infty) \mathbbm{1}(t < G_i) + Y_{it}(D_i,G_i)\mathbbm{1}(t \geq G_i)$. Assumption \ref{assump:ident-mtp}(c) is a positivity assumption where, conditional on covariates, treatment initiation has a positive probability of occurring during every cohort, including never-treatment. Additionally, the generalized propensity score $\pi^{\mathrm{MTP}}_{D,g}(d|\boldx)$, the dose density function conditional on covariates $\boldX=\boldx$ and being in treatment cohort $G=g$, is positive for every dose. For identification, the latter positivity assumption may be weakened as in Assumption \ref{assump:positivity-ident-c} to restrict positivity of the counterfactual dose distribution in regions where positivity is satisfied for the generalized propensity score $\pi^{\mathrm{MTP}}_{D,g}(d|\boldx)$. Finally, Assumption \ref{assump:ident-mtp}(d) is a set of conditional parallel trends assumption analogous to Assumptions \ref{assump:parallel-cond-a0} and \ref{assump:parallel-cond-d}.  

\begin{assumption}[Identification assumptions for multiple time periods] ~
\label{assump:ident-mtp}
    \begin{outline}
        \1[(a)] Causal consistency. If $G_i = g$ for $g \in \mathcal{G} \setminus \infty$ and $D_i = d$, then $Y_{it} = Y_{it}(g,d)$ for any $t$. If $G_i = \infty$, then $Y_{it} = Y_{it}(\infty)$ for any $t$. 
        \1[(b)] No anticipation. For all $t$ and $g$ such that $t < g$, $Y_{it}(d,g) = Y_{it}(\infty)$ for any $d$. 
        \1[(c)] Positivity. For some $0 < \epsilon < 1/2$, $\epsilon < \PP(G_i = g | \boldX_i) < 1 - \epsilon$ and $\pi^{\mathrm{MTP}}_{D,g}(d|\boldX_i) > \epsilon$ for all $g \in \mathcal{G} \setminus \infty$ and $d \in \mathcal{D}$. 
        \1[(d)] Conditional parallel trends. For $t \geq 1$, $\E[Y_{it}(\infty) - Y_{i,t-1}(\infty) | \boldX_i, G_i = g] = \E[Y_{it}(\infty) - Y_{i,t-1}(\infty) | \boldX_i, G_i > t]$. For $1 \leq g \leq t$ and all doses $d \in \mathcal{D}$, $\E[Y_{it}(d,g) - Y_{i,t-1}(d,g) | \boldX_i, G_i = g, D_i=d] = \E[Y_{it}(d,g) - Y_{i,t-1}(d,g) | \boldX_i, G_i = g]$.
    \end{outline}
\end{assumption}

\noindent
Theorem \ref{thm:identification-mtp} provides an identification result, where $\mu^{\mathrm{MTP}}_{d,g,t}(\boldx) = \E[Y_{it} - Y_{i,g-1} | \boldX_i=\boldx, G_i=g, D_i=d]$ is the expected outcome trend conditional on covariates and being in treatment cohort $g$ with dose $d$, and $\mu^{\mathrm{MTP}}_{0,g,t}(\boldx) = \E[Y_{it} - Y_{i,g-1} | \boldX_i=\boldx, G_i>t]$ is similarly the expected outcome trend conditional on covariates and being in the not-yet-treated group. Herein, let $Q_{\delta}(d|\boldx, G=g)$ be the distribution function imposed by the exponential tilt of the observed dose propensity score $\pi^{\mathrm{MTP}}_{D,g}(d|\boldx)$ with increment parameter $\delta$, and let $\ASDT(\delta, g, t)$ be the corresponding estimand.

\begin{theorem}[Identification with multiple time periods]
    \label{thm:identification-mtp}
    Under Assumption \ref{assump:ident-mtp},
    \begin{align*}
        \ASDT(\delta, g, t) &= \Psi^{\mathrm{MTP}}(\delta, g, t) \\
        &\equiv \E \left[ \int_{\mathcal{D}} \mu^{\mathrm{MTP}}_{d,g,t}(\boldX) \mathrm{d}Q_{\delta}(d|\boldX, G=g) \bigg| G=g \right] - \E[ \mu^{\mathrm{MTP}}_{0,g,t}(\boldX) | G=g].
    \end{align*}
\end{theorem}

The one-step EIF-based estimator of $\ASDT(\delta, g, t)$ follows a similar form as $\hat \psi$,
\begin{align*}
    &\hat \psi^{\mathrm{MTP}}(\delta,g,t) = \meanin \frac{\mathbbm{1}(G_i=g)}{\E_n G} \int_{\mathcal{D}} \hat \mu^{\mathrm{MTP}}_{b,g,t}(\boldX_i) \hat q_{\delta}(b|\boldX_i, G_i=g) \mathrm{d}b  \\
    &\hspace{1em} + \meanin \left\{ \frac{\mathbbm{1}(G_i=g)}{\E_n G} \frac{\hat q_{\delta}(D_i|\boldX_i, G_i=g)}{\hat \pi^{\mathrm{MTP}}_{D,g}(D_i|\boldX_i)}\left(Y_{it} - Y_{i,g-1} - \int_{\mathcal{D}} \hat \mu^{\mathrm{MTP}}_{b,g,t}(\boldX_i) \hat q_{\delta}(b|\boldX_i, G_i=g)\mathrm{d}b  \right)  \right\} \\
    &\hspace{1em} - \meanin \frac{ \mathbbm{1}(G_i=g)}{\E_n G} \hat \mu^{\mathrm{MTP}}_{0,g,t}(\boldX_i) \\
    &\hspace{1em} - \meanin \frac{\mathbbm{1}(G_i>t)}{\E_n \{\mathbbm{1}(G>t) \hat \pi^{\mathrm{MTP}}_{G=g,t}(\boldX) / [1 - \hat \pi^{\mathrm{MTP}}_{G=g,t}(\boldX)] \}} \left\{ \frac{\hat \pi^{\mathrm{MTP}}_{G=g,t}(\boldX_i)}{1 - \hat \pi^{\mathrm{MTP}}_{G=g,t}(\boldX_i)} \left(Y_{it} - Y_{i,g-1} - \hat \mu^{\mathrm{MTP}}_{0,g,t}(\boldX_i) \right) \right\},
\end{align*}
where $\hat \pi^{\mathrm{MTP}}_{G=g,t}(\boldx)$ is an estimator of $\PP(G_i = g | \boldX_i = \boldx, \mathbbm{1}(G_i > t) + \mathbbm{1}(G_i=g) = 1)$, the probability of being in treatment cohort $g$ in the set of units either untreated at time $t$ or which initiated treatment in cohort $g$, conditional on covariates. As before, Algorithm \ref{alg:cf} may be used to construct a cross-fit estimator $\hat \psi^{\mathrm{MTP, CF}}(\delta,g,t)$.  

Then, similar to Theorem \ref{thm:asymp-norm}, $\hat \psi^{\mathrm{MTP, CF}}(\delta,g,t)$ is a $\sqrt{n}$-CAN estimator with variance attaining the nonparametric efficiency bound in large samples, as shown in the first part of Theorem \ref{thm:asymp-norm-mtp}. For brevity, the conditions in Theorem \ref{thm:asymp-norm-mtp} and the expression for $\varphi_{\delta,g,t}^{\mathrm{MTP}}(\boldO)$, the EIF of $\Psi^{\mathrm{MTP}}(\delta,g,t)$, are left to the Appendix. The second part of Theorem \ref{thm:asymp-norm-mtp} shows the joint asymptotic normality of $\hat \psi^{\mathrm{MTP, CF}}(\delta,g,t)$ for fixed $\delta$ and varying $g$ and $t$. Let $\hat \psi^{\mathrm{MTP, CF}}_{t \geq g}(\delta)$ and $\Psi^{\mathrm{MTP}}_{t \geq g}(\delta)$ be vectors consisting of $\hat \psi^{\mathrm{MTP}}(\delta,g,t)$ and $\Psi^{\mathrm{MTP}}(\delta,g,t)$, respectively, for $1 \leq t \leq \mathcal{T}$, $g \in \mathcal{G} \setminus \infty$, and $t \geq g$. Similarly, $\varphi^{\mathrm{MTP}}_{t \geq g}(\delta; \boldO)$ is the corresponding vector of EIFs. 

\begin{theorem}[Asymptotic normality]
\label{thm:asymp-norm-mtp}
    Suppose Assumption \ref{assump:ident-mtp} and Assumptions \ref{assump:bounded-mtp} -- \ref{assump:conv-rates-mtp} in the Appendix hold. Then, as $n \rightarrow \infty$,
    \begin{align*}
        \sqrt{n}(\hat \psi^{\mathrm{MTP, CF}}(\delta,g,t) - \Psi^{\mathrm{MTP}}(\delta, g, t)) \rightarrow_d N(0, \sigma_{\mathrm{MTP}}^2),
    \end{align*}
    where $\sigma_{\mathrm{MTP}}^2 = \E[\varphi_{\delta,g,t}^{\mathrm{MTP}}(\boldO)^2]$ is the nonparametric efficiency bound. 
    
    Furthermore, as $n \rightarrow \infty$
    \begin{align*}
        \sqrt{n}(\hat \psi^{\mathrm{MTP, CF}}_{t \geq g}(\delta) - \Psi^{\mathrm{MTP}}_{t \geq g}(\delta)) \rightarrow_d N(0, \Sigma^{\mathrm{MTP}}),
    \end{align*}
    where $\Sigma^{\mathrm{MTP}} = \E[\varphi^{\mathrm{MTP}}_{t \geq g}(\delta; \boldO) \varphi^{\mathrm{MTP}}_{t \geq g}(\delta; \boldO)^{\top}] $. 
\end{theorem}

Theorem \ref{thm:asymp-norm-mtp} can be used to derive the asymptotic distribution of estimators \\
$\sum_{g=1}^{\mathcal{T}} \sum_{t=1}^{\mathcal{T}} \hat \omega(g,t) \hat \psi^{\mathrm{MTP, CF}}(\delta,g,t)$ of the aggregated parameters $\sum_{g=1}^{\mathcal{T}} \sum_{t=1}^{\mathcal{T}} \omega(g,t) \psi^{\mathrm{MTP, CF}}(\delta,g,t)$, where $\hat \omega(g,t)$ is an estimator of $\omega(g,t)$. Theorem \ref{thm:asymp-norm-agg} presents these results. For the aggregated parameters discussed earlier in this section, sample average plug-in estimators  $\hat \omega(g,t)$ fulfill the asymptotic linearity condition in Theorem \ref{thm:asymp-norm-agg}. 

\begin{theorem}[Asymptotic normality for aggregated parameters]
\label{thm:asymp-norm-agg}
    Assume that $\hat \omega(g,t)$ admits an asymptotic linear representation, i.e.,
    \begin{align*}
        \sqrt{n}(\hat \omega(g,t) - \omega(g,t)) = n^{-1/2} \sum_{i=1}^{n} \xi_{g,t}(\boldO_i) + o_{\mathbb{P}}(1),
    \end{align*}
    where $\xi_{g,t}(\boldO)$ is a mean-zero influence function with finite, positive definite variance. Then, as $n \rightarrow \infty$,
    \begin{align*}
        \sqrt{n}\left( \sum_{g=1}^{\mathcal{T}} \sum_{t=1}^{\mathcal{T}} \{ \hat \omega(g,t) \hat \psi^{\mathrm{MTP, CF}}(\delta,g,t) - \omega(g,t) \Psi^{\mathrm{MTP}}(\delta,g,t) \} \right) \rightarrow_d N(0, \sigma_{\mathrm{MTP, agg}}^2),
    \end{align*}
    where $\sigma_{\mathrm{MTP, agg}}^2 = \E \left[ \left( \sum_{g=1}^{\mathcal{T}} \sum_{t=1}^{\mathcal{T}} \{ \omega(g,t) \varphi_{\delta,g,t}^{\mathrm{MTP}}(\boldO) + \xi_{g,t}(\boldO) \Psi^{\mathrm{MTP}}(\delta,g,t) \} \right)^2 \right]$. Furthermore, the plug-in variance estimator $$\hat \sigma_{\mathrm{MTP, agg}}^2 = \meanin \left( \sum_{g=1}^{\mathcal{T}} \sum_{t=1}^{\mathcal{T}} \{ \hat \omega(g,t) \hat \varphi_{\delta,g,t}^{\mathrm{MTP}}(\boldO_i) + \hat \xi_{g,t}(\boldO_i) \hat \psi^{\mathrm{MTP, CF}}(\delta,g,t) \} \right)^2$$ is consistent for $\sigma_{\mathrm{MTP, agg}}^2$.
\end{theorem}

\section{Simulation} \label{sec:sim}

Simulation experiments were performed to assess finite sample performance of the proposed method. Simulation datasets were created according to the following data generating process: 
\begin{align*}
    \boldX = (X_{1},\dots,X_{10}) &\overset{iid}{\sim} \mathrm{Unif}(-1,1) \\
    A>0 | \boldX &\overset{iid}{\sim} \mathrm{Bernoulli}(\pi_{A>0}(\boldX)) \\
    D | \boldX, A>0 &\overset{iid}{\sim} \mathrm{Beta}(f^D_1(\boldX), f^D_2(\boldX)) \\
    \Delta Y | A, \boldX &\overset{iid}{\sim} N(\mu(A,\boldX), 1),
\end{align*}
where $\pi_{A>0}(\boldX) = 0.5\ \mathrm{logit}^{-1}\{-0.5 + 0.2\cos X_1^2 - 0.2 \exp(X_2) - 0.5 X_3^2 \} + 0.3\mathbbm{1}(X_4 < 0) | X_5|$, $\mu(A,\boldX) = \mathbbm{1}(A=0) f^{\mathrm{out}}(\boldX; \gamma) + \mathbbm{1}(A>0) (2D - D^2 + f^{\mathrm{out}}(\boldX; \gamma))$, and $f^{\mathrm{out}}(\boldX; \gamma) = [\mathrm{logit}^{-1}(X_1)]^2 + X_2 X_3 + X_4 - 2X_4^2 + 3 \sin(X_5^2)$. Two simulation scenarios were considered that differed in choice of dose distribution. In Scenario 1, $f^D_1(\boldX) = f^D_2(\boldX) = \exp(\boldX \lambda)$ where $\lambda = (-0.2,-0.156,\dots,0.2)$, which specifies a symmetric dose density function that concentrates density towards the middle of the support when $f^D_1(\boldX) > 1$ and concentrates density towards the edges of the support when $f^D_1(\boldX) < 1$. In Scenario 2, $f^D_1(\boldX) = 2$ and $f^D_2(\boldX) = 1 + 4 \mathrm{logit}^{-1}\left\{\sum_{j=1}^{10} (-0.11 + 0.11j)X_j \right\}$, which specifies a right-skewed density function when $f^D_2(\boldX) < 2$ and a left-skewed density function when $f^D_2(\boldX) > 2$. 

\begin{figure}[!h]
    \centering
    \begin{subfigure}{0.7\textwidth}
        \centering
        \includegraphics[width=1\textwidth]{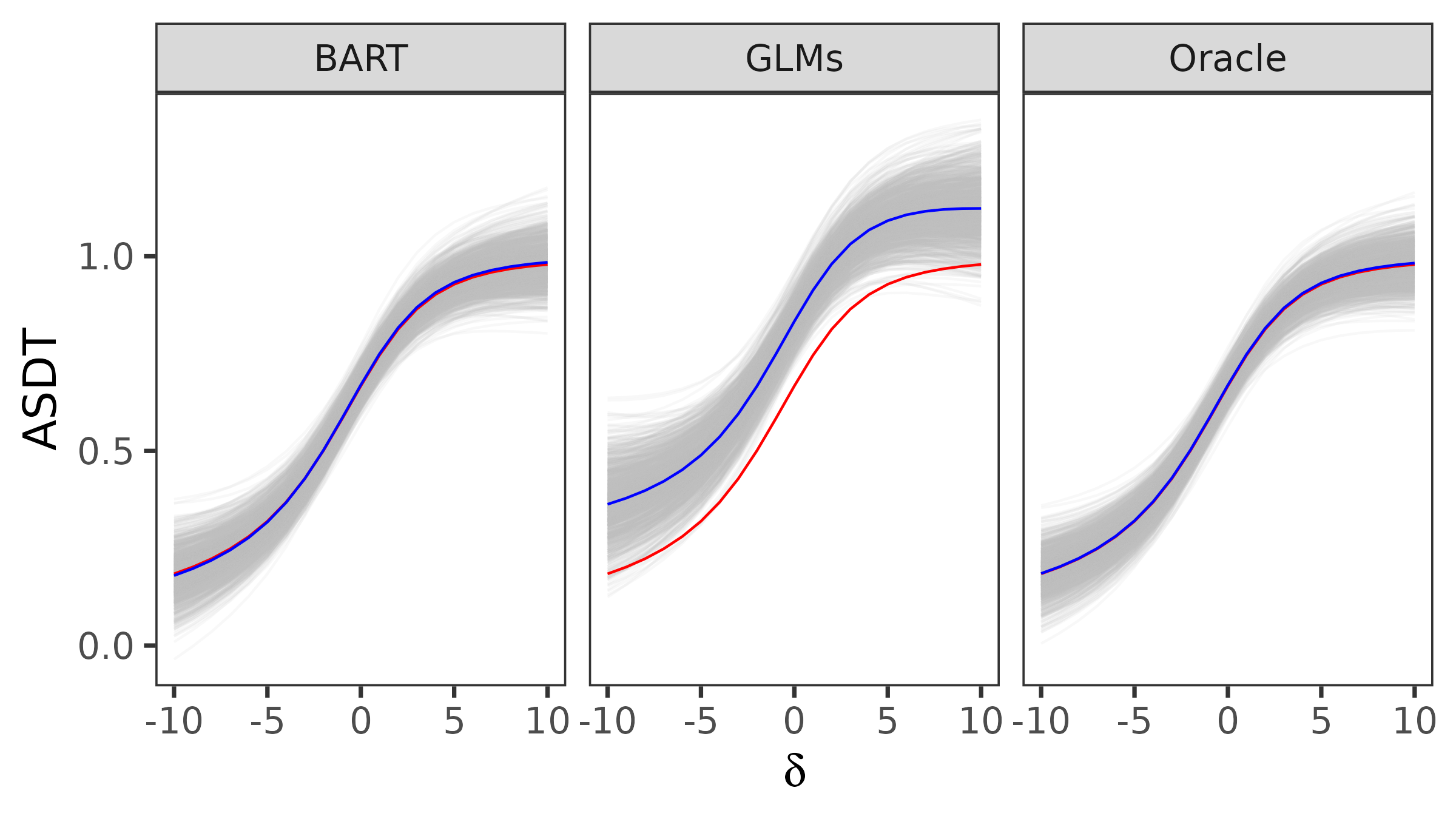}
        \caption{Scenario 1}
    \end{subfigure}
    \par \vspace{0.7cm}
    \begin{subfigure}{0.7\textwidth}
        \centering
        \includegraphics[width=1\textwidth]{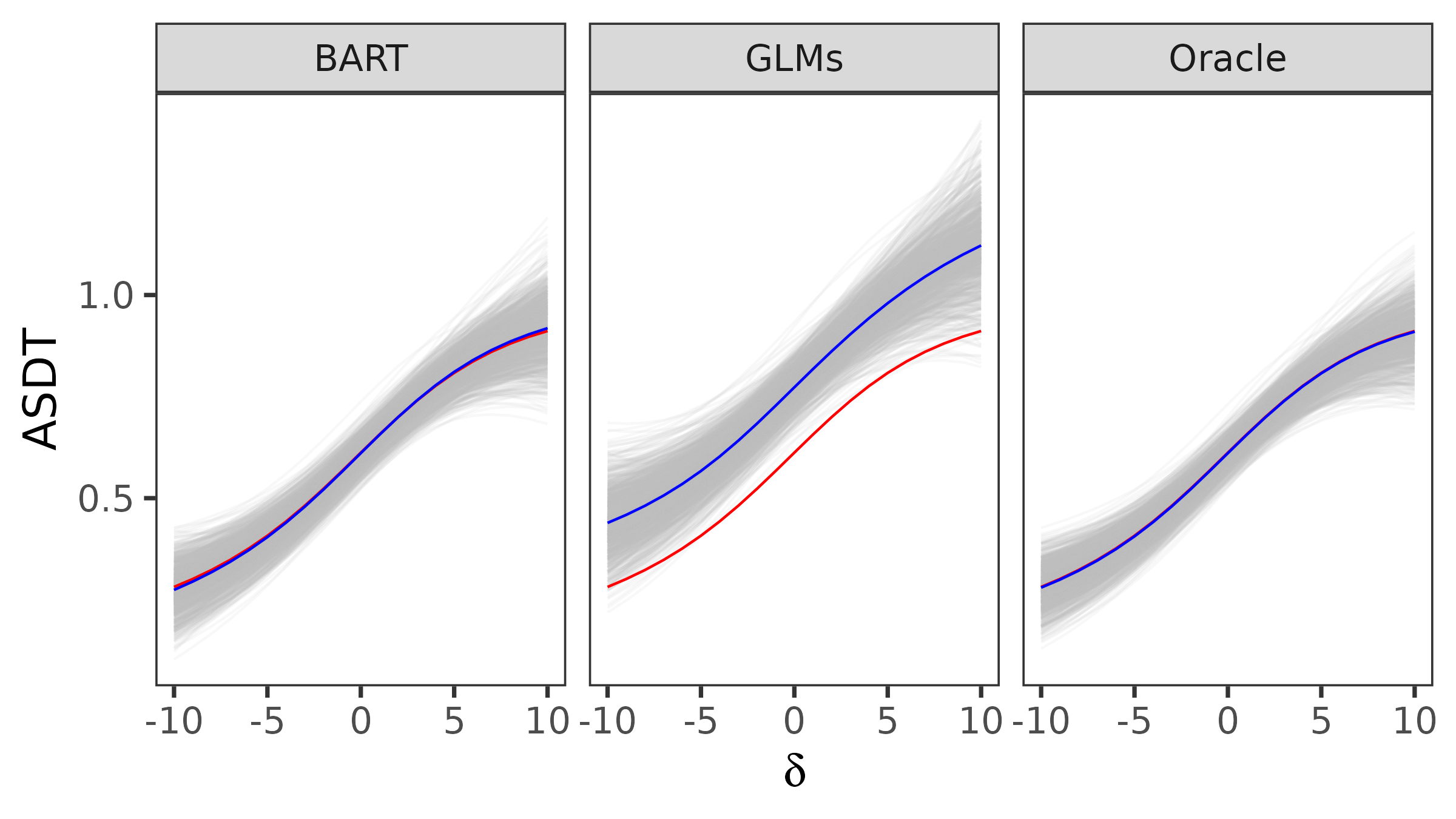}
        \caption{Scenario 2}
    \end{subfigure}
    \caption{Estimated ASDT under the exponential tilt stochastic policy with varying increments $\delta$ from 1000 simulations. Grey lines denote estimates from a single simulation; the blue line is the average of the grey lines; and the red line is the estimand. Facet labels refer to the estimator of all nuisance functions aside from the generalized dose propensity score.}
    \label{fig:sim-ests}
\end{figure}

In both scenarios the sample size was set to $n = 5000$ and $1000$ simulation datasets were generated. The proposed cross-fit estimator $\hat \psi^{\mathrm{CF}}$ with $K=5$ folds was used to estimate $\Psi^{\mathrm{tilt}}(\mathbb{P})$ with increments $\delta = (-10,-9,\dots,10)$. The conditional mean $\mu(A,\boldX)$ and dichotomized propensity score $\pi_{A>0}(\boldX)$ were estimated using BART, generalized linear models (GLMs) with maximum likelihood estimation, and an oracle estimator equal to the true nuisance functions. In all cases, the generalized propensity score $\pi_D(\boldX)$ was estimated using the kernel-transformed conditional mean method with linear models as discussed in Section \ref{sec:estimation}. 

\begin{figure}[!h]
    \centering
    \begin{subfigure}{0.7\textwidth}
        \centering
        \includegraphics[width=1\textwidth]{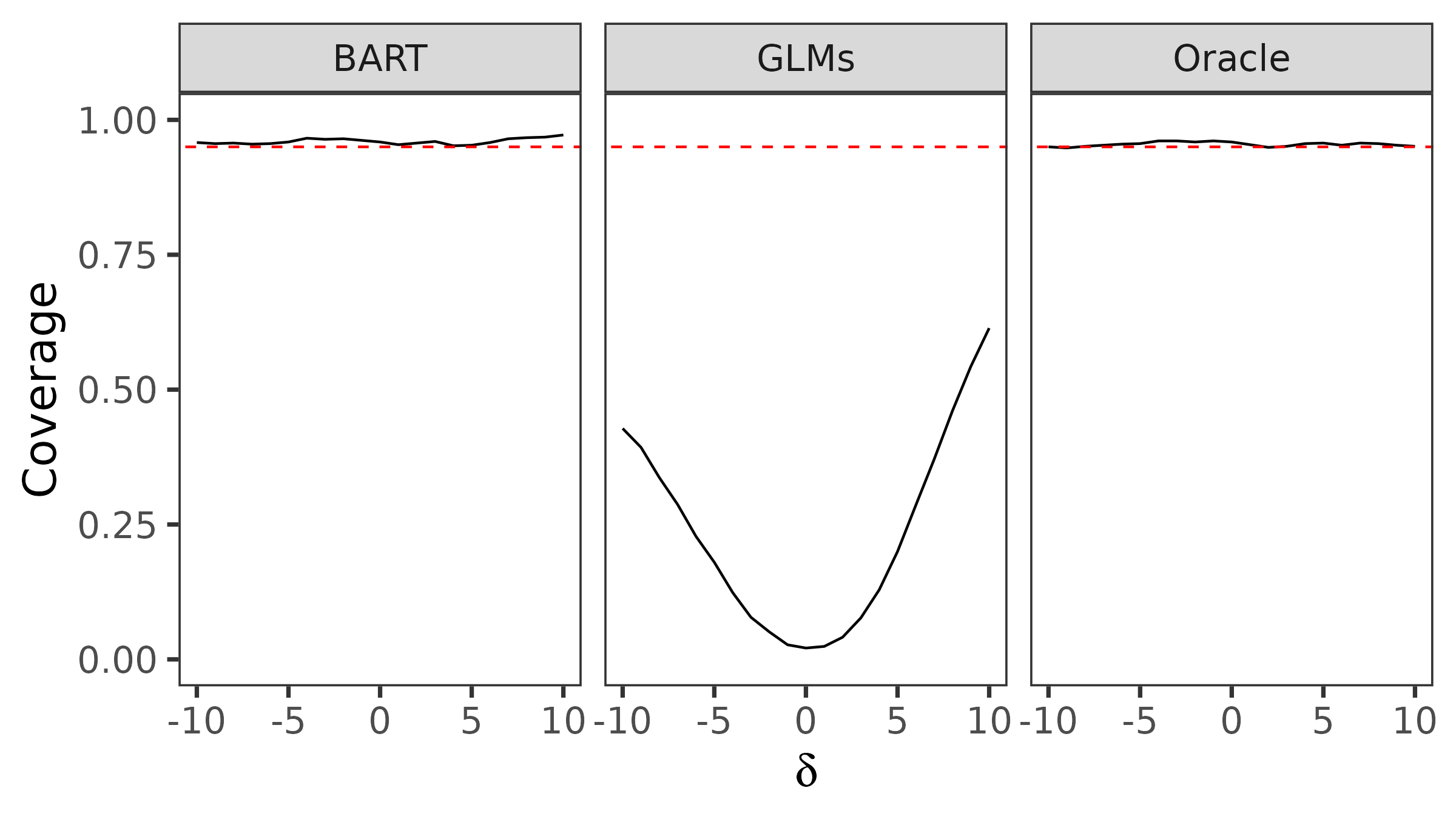}
        \caption{Scenario 1}
    \end{subfigure}
    \par \vspace{0.7cm}
    \begin{subfigure}{0.7\textwidth}
        \centering
        \includegraphics[width=1\textwidth]{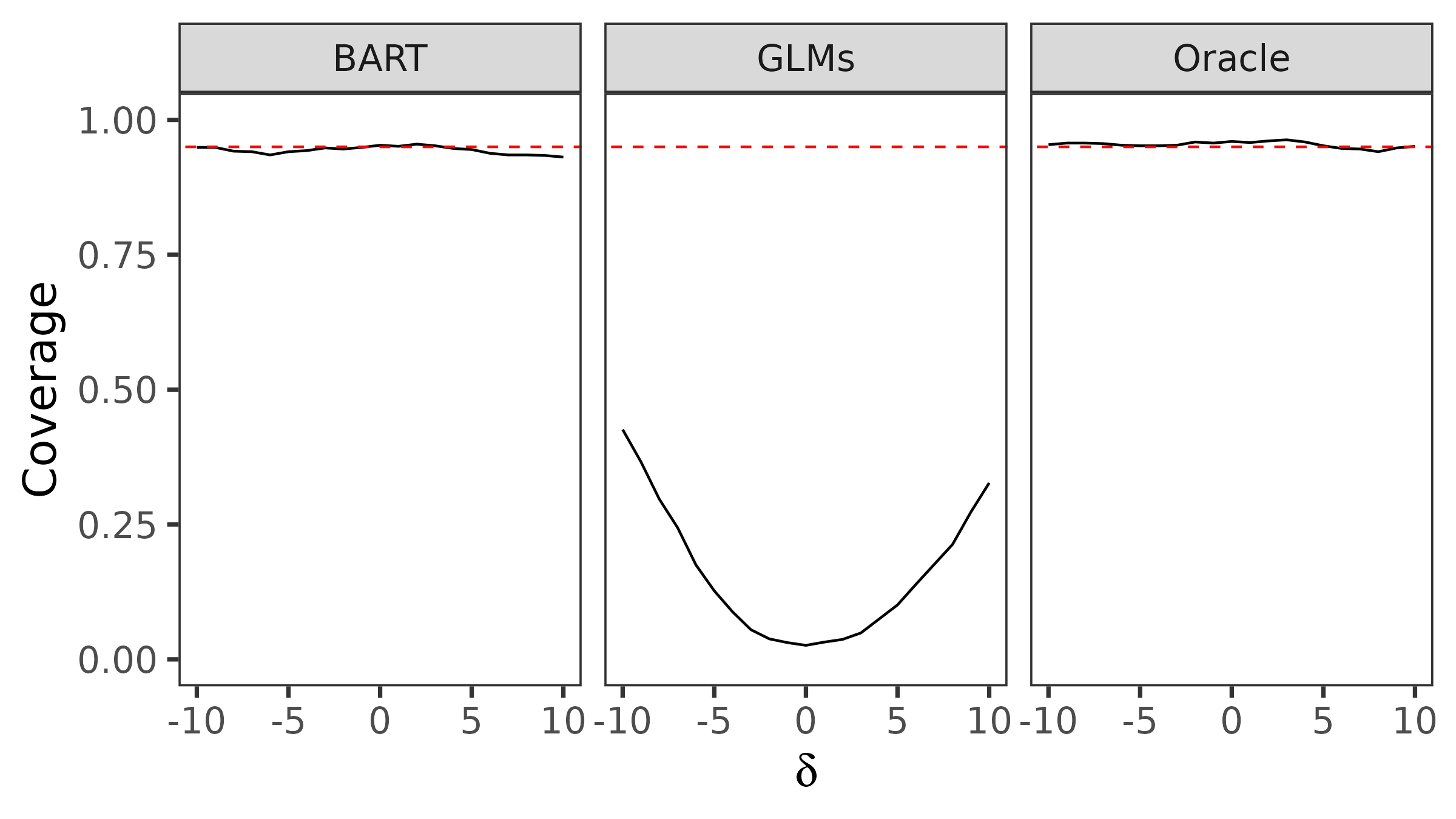}
        \caption{Scenario 2}
    \end{subfigure}
    \caption{Empirical coverage rate of 95\% confidence intervals from 1000 simulations. The black line shows the proportion of confidence intervals that contained the true estimand for each $\delta$ and the red dashed line is set at $0.95$. Facet labels refer to the estimator of all nuisance functions aside from the generalized dose propensity score.}
    \label{fig:sim-cov}
\end{figure}

The results are presented in Figure \ref{fig:sim-ests} where the grey lines represent estimates for each simulated dataset, the blue line is the average of the grey lines, and the red line represents the true target parameter. In both scenarios there was negligible average bias for all values of $\delta$ when BART was used to estimate $\mu(A,\boldX)$ and $\pi_{A>0}(\boldX)$, and performance matched that of the oracle estimator. However, when mis-specified GLMs were used to model the nuisance functions, there was substantial bias. Figure \ref{fig:sim-cov} displays the empirical coverage rate of Wald-like 95\% confidence intervals for each $\delta$ for both scenarios. Similar to the point estimates, using BART (or oracle estimators) to estimate $\mu(A,\boldX)$ and $\pi_{A>0}(\boldX)$ resulted in approximately nominal coverage for all values of $\delta$. Due to the bias, using GLM estimators resulted in poor coverage. A table with summary statistics of the simulation results is presented in the Appendix.

\section{Economic effects of hydraulic fracturing activity} \label{sec:application}

Hydraulic fracturing (``fracking") is an unconventional drilling technique used to increase the permeability of shale, allowing for extraction of oil and natural gas. Fracking involves drilling into shale formations (``plays") and pumping pressurized fracking fluid, a mixture of water and proprietary substances, to create cracks in the rock and stimulate the well. In the United States, the development of fracking has had substantial political and economic effects by increasing domestic energy production. However, fracking is controversial due to potential negative effects such as increased seismic activity, air and water pollution, and health risks \citep{black_economic_2021}. 

\begin{figure}[!h]
    \centering
    \includegraphics[width=1\linewidth]{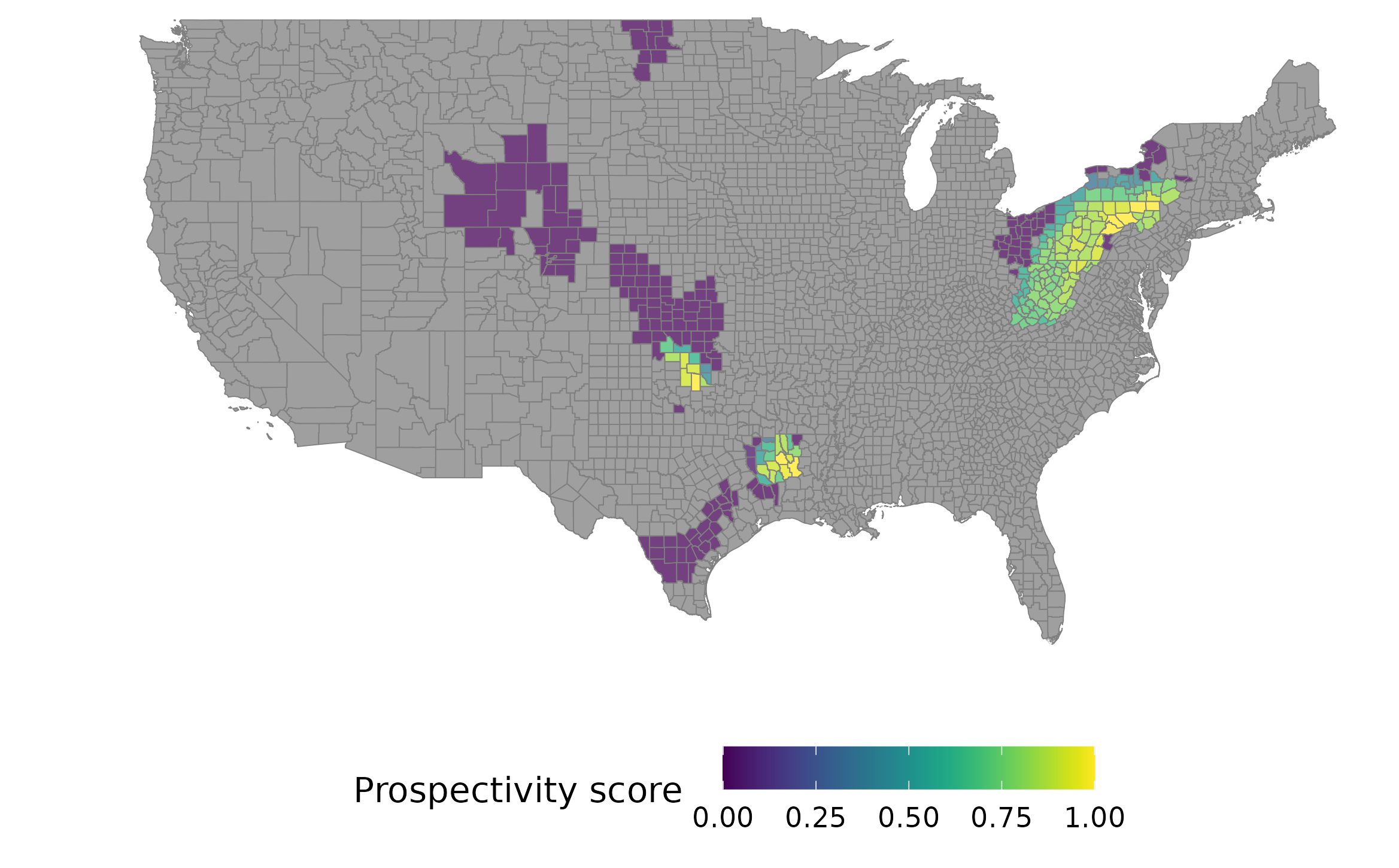}
    \caption{Map of prospectivity scores in 2008 where values of $0$ denote that fracking was not yet initiated. Grey areas represent counties that did not overlap with a shale play.}
    \label{fig:prosp_map_2008}
\end{figure}

\citet{bartik_local_2019} studied the local economic and welfare consequences of fracking in the United States from 1990 to 2012 using difference-in-differences to estimate deterministic causal effects. A number of outcomes, including employment, income, and housing prices, were considered, and county-level effects due to potential fracking activity were estimated. \citet{bartik_local_2019} purchased a ``prospectivity" index from Rystad Energy, an oil and gas consulting firm, which measures the potential productivity of different shale plays. \citet{bartik_local_2019} mapped these prospectivity scores to the county-level by taking the maximum score within each county. The continuous prospectivity score was then dichotomized with the top quartile of counties serving as the treated group and all other counties with fracking activity as the comparison group. \citet{callaway_event_2024} performed a re-analysis of (deterministic) employment effects but treated the prospectivity score as continuous. 

In this section, results are presented of an analysis using the proposed methods to examine how different distributions of potential fracking productivity affect economic outcomes using the county-level data provided by \citet{bartik_local_2019}. Specifically, the treatment was the mixture of the continuous prospectivity score and a binary indicator where zero indicates that no fracking occurred. Additionally, the original prospectivity scores were divided by the maximum within each shale play since prospectivity scores may not be exactly comparable between different shale plays. The outcomes were total employment (number of jobs) and total annual income (thousands of dollars), both on the log scale. Multiple time periods were observed and the treatment followed the staggered adoption pattern described in Section \ref{sec:mtp}. The analysis was restricted to the treatment initiation cohorts with relatively large sample sizes, 2008 and 2009, and the study period extended to 2012. There were 312 counties that overlapped with a shale play in 2008, where 156 were treated. In 2009, there were 85 counties that initiated treatment out of the 166 that were not treated in the previous year. Figure \ref{fig:prosp_map_2008} displays the geographical distribution of prospectivity scores in 2008. The Appendix contains further details on the prospectivity score distribution, sample sizes, and summary statistics.

County-level demographic covariates were gathered from the 2005-2009 5-year American Community Survey (ACS) \citep{us_census_bureau_american_2009}: total population (log), race, sex, and age. Data from the 2005-2009 ACS was used in order to most accurately capture demographic information at the timing of fracking initiation in 2008 or 2009. Alternatively, data from the 2000 Census could be used, though this data source would not capture changing demographics between 2000 and 2008 so we prefer the data from the 2005-2009 ACS.

Parallel trends were assumed conditional on the demographic covariates. In contrast, \citet{bartik_local_2019} and \citet{callaway_event_2024} assumed unconditional parallel trends. Though conditional and unconditional parallel trends assumptions are not nested, the former is arguably more plausible than the latter since economic growth may vary by the interaction between fracking initiation and county demographics. Furthermore, while estimating deterministic effects under conditional parallel trends typically does \textit{not} allow for positivity violations in the conditional dose prospectivity score distribution, the estimation of stochastic policy effects \textit{does} allow for such positivity violations. 

The ASDT was estimated yearly from treatment initiation to 2012, where the stochastic policy shift was the exponential tilt of the conditional prospectivity score distribution for increments $\delta \in \{-10,-9,\dots,10\}$. Additionally, a placebo test was performed where the ASDT was estimated yearly for three years prior to the reference outcome year (the treatment initiation year minus one), which is expected to be near zero provided the causal identification assumptions hold. The kernel-transformed outcome method with linear models was used to estimate the generalized propensity score. The other nuisance functions (expected conditional outcome trends and binary propensity score) were estimated using a Superlearner with generalized linear models, BART, and HAL as candidate prediction algorithms. The ASDT was also estimated under the Gaussian concentration policy with length parameter $l = 0.1$ (representing considerable concentration; see e.g., Figure \ref{fig:gauss-kernel}) and multiple concentration points $d' \in [0.5, 0.99]$; see the Appendix for the results. 

\begin{figure}[!h]
    \centering
    \includegraphics[width=1\linewidth]{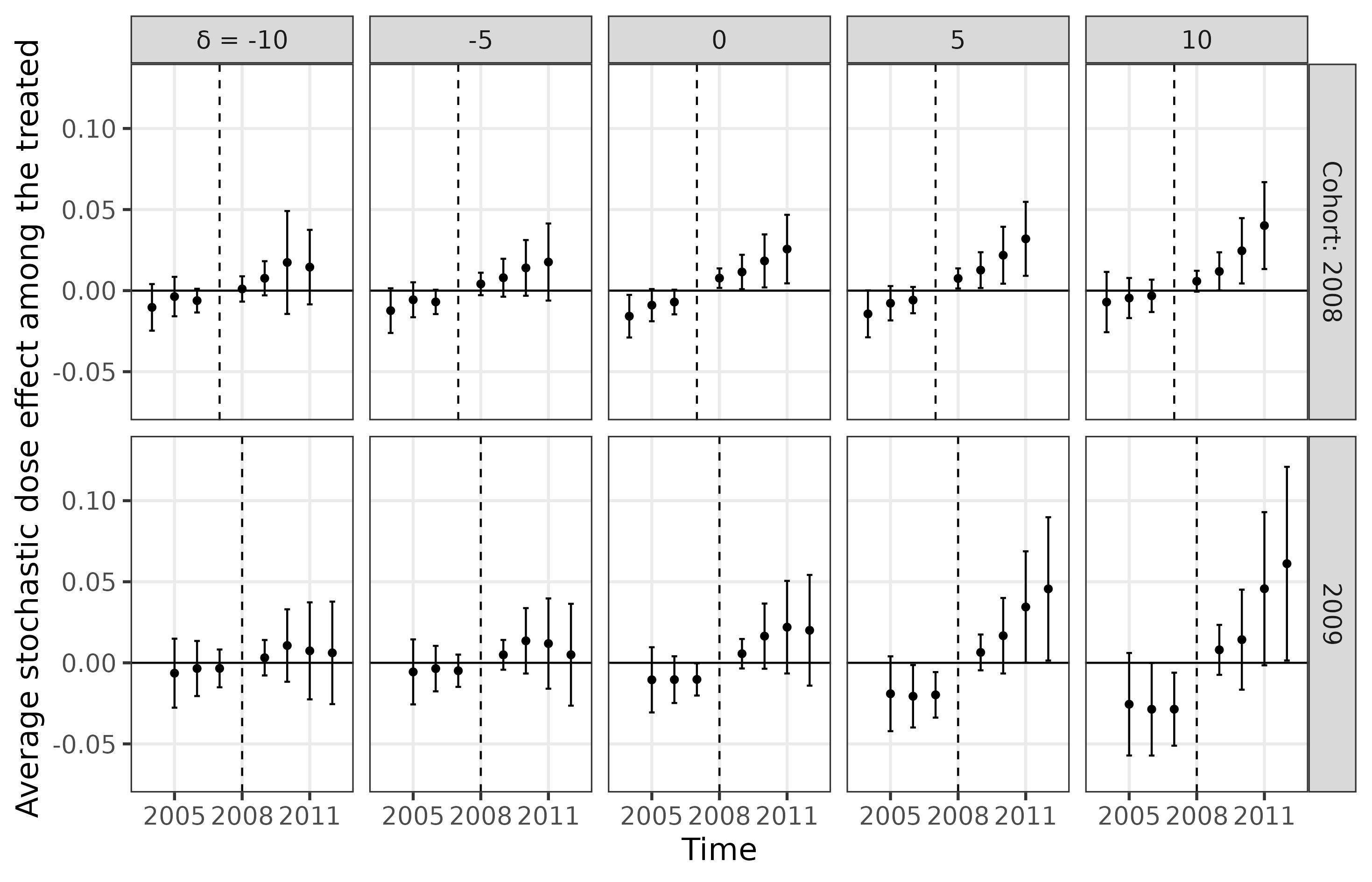}
    \caption{Employment effects due to shifts in probability distribution of fracking potential in 2008 and 2009. A larger $\delta$ denotes a shift of the probability distribution towards the maximal fracking potential value. The dashed line denotes the reference outcome year, circles represent point estimates, and vertical bars represent pointwise 95\% confidence intervals. Estimates to the left of the dashed lines are placebo tests.}
    \label{fig:results-emp}
\end{figure}

The results for employment are presented in Figure \ref{fig:results-emp}. When the prospectivity score distribution is shifted towards higher values, a modest positive effect is found, with effects increasing by treatment duration. Specifically, at the $\delta = 10$ value, representing a large shift in the prospectivity score distribution towards the maximum, the estimated ASDT for log employment in cohort 2008 and year 2011 was 0.04 (95\% confidence interval: [0.01, 0.07]), which corresponds to an approximately 4\% increase in employment. Additionally, when the increment parameter is smaller, corresponding to shifting the prospectivity score distribution towards the minimum of its support, effects tended to attenuate. Thus, if fracking potential were to tend to be less than what was actually observed, Figure \ref{fig:results-emp} shows small economic effects that do not pass the threshold of statistical significant at the 0.05 level. The results for income were similar and are presented in the Appendix. 

These results support the findings in \citet{bartik_local_2019} of positive economic effects due to fracking potential that increase with treatment duration. The results also align with \citet{callaway_event_2024}, who found that effects increased with dose. However, unlike those two papers, the analysis here utilized a conditional parallel trends assumption and targeted a causal estimand that may be more relevant for local policymakers' decision-making today. Additionally, the estimates here tended to be more modest than both \citet{bartik_local_2019} and \citet{callaway_event_2024}, which is particularly relevant in fracking applications since policymakers must decide on whether or not to allow fracking by weighing the benefits, such as the possible economic impacts discussed here, against the costs, such as the health risks and health costs linked to fracking. Thus, it is possible that including this paper's more modest effect estimates in a cost-benefit analysis would lead to different fracking recommendations, though we leave this to future work.

\section{Discussion} \label{sec:discussion}

This paper considers inference about the effect of modifying the distribution of a continuous dose when making parallel trends type assumptions. Under the exponential tilt policy, nonparametric efficient estimators with machine learning nuisance function estimation were proposed and shown to be $\sqrt{n}$-CAN under mild convergence rate conditions. Though the primary proposed estimator was based on the exponential tilt due to its generality and simplicity, the approach in this paper can be adapted to develop similar nonparametric efficient estimators for other stochastic policies of interest. Additionally, the framework developed in this paper could also be extended to the setting with multiple discrete treatments. In addition to considering other stochastic interventions, future work may involve further development of machine learning estimators for conditional densities like the generalized propensity score. Lastly, considering stochastic interventions with either binary or continuous treatments in other causal identification paradigms such as synthetic controls or regression discontinuity is a promising future direction.

\section*{Acknowledgments}

We thank Brian D. Richardson and Gary Hettinger for helpful comments. Michael Jetsupphasuk was supported by the National Institutes of Health (NIH) grant T32ES007018. Michael G. Hudgens was supported by NIH grant R01AI085073. The content in this article is solely the responsibility of the authors and does not necessarily represent the official views of the NIH. 

\section*{Disclosure statement}
The authors declare no potential conflict of interests.

\section*{Data availability}
The data used in the application is publicly available and can be found at \url{https://doi.org/10.1257/app.20170487} \citep{bartik_local_2019}. All code to re-create the results of the data application and simulation can be found on the following Github page: \url{https://github.com/mjetsupphasuk/did_stochastic}. 


\bibliography{references}

\newpage

\section{Appendix}


\subsection{Causal estimands and identification under unconditional parallel trends}

Consider the following causal estimands:
\begin{align*}
    \ASDT^{\mathrm{UPT}}(Q) &= \int_{\mathcal{D}} \E[Y_{i1}(d) - Y_{i1}(0) | A_i > 0] \mathrm{d}Q(d|A_i>0), \\
    \ASLD^{\mathrm{UPT}}(Q) &= \int_{\mathcal{D}} \E[Y_{i1}(d) - Y_{i1}(0) | D_i=d, A_i > 0] \mathrm{d}Q(d|A_i>0),
\end{align*}
which are versions of $\ASDT(Q)$ and $\ASLD(Q)$ that do not depend on covariates. For identification, Assumptions \ref{assump:parallel-a0} -- \ref{assump:parallel-alt} are unconditional parallel trends (UPT) versions of Assumptions \ref{assump:parallel-cond-a0} -- \ref{assump:parallel-cond-alt}, respectively.

\begin{assumption}[Unconditional untreated parallel trends between treated and untreated groups]
    \label{assump:parallel-a0}
    $\E[Y_{i1}(0) - Y_{i0}(0) | A_i>0] = \E[Y_{i1}(0) - Y_{i0}(0) | A_i=0]$.
\end{assumption}

\begin{assumption}[Unconditional dose-specific parallel trends between local and treated dose groups]
    \label{assump:parallel-d}
    For all doses $d \in \mathcal{D}$, $\E[Y_{i1}(d) - Y_{i0}(d) | A_i=d] = \E[Y_{i1}(d) - Y_{i0}(d) | A_i > 0]$.
\end{assumption}

\begin{assumption}[Unconditional untreated parallel trends between all treatment groups]
    \label{assump:parallel-alt}
    For all doses $d \in \mathcal{D}$, $\E[Y_{i1}(0) - Y_{i0}(0) | A_i=d] = \E[Y_{i1}(0) - Y_{i0}(0) | A_i = 0]$.
\end{assumption}

Similarly, Assumption \ref{assump:positivity-ident-u} is a version of Assumption \ref{assump:positivity-ident-c} that is unconditional on covariates. 

\begin{assumption}[Positivity unconditional on covariates]
    \label{assump:positivity-ident-u}
    (i) For some $0 < \epsilon < 1/2$, $\epsilon < \PP(A_i>0) < 1-\epsilon$. (ii) For any $d \in \mathcal{D}$, if $\pi_D(d) = 0$, then $q(d) = 0$.
\end{assumption}

Let $\mu_d = \E[\Delta Y | D=d, A>0]$ and $\mu_{A=0} = \E[\Delta Y | A=0]$. Theorem \ref{thm:identification-upt} states the identification results for $\ASDT^{\mathrm{UPT}}(Q)$ and $\ASLD^{\mathrm{UPT}}(Q)$ under unconditional parallel trends assumptions. 

\begin{theorem}[Identification]
    \label{thm:identification-upt}
    Under Assumptions \ref{assump:consistency}, \ref{assump:no_anticip}, \ref{assump:parallel-a0}, \ref{assump:parallel-d}, and \ref{assump:positivity-ident-u},
    \begin{align*}
    \ASDT^{\mathrm{UPT}}(Q) &= \Psi^{\mathrm{UPT}}(\mathbb{P}) \equiv \Psi^{\mathrm{UPT}, 1}(\mathbb{P}) - \Psi^{\mathrm{UPT}, 2}(\mathbb{P}),
    \end{align*}
    where $\Psi^{\mathrm{UPT}, 1}(\mathbb{P}) = \int_{\mathcal{D}} \mu_{d} \mathrm{d}Q(d|A>0)$ and $\Psi^{\mathrm{UPT}, 2}(\mathbb{P}) = \mu_{A=0}$.

    Under Assumptions \ref{assump:consistency}, \ref{assump:no_anticip}, \ref{assump:parallel-alt}, and \ref{assump:positivity-ident-u},
    \begin{align*}
        \ASLD^{\mathrm{UPT}}(Q) &= \Psi^{\mathrm{UPT}}(\mathbb{P}).
    \end{align*}
\end{theorem}

\subsection{Other stochastic interventions}

The Gaussian concentration policy shifts the generalized dose propensity score $\pi_D(d|\boldx)$ according to a concentration point parameter $d' \in \mathcal{D}$ and length parameter $l \in (0, \infty)$:
\begin{align*}
    q_{l, d'}(d|\boldx) &= \frac{\exp\{- (d-d')^2 / 2l ^2\} \pi_D(d|\boldx)}{\int \exp\{-(b-d')^2 / 2l^2\} \pi_D(b | \boldx) \mathrm{d}b}.
\end{align*}
Since the Gaussian concentration policy resembles the exponential tilt policy, the EIF and corresponding one-step EIF-based estimator are also similar. Let $\Psi^{\mathrm{gauss}, 1}(\mathbb{P})$ be the estimand $\Psi^{(1)}(\mathbb{P})$ under the Gaussian concentration policy. Then, following from Theorem 2, 
\begin{align*}
    \varphi^{\mathrm{gauss}, 1}(\boldO; \mathbb{P}) &= \frac{\mathbbm{1}(A>0)}{\PP(A>0)} \bigg \{ \frac{q_{., d'}(D|\boldX)}{\pi_D(D|\boldX)}\left(\Delta Y - \int_{\mathcal{D}} \mu_b(\boldX) q_{l,d'}(b|\boldX)\mathrm{d}b  \right) \\
    &\hspace{1em} + \int_{\mathcal{D}} \mu_b(\boldX) q_{l,d'}(b|\boldX)\mathrm{d}b - \Psi^{\mathrm{gauss}, 1}(\mathbb{P}) \bigg \}
\end{align*}
is the EIF of $\Psi^{\mathrm{gauss}, 1}(\mathbb{P})$. The corresponding one-step EIF-based estimator in constructed the same way as with the exponential tilt policy; in this case, the estimator is the same except with the estimated tilt policy $\hat q_{\delta}(d|\boldX_i)$ replaced with the estimated concentration policy $\hat q_{l,d'}(d|\boldX_i)$.

\subsection{Proofs}

Let $F_{\boldX}(\boldx)$ denote the distribution function for covariates $\boldX$. 

\begin{proof}[\textbf{Proof of Theorem \ref{thm:identification}}]

First, identification of $\ASDT(Q)$ is shown:
\begin{align}
    &\ASDT(Q) \equiv \int_{\mathcal{X} \times \mathcal{D}} \E[Y_{1}(d) - Y_{1}(0) | \boldX=\boldx, A > 0 ]\mathrm{d}Q(d | \boldx, A>0) \mathrm{d}F_{\boldX}(\boldx|A > 0) \nonumber \\
    &= \int_{\mathcal{X} \times \mathcal{D}} \E[Y_1(d) | \boldX=\boldx, A > 0 ]\mathrm{d}Q(d | \boldx, A>0) \mathrm{d}F_{\boldX}(\boldx|A > 0) \\
    &\hspace{1em} - \int_{\mathcal{X}} \E[Y_1(0) | \boldX=\boldx, A>0] \mathrm{d}F_{\boldX}(\boldx|A > 0) \nonumber \\
    &= \int_{\mathcal{X} \times \mathcal{D}} \E[Y_1(d) - Y_0(0) | \boldX=\boldx, A > 0 ]\mathrm{d}Q(d | \boldx, A>0) \mathrm{d}F_{\boldX}(\boldx|A > 0) \\
    &\hspace{1em} - \E[ \E[Y_1(0) - Y_0(0) | \boldX, A>0] | A>0] \nonumber \\
    &= \int_{\mathcal{X} \times \mathcal{D}} \E[Y_1(d) - Y_0(0) | \boldX=\boldx, A > 0 ]\mathrm{d}Q(d | \boldx, A>0) \mathrm{d}F_{\boldX}(\boldx|A > 0) \\
    &\hspace{1em} - \E[ \E[Y_1(0) - Y_0(0) | \boldX, A=0] | A>0] \nonumber \\
    &= \int_{\mathcal{X} \times \mathcal{D}} \E[Y_1(d) - Y_0(0) | \boldX=\boldx, A > 0 ]\mathrm{d}Q(d | \boldx, A>0) \mathrm{d}F_{\boldX}(\boldx|A > 0) \\
    &\hspace{1em} - \E[ \mu_{A=0}(\boldX) | A>0] \nonumber \\
    &= \int_{\mathcal{X} \times \mathcal{D}} \E[Y_1(d) - Y_0(0) | \boldX=\boldx, D=d, A>0] \mathrm{d}Q(d|\boldx, A>0)\mathrm{d}F_{\boldX}(\boldx|A>0) \\
    &\hspace{1em} - \E[ \mu_{A=0}(\boldX) | A>0] \nonumber \\
    &= \int_{\mathcal{X} \times \mathcal{D}} \E[\Delta Y | \boldX=\boldx, D=d, A>0] \mathrm{d}Q(d|\boldx, A>0)\mathrm{d}F_{\boldX}(\boldx|A>0) - \E[ \mu_{A=0}(\boldX) | A>0]  \\
    &= \int_{\mathcal{X} \times \mathcal{D}} \mu_d(\boldx) \mathrm{d}Q(d|\boldx, A>0)\mathrm{d}F_{\boldX}(\boldx|A>0) - \E[ \mu_{A=0}(\boldX) | A>0] \\
    &= \E \left[ \int_{\mathcal{D}} \{ \mu_d(\boldX) - \mu_{A=0}(\boldX) \} \mathrm{d}Q(d|\boldX, A>0) \bigg| A>0 \right],
\end{align}
where the first equality re-arranges terms, the second equality adds and subtracts $\E[Y_0(0) | \boldX=\boldx, A>0]$ and uses the definition of expectation, the third equality uses Assumption \ref{assump:parallel-cond-a0} (conditional parallel trends for the no treatment counterfactual), the fourth equality uses Assumption \ref{assump:consistency} (causal consistency) and the definition of $\mu_{A=0}(\boldX)$, the fifth equality uses Assumption \ref{assump:parallel-cond-d} (conditional parallel trends under dose counterfactuals), the sixth equality uses Assumption \ref{assump:consistency} (causal consistency) and Assumption \ref{assump:no_anticip} (no anticipation), the seventh equality uses the definition of $\mu_d(\boldX)$, and the eighth equality re-arranges terms. 

The proof for identifying the ASLD follows similarly,
\begin{align*}
    &\ASLD(Q) = \int_{\mathcal{X} \times \mathcal{D}} \E[Y_{1}(d) - Y_{1}(0) | \boldX=\boldx, D=d, A > 0 ]\mathrm{d}Q(d | \boldx, A>0) \mathrm{d}F_{\boldX}(\boldx|A > 0) \\
    &= \int_{\mathcal{X} \times \mathcal{D}} \E[Y_{1}(d) - Y_{0}(0) | \boldX=\boldx, D=d, A > 0 ]\mathrm{d}Q(d | \boldx, A>0) \mathrm{d}F_{\boldX}(\boldx|A > 0) \\
    &\hspace{1em} - \int_{\mathcal{X} \times \mathcal{D}} \E[Y_{1}(0) - Y_{0}(0) | \boldX=\boldx, D=d, A > 0 ]\mathrm{d}Q(d | \boldx, A>0) \mathrm{d}F_{\boldX}(\boldx|A > 0) \\
    &= \int_{\mathcal{X} \times \mathcal{D}} \E[Y_{1}(d) - Y_{0}(0) | \boldX=\boldx, D=d, A > 0 ]\mathrm{d}Q(d | \boldx, A>0) \mathrm{d}F_{\boldX}(\boldx|A > 0) \\
    &\hspace{1em} - \int_{\mathcal{X} \times \mathcal{D}} \E[Y_{1}(0) - Y_{0}(0) | \boldX=\boldx, A=0 ]\mathrm{d}Q(d | \boldx, A>0) \mathrm{d}F_{\boldX}(\boldx|A > 0) \\
    &= \int_{\mathcal{X} \times \mathcal{D}} \E[Y_{1}(d) - Y_{0}(0) | \boldX=\boldx, D=d, A > 0 ]\mathrm{d}Q(d | \boldx, A>0) \mathrm{d}F_{\boldX}(\boldx|A > 0) \\
    &\hspace{1em} - \E[ \E[Y_1(0) - Y_0(0) | \boldX, A=0] | A>0] \\
    &= \int_{\mathcal{X} \times \mathcal{D}} \E[Y_{1}(d) - Y_{0}(0) | \boldX=\boldx, D=d, A > 0 ]\mathrm{d}Q(d | \boldx, A>0) \mathrm{d}F_{\boldX}(\boldx|A > 0) \\
    &\hspace{1em} - \E[ \mu_{A=0}(\boldX) | A>0] \\
    &= \int_{\mathcal{X} \times \mathcal{D}} \E[\Delta Y | \boldX=\boldx, D=d, A > 0 ]\mathrm{d}Q(d | \boldx, A>0) \mathrm{d}F_{\boldX}(\boldx|A > 0) - \E[ \mu_{A=0}(\boldX) | A>0] \\
    &= \int_{\mathcal{X} \times \mathcal{D}} \mu_{d}(\boldx) \mathrm{d}Q(d | \boldx, A>0) \mathrm{d}F_{\boldX}(\boldx|A > 0) - \E[ \mu_{A=0}(\boldX) | A>0]. 
\end{align*}

\end{proof}

\begin{proof}[\textbf{Proof of Theorem \ref{thm:eif}}]
    Recall that $\ASDT(Q) = \int \mu_{d}(\boldx) \mathrm{d}Q(d|\boldx, A>0) \mathrm{d}F_{\boldX}(\boldx|A>0) - \E[\mu_{A=0}(\boldX)|A>0]$. Let $\Psi^{(1)}(\mathbb{P}) = \int \mu_{d}(\boldx) \mathrm{d}Q(d|\boldx, A>0) \mathrm{d}F_{\boldX}(\boldx|A>0)$, $\Psi^{(2)}(\mathbb{P}) = \E[\mu_{A=0}(\boldX)|A>0]$, and $\Psi(\mathbb{P}) = \ASDT(Q) = \Psi^{(1)}(\mathbb{P}) - \Psi^{(2)}(\mathbb{P})$. Notice that $\E[\mu_{A=0}(\boldX)|A>0]$ is the functional of one half of the ATT under ignorability, i.e., $\E[\mu_{A=0}(\boldX)|A>0] = \E[Y(0)|A>0]$ and its EIF has been derived in other work \citep{renson_pulling_2025}:
    \begin{align*}
        \varphi^{(2)}(\boldO; \mathbb{P}) &= \frac{\mathbbm{1}(A=0)}{\PP(A>0)} \left\{ \frac{\pi_{A>0}(\boldX)}{1-\pi_{A>0}(\boldX)} \left(\Delta Y - \mu_{A=0}(\boldX) \right) \right\} + \frac{\mathbbm{1}(A>0)}{\PP(A>0)} \left \{ \mu_{A=0}(\boldX) - \Psi^{(2)}(\mathbb{P}) \right \}.
    \end{align*}
    Further, $\Psi^{(1)}(\mathbb{P})$ is the same as the estimand considered in \citet{schindl_incremental_2026} except with all expectations conditioning on $A>0$. Like \citet{schindl_incremental_2026}, consider a general tilt for the distribution of $Q$ with finite increment parameter $\nu$ and a generic smooth function $f(\cdot)$,
    \begin{align*}
        q_{\nu}(d|\boldx) &= \frac{f_{\nu}(\pi_D(d|\boldx,A>0))}{\int_b f_{\nu}(\pi_D(b|\boldx,A>0)) \mathrm{d}b}.
    \end{align*}
    To ease notation, let $\mu_d = \mu_d(\boldX)$ and $\pi_d = \pi_D(d|\boldX)$. Then, the conjectured EIF is:
    \begin{align*}
        \varphi^{(1)}(\boldO; \mathbb{P}) &= \frac{\mathbbm{1}(A>0)}{\PP(A>0)} \bigg \{ \frac{f_{\nu}(\pi_D(D|\boldX))}{\pi_D(d|\boldX) \int_b f_{\nu}(\pi_D)\mathrm{d}b} (\nu Y - \mu_D(\boldX)) + \frac{f'_{\nu}(\pi_D) \mu_D(\boldX)}{\int_b f_{\nu}(\pi_D)\mathrm{d}b} - \frac{\int_b f'_{\nu}(\pi_D) \pi_D \mu_b \mathrm{d}b}{\int_b f_{\nu}(\pi_D)\mathrm{d}b}  \\
        &\hspace{4em} - \frac{\int_b \mu_b f_{\nu}(\pi_D) \mathrm{d}b}{[\int_b f_{\nu}(\pi_D)\mathrm{d}b]^2}\left(f'_{\nu}(\pi_D) - \int_b f'_{\nu}(\pi_D) \pi_D \mathrm{d}b \right) + \frac{\int_b f_{\nu}(\pi_D) \mu_b \mathrm{d}b}{\int_b f_{\nu}(\pi_D) \mathrm{d}b} - \Psi^{(1)}(\mathbb{P}) \bigg \}.
    \end{align*}
    To verify that the conjectured EIF is indeed the EIF, Lemma 2 of \citet{kennedy_semiparametric_2023-1} is used, which proves that if the von Mises expansion of $\Psi(\mathbb{P})$ holds for the conjectured EIF, then the conjectured EIF is the true EIF. The von Mises expansion states that
    \begin{align*}
        \Psi(\Tilde{\mathbb{P}}) - \Psi(\mathbb{P}) &= \int \varphi(\boldO; \Tilde{\mathbb{P}})d(\Tilde{\mathbb{P}}-\mathbb{P}) + R_2(\Tilde{\mathbb{P}}, \mathbb{P}),
    \end{align*}
    where $\Tilde{\mathbb{P}}$ and $\mathbb{P}$ are in the nonparametric model $\mathcal{P}$ and $R_2(\Tilde{\mathbb{P}}, \mathbb{P})$ is a remainder term \citep{kennedy_semiparametric_2023}. The von Mises expansion holds if $R_2(\Tilde{\mathbb{P}}, \mathbb{P})$ is a second-order remainder term in the sense that it consists of products and squares of differences of $\mathbb{P}$ and $\Tilde{\mathbb{P}}$. 

    To simplify notation, let $p = \PP(A>0)$ and $\hat p = \meanin \mathbbm{1}(A_i > 0)$. Further, choose $\Tilde{\mathbb{P}} = \hat{\mathbb{P}}$ such that $\hat \psi = \Psi(\hat{\mathbb{P}})$. Decompose $\Psi(\hat{\mathbb{P}}) = \Psi^{(1)}(\hat{\mathbb{P}}) - \Psi^{(2)}(\hat{\mathbb{P}})$ where,
    \begin{align*}
        \Psi^{(1)}(\hat{\mathbb{P}}) &= \meanin \frac{\mathbbm{1}(A_i>0)}{\hat p} \int_{\mathcal{D}} \hat \mu_d(\boldX_i) \mathrm{d}\hat{Q}(d|\boldX_i, A_i>0) \\
        &\hspace{2em} + \meanin \left\{ \frac{\mathbbm{1}(A_i>0)}{\hat p} \frac{\hat q_{\delta}(D_i|\boldX_i)}{\hat \pi_D(\boldX_i)}\left(\Delta Y_i - \int_{\mathcal{D}} \hat \mu_b(\boldX_i) \hat q_{\delta}(b|\boldX_i)\mathrm{d}b  \right)  \right\} \\
        \Psi^{(2)}(\hat{\mathbb{P}}) &= \meanin \frac{ \mathbbm{1}(A_i>0)}{\hat p} \hat \mu_{A=0}(\boldX_i) + \meanin \frac{\mathbbm{1}(A_i=0)}{\hat p} \left\{ \frac{\hat \pi_{A>0}(\boldX_i)}{1 - \hat \pi_{A>0}(\boldX_i)} \left(\Delta Y_i - \hat \mu_{A=0}(\boldX_i) \right) \right\},
    \end{align*}
    i.e., $\{\hat \mu(\boldX), \hat \pi(\boldX) \}$ are in the distribution $\hat{\mathbb{P}}$ and $\Psi^{(r)}(\hat{\mathbb{P}}), r \in \{1,2\}$ are one-step estimators.

    The remainder $R_2(\hat{\mathbb{P}}, \mathbb{P})$ can be decomposed into two components that correspond to the two estimands that make up $\Psi(\mathbb{P})$, i.e., $R_2^{(1)}(\hat{\mathbb{P}}, \mathbb{P})$ is the remainder term for $\Psi^{(1)}(\mathbb{P})$ and $R_2^{(2)}(\hat{\mathbb{P}}, \mathbb{P})$ is the remainder term for $\Psi^{(2)}(\mathbb{P})$. \citet{renson_pulling_2025} proved that $R_2^{(2)}(\hat{\mathbb{P}}, \mathbb{P})$ is second-order so it remains to be shown the same for $R_2^{(1)}(\hat{\mathbb{P}}, \mathbb{P})$. 

    Let $\phi^{(1)}(\boldO; \mathbb{P}) = \varphi^{(1)}(\boldO; \mathbb{P}) + \frac{\mathbbm{1}(A>0)}{\PP(A>0)} \Psi^{(1)}(\mathbb{P})$, i.e., the un-centered EIF. Then,
    \begin{align*}
        R_2^{(1)}(\hat{\mathbb{P}}, \mathbb{P}) &= \Psi^{(1)}(\hat{\mathbb{P}}) - \Psi^{(1)}(\mathbb{P}) + \int [\phi^{(1)}(\boldO; \hat{\mathbb{P}}) - \frac{\mathbbm{1}(A>0)}{\hat p} \Psi^{(1)}(\hat{\mathbb{P}})] \mathrm{d}\mathbb{P}(\boldo) \\
        &= \Psi^{(1)}(\hat{\mathbb{P}}) - \Psi^{(1)}(\hat{\mathbb{P}}) - \frac{p}{\hat p} \Psi^{(1)}(\mathbb{P}) + \int \phi^{(1)}(\boldO; \hat{\mathbb{P}}) \mathrm{d}\mathbb{P}(\boldo) \\
        &= \Psi^{(1)}(\hat{\mathbb{P}}) \left[ \frac{\hat p - p}{\hat p} \right] + \E[\phi^{(1)}(\boldO; \hat{\mathbb{P}}) - \Psi^{(1)}(\mathbb{P})] \\
        &= [\Psi^{(1)}(\hat{\mathbb{P}}) - \Psi^{(1)}(\mathbb{P})] \left[ \frac{\hat p - p}{\hat p} \right] + \E \left[\phi^{(1)}(\boldO; \hat{\mathbb{P}}) - \frac{\mathbbm{1}(A>0)}{\hat p}\Psi^{(1)}(\mathbb{P}) \right] .
    \end{align*}
    The first term, $[\Psi^{(1)}(\hat{\mathbb{P}}) - \Psi^{(1)}(\mathbb{P})] \left[ \frac{\hat p - p}{\hat p} \right]$, is second-order. The remaining term \\
    $\E \left[\phi^{(1)}(\boldO; \hat{\mathbb{P}}) - \frac{\mathbbm{1}(A>0)}{\hat p}\Psi^{(1)}(\mathbb{P}) \right]$ is analogous to the remainder term in \citet{schindl_incremental_2026} and was shown to be second-order. Specifically, the result follows from the proofs of Proposition 3 and Theorem 3 in \citet{schindl_incremental_2026}.
\end{proof}

\begin{proof}[\textbf{Proof of Theorem \ref{thm:asymp-norm}}]

    To begin, the interim result of consistency is proved, i.e., $\Psi(\hat{\mathbb{P}}) \rightarrow_p \Psi(\mathbb{P})$. Since $\Psi^{(2)}(\hat{\mathbb{P}}) \rightarrow_p \Psi^{(2)}(\mathbb{P})$ has been shown in previous work, it suffices to show $\Psi^{(1)}(\hat{\mathbb{P}}) \rightarrow_p \Psi^{(1)}(\mathbb{P})$. Further, since $\Psi^{(1)}(\hat{\mathbb{P}}) - \E[\Psi^{(1)}(\hat{\mathbb{P}})] \rightarrow_p 0$ by the law of large numbers, it suffices to show $\E[\Psi^{(1)}(\hat{\mathbb{P}}) - \Psi^{(1)}(\mathbb{P})] \rightarrow_p 0$. Then,
    \begin{align*}
        &\E[\Psi^{(1)}(\hat{\mathbb{P}}) - \Psi^{(1)}(\mathbb{P})] \\
        &= \E \left[ \meanin \frac{\mathbbm{1}(A_i>0)}{\hat p} \int_{\mathcal{D}} \hat \mu_d(\boldX_i) \mathrm{d}\hat{Q}(d|\boldX_i, A_i>0) - \frac{\mathbbm{1}(A_i>0)}{ p} \int_{\mathcal{D}} \mu_d(\boldX_i) d{Q}(d|\boldX_i, A_i>0) \right] \\
        &\hspace{2em} + \E \left[\meanin \left\{ \frac{\mathbbm{1}(A_i>0)}{\hat p} \frac{\hat q_{\delta}(D_i|\boldX_i)}{\hat \pi_D(\boldX_i)}\left(\Delta Y_i  \right)  \right\} - \left\{ \frac{\mathbbm{1}(A_i>0)}{ p} \frac{ q_{\delta}(D_i|\boldX_i)}{ \pi_D(\boldX_i)}\left(\Delta Y_i  \right)  \right\} \right] \\
        &\hspace{2em} - \E \bigg[\meanin \left\{ \frac{\mathbbm{1}(A_i>0)}{\hat p} \frac{\hat q_{\delta}(D_i|\boldX_i)}{\hat \pi_D(\boldX_i)}\left(\int_{\mathcal{D}} \hat \mu_b(\boldX_i) \hat q_{\delta}(b|\boldX_i)\mathrm{d}b  \right)  \right\} \\
        &\hspace{4em} - \left\{ \frac{\mathbbm{1}(A_i>0)}{ p} \frac{ q_{\delta}(D_i|\boldX_i)}{ \pi_D(\boldX_i)}\left(\int_{\mathcal{D}}  \mu_b(\boldX_i)  q_{\delta}(b|\boldX_i)\mathrm{d}b  \right)  \right\} \bigg].
    \end{align*}

    Using the results $\hat p - p \rightarrow_p 0$ by the law of large numbers, positivity, and boundedness of the data,
    \begin{align*}
        & \E \left[ \meanin \frac{\mathbbm{1}(A_i>0)}{\hat p} \int_{\mathcal{D}} \hat \mu_d(\boldX_i) \mathrm{d}\hat{Q}(d|\boldX_i, A_i>0) - \frac{\mathbbm{1}(A_i>0)}{ p} \int_{\mathcal{D}} \mu_d(\boldX_i) d{Q}(d|\boldX_i, A_i>0) \right] \\
        &\lesssim \E\left[\int_{\mathcal{D}} (\hat q(b|\boldX_i) - q(b|\boldX_i))^2 \pi_D(b|\boldX_i) \mathrm{d}b \right]^{1/2} + \E\left[\int_{\mathcal{D}} (\hat \mu_b(\boldX_i) - \mu_b(\boldX_i))^2 \pi_D(b|\boldX_i) \mathrm{d}b \right]^{1/2} + o_{\mathbb{P}}(1) \\
        &\lesssim \E\left[\int_{\mathcal{D}} (\hat q(b|\boldX_i) - q(b|\boldX_i))^2 \mathrm{d}b \right]^{1/2} + \E\left[\int_{\mathcal{D}} (\hat \mu_b(\boldX_i) - \mu_b(\boldX_i))^2 \mathrm{d}b \right]^{1/2} + o_{\mathbb{P}}(1),
    \end{align*}
    where $a \lesssim b$ denotes $a \leq Cb$ where $C$ is a finite constant. Similarly,
    \begin{align*}
        &\E \left[\meanin \left\{ \frac{\mathbbm{1}(A_i>0)}{\hat p} \frac{\hat q_{\delta}(D_i|\boldX_i)}{\hat \pi_D(\boldX_i)}\left(\Delta Y_i  \right)  \right\} - \left\{ \frac{\mathbbm{1}(A_i>0)}{ p} \frac{ q_{\delta}(D_i|\boldX_i)}{ \pi_D(\boldX_i)}\left(\Delta Y_i  \right)  \right\} \right] \\
        &\lesssim \E \left[ \pi(\hat q - q) - q(\hat \pi - \pi) \right] + o_{\mathbb{P}}(1),
    \end{align*}
    and,
    \begin{align*}
        &\E \bigg[\meanin \left\{ \frac{\mathbbm{1}(A_i>0)}{\hat p} \frac{\hat q_{\delta}(D_i|\boldX_i)}{\hat \pi_D(\boldX_i)}\left(\int_{\mathcal{D}} \hat \mu_b(\boldX_i) \hat q_{\delta}(b|\boldX_i)\mathrm{d}b  \right)  \right\} \\
        &\hspace{2em} - \left\{ \frac{\mathbbm{1}(A_i>0)}{ p} \frac{ q_{\delta}(D_i|\boldX_i)}{ \pi_D(\boldX_i)}\left(\int_{\mathcal{D}}  \mu_b(\boldX_i)  q_{\delta}(b|\boldX_i)\mathrm{d}b  \right)  \right\} \bigg] \\
        &\lesssim \E \left[ \left(\frac{\hat q}{\hat \pi} - \frac{q}{\pi} \right) \int_{\mathcal{D}} \hat \mu_b \hat q_b \mathrm{d}b \right] + \E \left[\frac{q}{\pi} \int_{\mathcal{D}} (\hat \mu_b \hat q_b - \mu_b q_b)\mathrm{d}b \right] + o_{\mathbb{P}}(1) \\
        &\lesssim \E \left[ \left(\frac{\hat q}{\hat \pi} - \frac{q}{\pi} \right) \int_{\mathcal{D}} \hat \mu_b \hat q_b \mathrm{d}b \right] + \E \left[\int_{\mathcal{D}} (\hat \mu_b \hat q_b - \mu_b q_b)\mathrm{d}b \right] + o_{\mathbb{P}}(1) \\
        &= \E \left[ \left(\frac{\hat q \pi - q \hat \pi}{\hat \pi \pi} \right) \int_{\mathcal{D}} \hat \mu_b \hat q_b \mathrm{d}b \right] + \E \left[\int_{\mathcal{D}} \{ (\hat \mu_b - \mu_b) \hat q_b - \mu_b (q_b - \hat q_b) \} \mathrm{d}b \right] + o_{\mathbb{P}}(1) \\
        &= \E \left[ \left(\frac{(\hat q - q) \pi - q (\hat \pi - \pi)}{\hat \pi \pi} \right) \int_{\mathcal{D}} \hat \mu_b \hat q_b \mathrm{d}b \right] + \E \left[\int_{\mathcal{D}} \{ (\hat \mu_b - \mu_b) \hat q_b - \mu_b (q_b - \hat q_b) \} \mathrm{d}b \right] + o_{\mathbb{P}}(1),
    \end{align*}
    where $q$ is shorthand for $q(d|\boldX)$ and similarly for $\pi$, $\mu$ and the corresponding estimators. Then, sufficient conditions for convergence are $\| \hat \pi - \pi \| = o_{\mathbb{P}}(1)$ and $\| \hat \mu - \mu \| = o_{\mathbb{P}}(1)$, where $\| \cdot \|$ denotes the squared $L_2({\mathbb{P}})$ norm and noting that $\| \hat \pi - \pi \| = o_{\mathbb{P}}(1)$ implies $\| \hat q - q \| = o_{\mathbb{P}}(1)$ by Lipschitz continuity of $q$ with respect to $\pi$. Thus, all terms of $\E[\Psi^{(1)}(\hat{\mathbb{P}}) - \Psi^{(1)}(\mathbb{P})]$ have been shown to converge to zero so $\Psi(\hat{\mathbb{P}}) \rightarrow_p \Psi(\mathbb{P})$, which completes the consistency proof. 

    To prove asymptotic normality, recall the von Mises expansion of $\Psi(\mathbb{P})$:
    \begin{align*}
        \Psi(\hat{\mathbb{P}}) - \Psi(\mathbb{P}) &= \int \varphi(\boldO; \hat{\mathbb{P}})d(\hat{\mathbb{P}}-\mathbb{P}) + R_2(\hat{\mathbb{P}}, \mathbb{P}) \\
        &= -\mathbb{P}[\varphi(\boldO; \hat{\mathbb{P}})] + R_2(\hat{\mathbb{P}}, \mathbb{P}) \\
        &= (\mathbb{P}_n - \mathbb{P})\varphi(\boldO; \mathbb{P}) + (\mathbb{P}_n - \mathbb{P})(\varphi(\boldO; \hat{\mathbb{P}}) - \varphi(\boldO; \mathbb{P})) + R_2(\hat{\mathbb{P}}, \mathbb{P}),
    \end{align*}
    where $\mathbb{P}[f(\boldO)] = \E_{\mathbb{P}}[f(\boldO)]$ and $\mathbb{P}_n[f(\boldO)] = \meanin f(\boldO_i)$. Each term, scaled by $\sqrt{n}$, in the last expression above will be analyzed in turn.

    First, $\sqrt{n}(\mathbb{P}_n - \mathbb{P})\varphi(\boldO; \mathbb{P}) \rightarrow_d N(0, \sigma^2)$ by the central limit theorem, where
    \begin{align*}
        \sigma^2 &= \Var(\varphi(\boldO; \mathbb{P})) \\
        &= \Var(\varphi^{(1)}(\boldO; \mathbb{P}) + \varphi^{(2)}(\boldO; \mathbb{P})) \\
        &= \Var(\varphi^{(1)}(\boldO; \mathbb{P})) + \Var(\varphi^{(2)}(\boldO; \mathbb{P})) + 2 \Cov(\varphi^{(1)}(\boldO; \mathbb{P}), \varphi^{(2)}(\boldO; \mathbb{P})) \\
        &= \E[\varphi^{(1)}(\boldO; \mathbb{P})^2] + \E[\varphi^{(2)}(\boldO; \mathbb{P})^2] + 2 \E[\varphi^{(1)}(\boldO; \mathbb{P}) \varphi^{(2)}(\boldO; \mathbb{P})],
    \end{align*}
    where we use that the EIF is mean-zero in the last equality. 

    Second, we show $\sqrt{n}(\mathbb{P}_n - \mathbb{P})(\varphi(\boldO; \hat{\mathbb{P}}) - \varphi(\boldO; \mathbb{P})) = o_{\mathbb{P}}(1)$. Recall that the proposed estimator is the cross-fit estimator $\hat \psi^{\mathrm{CF}}$, which averages over the estimators $\hat \psi^{\mathrm{CF}}_k$ for each fold $k$. Since the data are iid, $\hat \psi^{\mathrm{CF}}_k$ has a similar decomposition as given above for $\Psi(\hat{\mathbb{P}})$. Lemma 1 of \citet{kennedy_semiparametric_2023} shows that a sufficient condition for the empirical process term to be $o_{\mathbb{P}}(1)$ when cross-fitting is used with finitely many folds $K$ is $\| \varphi(\boldO; \hat{\mathbb{P}}) - \varphi(\boldO; \mathbb{P}) \| = o_{\mathbb{P}}(1)$. Thus,
    \begin{align*}
        &\| \varphi(\boldO; \hat{\mathbb{P}}) - \varphi(\boldO; \mathbb{P}) \| \\
        &= \E[ \{\varphi^{(1)}(\boldO; \hat{\mathbb{P}}) - \varphi^{(1)}(\boldO; \mathbb{P}) + \varphi^{(1)}(\boldO; \hat{\mathbb{P}}) - \varphi^{(1)}(\boldO; \mathbb{P}) \}^2] \\
        &= \| \varphi^{(1)}(\boldO; \hat{\mathbb{P}}) - \varphi^{(1)}(\boldO; \mathbb{P}) \| \\
        &\hspace{2em} + \| \varphi^{(2)}(\boldO; \hat{\mathbb{P}}) - \varphi^{(2)}(\boldO; \mathbb{P}) \| \\
        &\hspace{2em} + 2 \E[\{\varphi^{(1)}(\boldO; \hat{\mathbb{P}}) - \varphi^{(1)}(\boldO; \mathbb{P}) \} \{ \varphi^{(2)}(\boldO; \hat{\mathbb{P}}) - \varphi^{(2)}(\boldO; \mathbb{P})\}] \\
        &\leq \| \varphi^{(1)}(\boldO; \hat{\mathbb{P}}) - \varphi^{(1)}(\boldO; \mathbb{P}) \| + \| \varphi^{(2)}(\boldO; \hat{\mathbb{P}}) - \varphi^{(2)}(\boldO; \mathbb{P}) \| \\
        &\hspace{2em} + 2 \| \varphi^{(1)}(\boldO; \hat{\mathbb{P}}) - \varphi^{(1)}(\boldO; \mathbb{P}) \| \| \varphi^{(2)}(\boldO; \hat{\mathbb{P}}) - \varphi^{(2)}(\boldO; \mathbb{P}) \|,
    \end{align*}
    where the last inequality follows from Cauchy-Schwarz. From previous work, $\| \varphi^{(1)}(\boldO; \hat{\mathbb{P}}) - \varphi^{(1)}(\boldO; \mathbb{P}) \| = o_{\mathbb{P}}(1)$ and $\| \varphi^{(2)}(\boldO; \hat{\mathbb{P}}) - \varphi^{(2)}(\boldO; \mathbb{P}) \| = o_{\mathbb{P}}(1)$ under the convergence rate conditions provided in the theorem. Therefore, $\| \varphi(\boldO; \hat{\mathbb{P}}) - \varphi(\boldO; \mathbb{P}) \| = o_{\mathbb{P}}(1)$.

    Finally, the remainder term $R_2(\hat{\mathbb{P}}, \mathbb{P})$ is shown to be $o_{\mathbb{P}}(n^{-1/2})$. Recall the decomposition of $R_2(\hat{\mathbb{P}}, \mathbb{P})$,
    \begin{align*}
        R_2^{(1)}(\hat{\mathbb{P}}, \mathbb{P}) &= \Psi^{(1)}(\hat{\mathbb{P}}) - \Psi^{(1)}(\mathbb{P}) + \int [\phi^{(1)}(\boldO; \hat{\mathbb{P}}) - \frac{\mathbbm{1}(A>0)}{\hat p} \Psi^{(1)}(\hat{\mathbb{P}})] \mathrm{d}\mathbb{P}(\boldo) \\
        &= \Psi^{(1)}(\hat{\mathbb{P}}) - \Psi^{(1)}(\hat{\mathbb{P}}) - \frac{p}{\hat p} \Psi^{(1)}(\mathbb{P}) + \int \phi^{(1)}(\boldO; \hat{\mathbb{P}}) \mathrm{d}\mathbb{P}(\boldo) \\
        &= \Psi^{(1)}(\hat{\mathbb{P}}) \left[ \frac{\hat p - p}{\hat p} \right] + \E[\phi^{(1)}(\boldO; \hat{\mathbb{P}}) - \Psi^{(1)}(\mathbb{P})] \\
        &= [\Psi^{(1)}(\hat{\mathbb{P}}) - \Psi^{(1)}(\mathbb{P})] \left[ \frac{\hat p - p}{\hat p} \right] + \E \left[\phi^{(1)}(\boldO; \hat{\mathbb{P}}) - \frac{\mathbbm{1}(A>0)}{\hat p}\Psi^{(1)}(\mathbb{P}) \right].
    \end{align*}
    The first term, $[\Psi^{(1)}(\hat{\mathbb{P}}) - \Psi^{(1)}(\mathbb{P})] \left[ \frac{\hat p - p}{\hat p} \right]$, is $o_{\mathbb{P}}(n^{-1/2})$ by the law of large numbers and consistency result above. The term $\E \left[\phi^{(1)}(\boldO; \hat{\mathbb{P}}) - \frac{\mathbbm{1}(A>0)}{\hat p}\Psi^{(1)}(\mathbb{P}) \right]$ was shown to be $o_{\mathbb{P}}(n^{-1/2})$ in the proofs of Proposition 3 and Theorem 3 in \citet{schindl_incremental_2026} and adjusting Lemma 3 to use our version of positivity (Assumption 7). 
    
\end{proof}

\begin{proof}[\textbf{Proof of Theorem \ref{thm:var-est}}]
    In the proof of Theorem \ref{thm:asymp-norm}, it was shown that the asymptotic variance was
    \begin{align*}
        \sigma^2 &= \E[\varphi^{(1)}(\boldO; \mathbb{P})^2] + \E[\varphi^{(2)}(\boldO; \mathbb{P})^2] - 2 \E[\varphi^{(1)}(\boldO; \mathbb{P}) \varphi^{(2)}(\boldO; \mathbb{P})].
    \end{align*}
    Let the corresponding plug-in estimator be
    \begin{align*}
        \hat \sigma^2 &= \mathbb{P}_n[\varphi^{(1)}(\boldO; \hat{\mathbb{P}})^2] + \mathbb{P}_n[\varphi^{(2)}(\boldO; \hat{\mathbb{P}})^2] - 2 \mathbb{P}_n[\varphi^{(1)}(\boldO; \hat{\mathbb{P}}) \varphi^{(2)}(\boldO; \hat{\mathbb{P}})].
    \end{align*}
    \citet{schindl_incremental_2026} and \citet{renson_pulling_2025}, respectively, showed that the first two terms are consistent so it remains to show that the last term in $\hat \sigma^2$ is consistent. To show this, observe that
    \begin{align*}
        &\E \left[\meanin \varphi_i^{(1)}(\hat{\mathbb{P}}) \varphi_i^{(2)}(\hat{\mathbb{P}}) - \varphi_i^{(1)}(\mathbb{P}) \varphi_i^{(2)}(\mathbb{P}) \right] \\
        &= \E \bigg[ \meanin \frac{\mathbbm{1}(A_i>0)}{p^2} \bigg\{ \frac{\hat q}{\hat \pi}\left(\Delta Y_i - \int_{\mathcal{D}} \hat \mu \hat q \right)\hat \mu_{A=0} - \frac{q}{\pi} \left(\Delta Y_i - \int_{\mathcal{D}} \mu q \right)\mu_{A=0} \\
        &\hspace{2em} - \frac{\hat q}{\hat \pi}\left(\Delta Y_i - \int_{\mathcal{D}} \hat \mu \hat q \right)\hat \psi^{(2)} - \frac{q}{\pi} \left(\Delta Y_i - \int_{\mathcal{D}} \mu q \right)\psi^{(2)}  - \hat \psi^{(2)}\int_{\mathcal{D}} \hat \mu \hat q - \psi^{(2)}\int_{\mathcal{D}} \mu q \\
        &\hspace{2em} - \hat \psi^{(1)} \hat \mu_{A=0} - \psi^{(1)} \mu_{A=0} + \hat \psi^{(1)} \hat \psi^{(2)} - \psi^{(1)} \psi^{(2)} \bigg\} \bigg] + o_{\mathbb{P}}(1).
    \end{align*}
    Then, observe,
    \begin{align*}
        &\E \left[ \frac{\hat q}{\hat \pi}\left(\Delta Y_i - \int_{\mathcal{D}} \hat \mu \hat q \right)\hat \mu_{A=0} - \frac{q}{\pi} \left(\Delta Y_i - \int_{\mathcal{D}} \mu q \right)\mu_{A=0} \right] \\
        &\lesssim \E \left[\frac{\hat q}{\hat \pi}\left(\Delta Y_i - \int_{\mathcal{D}} \hat \mu \hat q \right) - \frac{q}{\pi} \left(\Delta Y_i - \int_{\mathcal{D}} \mu q \right)  \right] + o_{\mathbb{P}}(1),
    \end{align*}
    which is $o_{\mathbb{P}}(1)$ from the consistency proof provided in the Proof of Theorem \ref{thm:asymp-norm}. It can also be shown that,
    \begin{align*}
        &\E \left[ \hat \mu_{A=0} \int_{\mathcal{D}} \hat \mu \hat q - \mu_{A=0} \int_{\mathcal{D}} \mu q \right] \\
        &\lesssim \E \left[ (\hat \mu_{A=0} - \mu_{A=0})  \int_{\mathcal{D}} \hat \mu \hat q - \mu_{A=0} \int_{\mathcal{D}} (\hat \mu \hat q - \mu q) \right] + o_{\mathbb{P}}(1) \\
        &= o_{\mathbb{P}}(1).
    \end{align*}
    Also,
    \begin{align*}
        &\E \left[ \hat \psi^{(1)} \hat \mu_{A=0} - \psi^{(1)} \mu_{A=0} \right] \\
        &= \E \left[ \hat \psi^{(1)} ( \hat \mu_{A=0} - \mu_{A=0}) + \mu_{A=0}(\hat \psi^{(1)} - \psi^{(1)}) \right] \\
        &= o_{\mathbb{P}}(1),
    \end{align*}
    and by a similar argument $\E \left[\hat \psi^{(1)} \hat \psi^{(2)} - \psi^{(1)} \psi^{(2)} \right] = o_{\mathbb{P}}(1)$. Combining these arguments together, we arrive at the result,
    \begin{align*}
        \E \left[\meanin \varphi_i^{(1)}(\hat{\mathbb{P}}) \varphi_i^{(2)}(\hat{\mathbb{P}}) - \varphi_i^{(1)}(\mathbb{P}) \varphi_i^{(2)}(\mathbb{P}) \right] = o_{\mathbb{P}}(1),
    \end{align*}
    which completes this proof.
\end{proof}

\begin{proof}[\textbf{Proof of Theorem \ref{thm:identification-mtp}}]
The proof of Theorem \ref{thm:identification-mtp} follows similarly as the proof of Theorem \ref{thm:identification},
\begin{align*}
    &\ASDT(Q, g, t) \\
    &= \int_{\mathcal{X} \times \mathcal{D}} \E[Y_t(d,g) | \boldX=\boldx, G=g] \mathrm{d}Q(d|\boldx, G=g) \mathrm{d}F_{\boldX}(\boldx|G=g)  \\
    &\hspace{1em} - \int_{\mathcal{X}} \E[Y_t(\infty) | \boldX=\boldx, G=g] \mathrm{d}F_{\boldX}(\boldx|G=g)  \\
    &= \int_{\mathcal{X} \times \mathcal{D}}\E[Y_t(d,g) - Y_{g-1}(d,g) | \boldX=\boldx, G=g] \mathrm{d}Q(d|\boldx, G=g) \mathrm{d}F_{\boldX}(\boldx|G=g)  \\
    &\hspace{1em} - \E[ \E[Y_t(\infty) - Y_{g-1}(d,g) | \boldX, G=g] | G=g] \\
    &= \int_{\mathcal{X} \times \mathcal{D}}\E[Y_t(d,g) - Y_{g-1}(d,g) | \boldX=\boldx, G=g] \mathrm{d}Q(d|\boldx, G=g) \mathrm{d}F_{\boldX}(\boldx|G=g)  \\
    &\hspace{1em} - \E[ \E[Y_t(\infty) - Y_{g-1}(\infty) | \boldX, G=g] | G=g] \\
    &= \int_{\mathcal{X} \times \mathcal{D}}\E[Y_t(d,g) - Y_{g-1}(d,g) | \boldX=\boldx, G=g] \mathrm{d}Q(d|\boldx, G=g) \mathrm{d}F_{\boldX}(\boldx|G=g)  \\
    &\hspace{1em} - \E \left[ \sum_{l=0}^{t-g} \E[Y_{t-l}(\infty) - Y_{t-l-1}(\infty) | \boldX, G=g] \bigg| G=g \right] \\
    &= \int_{\mathcal{X} \times \mathcal{D}}\E[Y_t(d,g) - Y_{g-1}(d,g) | \boldX=\boldx, G=g] \mathrm{d}Q(d|\boldx, G=g) \mathrm{d}F_{\boldX}(\boldx|G=g)  \\
    &\hspace{1em} - \E \left[ \sum_{l=0}^{t-g} \E[Y_{t-l}(\infty) - Y_{t-l-1}(\infty) | \boldX, G>t] \bigg| G=g \right] \\
    &= \int_{\mathcal{X} \times \mathcal{D}}\E[Y_t(d,g) - Y_{g-1}(d,g) | \boldX=\boldx, G=g] \mathrm{d}Q(d|\boldx, G=g) \mathrm{d}F_{\boldX}(\boldx|G=g)  - \E[ \mu^{\mathrm{MTP}}_{0,g,t}(\boldX) | G=g] \\
    &= \int_{\mathcal{X} \times \mathcal{D}} \sum_{l=0}^{t-g} \E[Y_{t-l}(d,g) - Y_{t-l-1}(d,g) | \boldX=\boldx, G=g] \mathrm{d}Q(d|\boldx, G=g) \mathrm{d}F_{\boldX}(\boldx|G=g)  - \E[ \mu^{\mathrm{MTP}}_{0,g,t}(\boldX) | G=g] \\
    &= \int_{\mathcal{X} \times \mathcal{D}} \sum_{l=0}^{t-g} \E[Y_{t-l}(d,g) - Y_{t-l-1}(d,g) | \boldX=\boldx, G=g, D=d] \mathrm{d}Q(d|\boldx, G=g) \mathrm{d}F_{\boldX}(\boldx|G=g)  \\
    &\hspace{2em} - \E[ \mu^{\mathrm{MTP}}_{0,g,t}(\boldX) | G=g] \\
    &= \int_{\mathcal{X} \times \mathcal{D}} \E[Y_{t}(d,g) - Y_{g-1}(d,g) | \boldX=\boldx, G=g, D=d] \mathrm{d}Q(d|\boldx, G=g) \mathrm{d}F_{\boldX}(\boldx|G=g)  - \E[ \mu^{\mathrm{MTP}}_{0,g,t}(\boldX) | G=g] \\
    &= \int_{\mathcal{X} \times \mathcal{D}} \E[Y_{t} - Y_{g-1} | \boldX=\boldx, G=g, D=d] \mathrm{d}Q(d|\boldx, G=g) \mathrm{d}F_{\boldX}(\boldx|G=g)  - \E[ \mu^{\mathrm{MTP}}_{0,g,t}(\boldX) | G=g] \\
    &= \int_{\mathcal{X} \times \mathcal{D}} \mu^{\mathrm{MTP}}_{d,g,t}(\boldx) \mathrm{d}Q(d|\boldx, G=g) \mathrm{d}F_{\boldX}(\boldx|G=g)  - \E[ \mu^{\mathrm{MTP}}_{0,g,t}(\boldX) | G=g].
\end{align*}

\end{proof}

\begin{proof}[\textbf{Proof of Theorem \ref{thm:asymp-norm-mtp}}]

    The expression for the EIF $\varphi_{\delta,g,t}^{\mathrm{MTP}}(\boldO)$ is 
    \begin{align*}
        \varphi_{\delta,g,t}^{\mathrm{MTP}}(\boldO) &= \frac{\mathbbm{1}(G=g)}{\PP(G=g)} \bigg \{ \frac{q_{\delta}(D|\boldX,G=g)}{\pi_{D,g}^{\mathrm{MTP}}(D|\boldX)}\left(Y_{t} - Y_{g-1} - \int_{\mathcal{D}} \mu_{b,g,t}^{\mathrm{MTP}}(\boldX) q_{\delta}(b|\boldX, G=g)\mathrm{d}b  \right) \\
        &\hspace{1em} + \int_{\mathcal{D}} \mu_{b,g,t}^{\mathrm{MTP}}(\boldX) q_{\delta}(b|\boldX, G=g)\mathrm{d}b - \Psi^{\mathrm{MTP, tilt}, 1}(\mathbb{P}) \bigg \} \\
        &\hspace{1em}- \frac{\mathbbm{1}(G>t)}{\PP(G=g)} \left\{ \frac{\pi_{G=g,t}^{\mathrm{MTP}}(\boldX)}{1-\pi_{G=g,t}^{\mathrm{MTP}}(\boldX)} \left(\Delta Y - \mu_{0,g,t}^{\mathrm{MTP}}(\boldX) \right) \right\} \\
        &\hspace{1em}+ \frac{\mathbbm{1}(G=g)}{\PP(G=g)} \left \{ \mu_{0,g,t}^{\mathrm{MTP}}(\boldX) - \Psi^{\mathrm{MTP}, 2}(\mathbb{P}) \right \},
    \end{align*}
    where $$\Psi^{\mathrm{MTP, tilt}, 1}(\mathbb{P}) = \int_{\mathcal{D} \times \mathcal{X}} \mu^{\mathrm{MTP}}_{d,g,t}(\boldx) \mathrm{d}Q(d|\boldx, G=g)\mathrm{d}F_{\boldX}(\boldx|G=g)$$ and $$\Psi^{\mathrm{MTP}, 2}(\mathbb{P}) = \E[ \mu^{\mathrm{MTP}}_{0,g,t}(\boldX) | G=g].$$

    Suppose Assumptions \ref{assump:bounded-mtp} and \ref{assump:conv-rates-mtp} hold, where Assumption \ref{assump:bounded-mtp} is the multiple time periods version of Assumption \ref{assump:bounded}, and Assumption \ref{assump:conv-rates-mtp} is the multiple time periods version of Assumption \ref{assump:conv-rates}. Further assume that positivity as in Assumption \ref{assump:ident-mtp} holds.

    \begin{assumption}[Bounded data and propensity scores - multiple time periods]
        \label{assump:bounded-mtp}
        The outcomes $Y_{it}$ and covariates $\boldX$ are bounded random variables. Also, there exists $\epsilon_{\pi}^{\mathrm{max}} \in (0,1)$ such that $\pi_{D,g}^{\mathrm{MTP}}(d|\boldx) \leq \epsilon_{\pi}^{\mathrm{max}}$ for all $d \in \mathcal{D}$ and $\boldx \in \mathcal{X}$.
    \end{assumption}
    
    \begin{assumption}[Convergence rates of nuisance function estimators - multiple time periods]
        \label{assump:conv-rates-mtp}
        \begin{align*}
            \| \hat \mu_{d,g,t} - \mu_{d,g,t} \| &= o_{\mathbb{P}}(1) \\
            \| \hat \pi_D - \pi_D \|_{L^2_x, L^{\infty}_d} \times \| \hat \mu_{d,g,t} - \mu_{d,g,t} \|_{L^2_x, L^{\infty}_d} &= o_{\mathbb{P}}(n^{-1/2}) \\
            \| \hat \pi_{D,g}^{\mathrm{MTP}} - \pi_{D,g}^{\mathrm{MTP}} \|^2_{L^2_x, L^{\infty}_d} &= o_{\mathbb{P}}(n^{-1/2}) \\
            \| \hat \pi_{G=g,t}^{\mathrm{MTP}} - \pi_{G=g,t}^{\mathrm{MTP}} \| &= o_{\mathbb{P}}(1) \\
            \| \hat \mu_{0,g,t} - \mu_{0,g,t} \| &= o_{\mathbb{P}}(1) \\
            \| \hat \pi_{G=g,t}^{\mathrm{MTP}} - \pi_{G=g,t}^{\mathrm{MTP}} \| \times \| \hat \mu_{0,g,t} - \mu_{0,g,t} \| &= o_{\mathbb{P}}(n^{-1/2}).
        \end{align*}
    \end{assumption}

    The first part of the theorem stating pointwise convergence of the estimator for fixed $\delta$, $g$, and $t$ follows immediately from the proof of Theorem \ref{thm:asymp-norm} after noticing that the estimands and estimators in the multiple time periods setting can be expressed as corresponding estimands and estimators in the two time period setting. Thus, the formal proof is omitted for brevity. 

    The second part of the theorem regarding convergence of the vector of estimators for fixed $\delta$ and varying $t$ and $g$ follows directly from the asymptotic linear representations of $\hat \psi^{\mathrm{MTP,CF}}(\delta, g, t)$ and the multivariate central limit theorem. 
\end{proof}

\begin{proof}[\textbf{Proof of Theorem \ref{thm:asymp-norm-agg}}]
    To ease notation, the arguments for parameters and estimators will be omitted here, e.g., $\hat \omega(g,t)$ will be written $\hat \omega$. Decompose the summand as,
    \begin{align*}
        \hat \omega \hat \psi^{\mathrm{MTP, CF}} - \omega \Psi^{\mathrm{MTP}} &= \hat \omega (\hat \psi^{\mathrm{MTP, CF}} - \Psi^{\mathrm{MTP}}) + \Psi^{\mathrm{MTP}}(\hat \omega - \omega) \\
        &= \omega (\hat \psi^{\mathrm{MTP, CF}} - \Psi^{\mathrm{MTP}}) + \Psi^{\mathrm{MTP}}(\hat \omega - \omega) + (\hat \omega - \omega)(\hat \psi^{\mathrm{MTP, CF}} - \Psi^{\mathrm{MTP}}).
    \end{align*}

    By the asymptotic linear representations of $\hat \omega$ and $\hat \psi^{\mathrm{MTP, CF}}$, the term $\sqrt{n} (\hat \omega - \omega)(\hat \psi^{\mathrm{MTP, CF}} - \Psi^{\mathrm{MTP}}) = o_{\mathbb{P}}(1)$. Then, again using the asymptotic linear representations of $\hat \omega$ and $\hat \psi^{\mathrm{MTP, CF}}$ and central limit theorem arguments, the final result immediately follows.
\end{proof}

\begin{proof}[\textbf{Proof of Theorem \ref{thm:identification-upt}}]

The proof of Theorem \ref{thm:identification-upt} follows similarly as the proof of Theorem \ref{thm:identification}. First, $\ASDT^{\mathrm{UPT}}(Q)$ is identified according to the following:
\begin{align*}
    &\ASDT^{\mathrm{UPT}}(Q) = \int_{\mathcal{D}} \E[Y_1(d) | A > 0 ]\mathrm{d}Q(d | A>0) - \E[Y_1(0) | A>0] \\
    &= \int_{\mathcal{D}} \E[Y_1(d) - Y_0(0) | A > 0 ]\mathrm{d}Q(d | A>0) - \E[Y_1(0) - Y_0(0) | A=0] \\
    &= \int_{\mathcal{D}} \E[Y_1(d) - Y_0(0) | A > 0 ]\mathrm{d}Q(d | A>0) - \mu_{A=0} \\
    &= \int_{\mathcal{D}} \E[Y_1(d) - Y_0(0) | D=d, A > 0 ]\mathrm{d}Q(d | A>0) - \mu_{A=0} \\
    &= \int_{\mathcal{D}} \mu_d \mathrm{d}Q(d| A>0) - \mu_{A=0}.
\end{align*}

Second, $\ASLD^{\mathrm{UPT}}(Q)$ is identified according to the following:
\begin{align*}
    &\ASLD^{\mathrm{UPT}}(Q) \\
    &= \int_{\mathcal{D}} \E[Y_1(d) | D=d, A > 0 ]\mathrm{d}Q(d | A>0) - \int_{\mathcal{D}} \E[Y_1(0) | D=d, A>0] \mathrm{d}Q(d | A>0) \\
    &= \int_{\mathcal{D}} \E[Y_1(d) - Y_0(0) | D=d, A > 0 ]\mathrm{d}Q(d | A>0) - \int_{\mathcal{D}} \E[Y_1(0) - Y_0(0) | D=d, A>0] \mathrm{d}Q(d | A>0) \\
    &= \int_{\mathcal{D}} \E[Y_1(d) - Y_0(0) | D=d, A > 0 ]\mathrm{d}Q(d | A>0) - \int_{\mathcal{D}} \E[Y_1(0) - Y_0(0) | A=0] \mathrm{d}Q(d | A>0) \\
    &= \int_{\mathcal{D}} \E[Y_1(d) - Y_0(0) | D=d, A > 0 ]\mathrm{d}Q(d | A>0) - \mu_{A=0} \\
    &= \int_{\mathcal{D}} \E[\Delta Y | D=d, A > 0 ]\mathrm{d}Q(d | A>0) - \mu_{A=0} \\
    &= \int_{\mathcal{D}} \mu_d \mathrm{d}Q(d| A>0) - \mu_{A=0}.
\end{align*}
    
\end{proof}

\subsection{Additional simulation results}

Table \ref{tab:sim_results} compares the results of estimating $\ASDT(\delta)$ with $\hat \psi^{\mathrm{CF}}$ using either BART or GLMs to estimate nuisance functions, from 1000 simulated datasets using the data generating process specified in Scenario 2. The estimators were evaluated at five increment values $\delta \in \{-10, -5, 0, 5, 10\}$ of the exponential tilt stochastic intervention. These results show that BART is able to correctly model the nonlinear nuisance functions while GLMs fail. Specifically, using BART to estimate nuisance functions resulted in low bias, low mean squared error, and nominal or near nominal coverage. In contrast, estimation using GLMs performed poorly, resulting in large bias, high mean squared error, and extreme under-coverage. Results from Scenario 1 are similar and are thus omitted. 

\begin{table}[!h]
\caption{Results from 1000 simulations.}
\label{tab:sim_results}
\begin{tabular}{ccccccccccccc}
\hline
 &  & \multicolumn{5}{c}{BART} &  & \multicolumn{5}{c}{GLMs} \\ \cline{3-7} \cline{9-13} 
$\delta$ &  & Bias & MSE & ESE & ASE & Coverage &  & Bias & MSE & ESE & ASE & Coverage \\ \hline
-10 &  & -0.7 & 0.3 & 5.4 & 5.4 & 94.9 &  & 15.8 & 3.0 & 7.3 & 7.3 & 42.6 \\
-5 &  & -0.3 & 0.1 & 3.8 & 3.8 & 94.1 &  & 15.9 & 2.8 & 5.2 & 5.1 & 12.7 \\
0 &  & 0.0 & 0.1 & 3.1 & 3.1 & 95.3 &  & 16.1 & 2.8 & 4.1 & 4.2 & 2.6 \\
5 &  & 0.2 & 0.2 & 4.0 & 3.9 & 94.5 &  & 17.2 & 3.2 & 5.4 & 5.3 & 10.1 \\
10 &  & 0.7 & 0.5 & 7.3 & 6.4 & 93.1 &  & 21.1 & 5.3 & 9.6 & 8.8 & 32.7 \\ \hline
\end{tabular} \par
\smallskip
Bias: average bias ($\times 100$), MSE: mean squared error ($\times 100$), ESE: empirical standard error ($\times 100$), ASE: average standard error estimates ($\times 100$), Coverage: 95\% confidence interval coverage (\%).
\end{table}

\subsubsection{Simulations with multiple time periods}

Simulation experiments were performed to assess finite sample performance of the proposed method in the setting with multiple post-treatment time periods. Simulation datasets were created using a staggered treatment adoption design with continuous treatment doses. Each simulated dataset contained $n=5000$ units with time periods $t \in \{0, 1, \dots, \mathcal{T}\}$ where $\mathcal{T}=3$. The data generating process was specified as follows:
\begin{align*}
\mathbf{X}_i = (X_{i1}, \ldots, X_{i10}) &\stackrel{iid}{\sim} \text{Unif}(-1, 1) \\
G_i=g | \boldX_i &\overset{iid}{\sim} \mathrm{Categorical}(\pi^{\mathrm{MTP}}_{G=g}(\mathbf{X}_i)) \\
D_i \mid \mathbf{X}_i, (G_i \leq t) &\stackrel{iid}{\sim} \text{Beta}(f_1^D(\mathbf{X}_i), f_2^D(\mathbf{X}_i)) \\
 Y_{it} \mid \mathbf{X}_i, G_i &\stackrel{iid}{\sim} N(\mu^*(\mathbf{X}_i, G_i, t), 1),
\end{align*}
where $\pi^{\mathrm{MTP}}_{G=g}(\mathbf{X}_i) = P(G_i = g \mid \mathbf{X}_i)$ are the cohort selection probabilities. The functions in the data generating process were specified to closely align with the non-linear data generating processes established in the two time period simulation setting (Section 5). The mean outcome model was specified to be $\mu^*(\mathbf{X}_i, G_i, t) = t \times f^{\mathrm{out}}(\mathbf{X}_i) + \mathbbm{1}(t \geq G_i) \zeta(D_i,t,G_i)$, where $f^{\mathrm{out}}(\mathbf{X}_i) = \text{logit}^{-1}(X_{i1})^2 X_{i2} X_{i3} + X_{i4} - 2X_{i4}^2 + 3\sin(X_{i5}^2)$ as in Section 5 and $\zeta(D_i,t,G_i)$ encodes the treatment effect. This model implies $\mu^{\mathrm{MTP}}_{0,g,t}(\mathbf{x}) \coloneqq \mathrm{E}[Y_{it} - Y_{i,g-1} | \mathbf{X}_i=\mathbf{x}, G_i>t] = (t-g+1) f^{\mathrm{out}}(\mathbf{X}_i)$ and $\mu^{\mathrm{MTP}}_{d,g,t}(\mathbf{x}) \coloneqq \mathrm{E}[Y_{it} - Y_{i,g-1} | \mathbf{X}_i=\mathbf{x}, G_i=g, D_i=d] = \mu^{\mathrm{MTP}}_{0,g,t}(\mathbf{x}) + \mathbbm{1}(t \geq g) \zeta(d,e,g)$. The treatment effect $\zeta(d,t,g)$ was specified as a polynomial response surface depending on the dose $d$ and the time exposed $e = t - g$: $\zeta(d,t,g) = (0.2 + 0.1e)d + 0.5d^2 + 0.1d^3$. The cohort propensity score $\pi^{\mathrm{MTP}}_{G=g}(\mathbf{X}_i)$ was specified 
\begin{align*}
    \pi^{\mathrm{MTP}}_{G=g}(\mathbf{X}_i) &= \frac{\exp\left( \frac{g}{2\mathcal{T}} h(\mathbf{X}_i) \right)}{\sum_{k=1}^{\mathcal{T}+1} \exp\left( \frac{k}{2\mathcal{T}} h(\mathbf{X}_i) \right)}
\end{align*}
for $g \in \{1,\dots,\mathcal{T}\}$ and
\begin{align*}
    \pi^{\mathrm{MTP}}_{G=\infty}(\mathbf{X}_i) &= \frac{\exp\left( \frac{\mathcal{T} + 1}{2\mathcal{T}} h(\mathbf{X}_i) \right)}{\sum_{k=1}^{\mathcal{T}+1} \exp\left( \frac{k}{2\mathcal{T}} h(\mathbf{X}_i) \right)},
\end{align*}
with the non-linear predictor $h(\mathbf{X}_i) = 3\cos(X_{i1}^2) - 0.5\exp(X_{i2}) - 1.5X_{i3}^2 + 4 \mathbbm{1}(X_{i4} < 0)|X_{i5}|$. Finally, the Beta distribution parameters, $f_1^D(\mathbf{X})$ and $f_2^D(\mathbf{X})$, for the conditional dose distribution were the same as those detailed in Section 5 of the main text. Like Section 5, two simulation scenarios were considered that varied in the specification of the dose density functions $f_1^D(\mathbf{X})$ and $f_2^D(\mathbf{X})$.

For estimation, the proposed cross-fit estimator with $K = 5$ folds was used to evaluate the average shifted dose effects across a sequence of exponential tilt increments $\delta \in \{-10, -9, \dots, 10 \}$. The nuisance functions---specifically the conditional mean outcome functions $\mu^{\mathrm{MTP}}_{0,g,t}(\mathbf{x})$ and $\mu^{\mathrm{MTP}}_{d,g,t}(\mathbf{x})$ and the cohort propensity score $\pi^{\mathrm{MTP}}_{G=g}(\mathbf{x})$---were estimated using three separate approaches: BART, GLMs with maximum likelihood estimation, and an oracle estimator utilizing the true data generating functions. The dose propensity scores were estimated using the kernel-transformed outcome method as described in the main text. The estimator $\hat \psi^{\mathrm{MTP, CF}}(\delta,g,t)$ was computed for each valid cohort-time pair along with estimators of the aggregated parameters: $\ASDT^{\mathrm{es}}(Q, e)$, $\ASDT^{\mathrm{es,bal}}(Q, e; e')$, $ \ASDT^{\mathrm{cohort}}(Q, g; e_1, e_2)$, $\ASDT^{\mathrm{overall}}(Q)$.

\begin{figure}[p!]
    \centering
    \begin{subfigure}{\textwidth}
        \centering
        \includegraphics[width=0.95\textwidth]{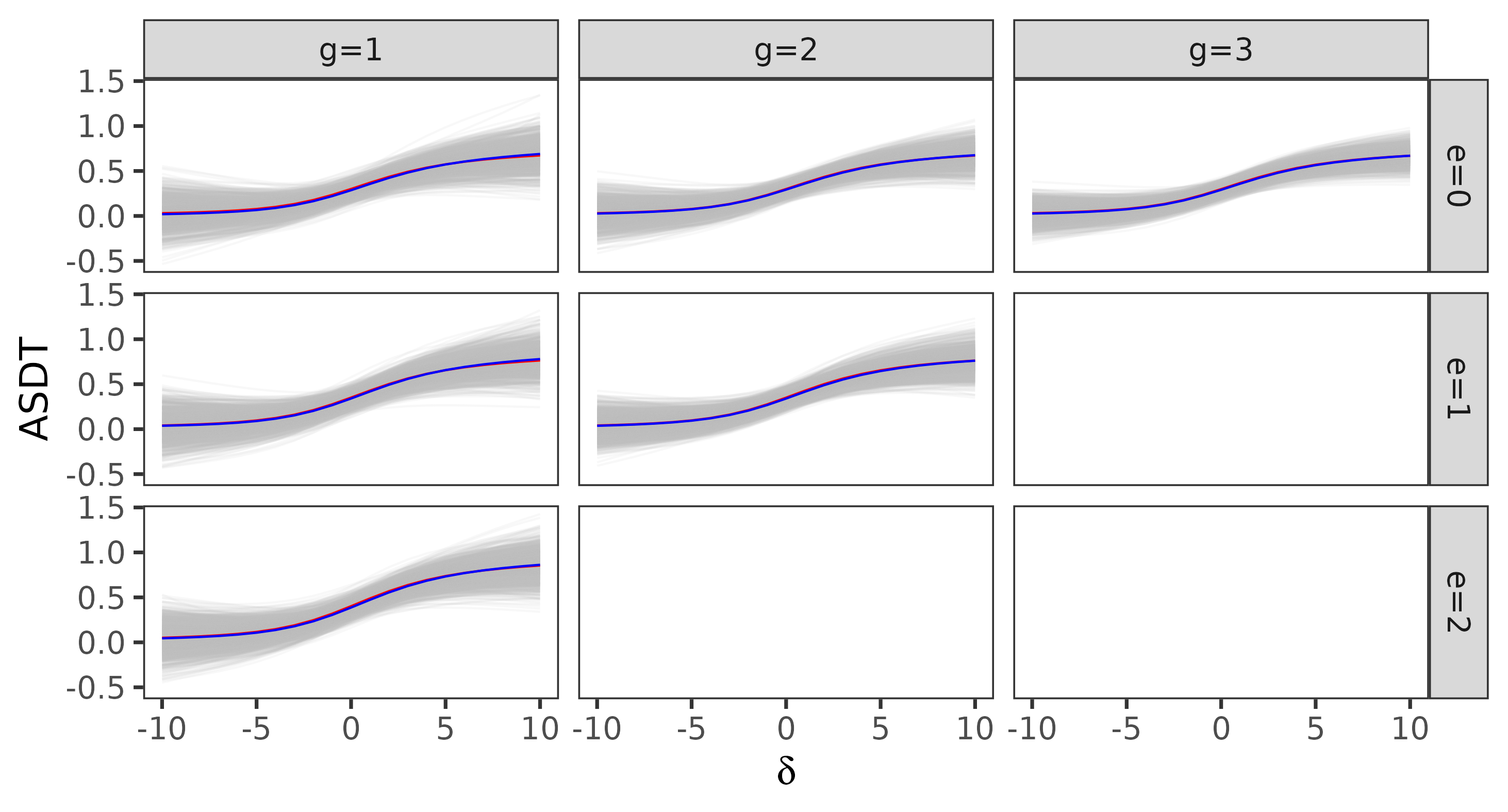}
        \caption{BART}
    \end{subfigure}
    \par \vspace{0.7cm}
    \begin{subfigure}{\textwidth}
        \centering
        \includegraphics[width=0.95\textwidth]{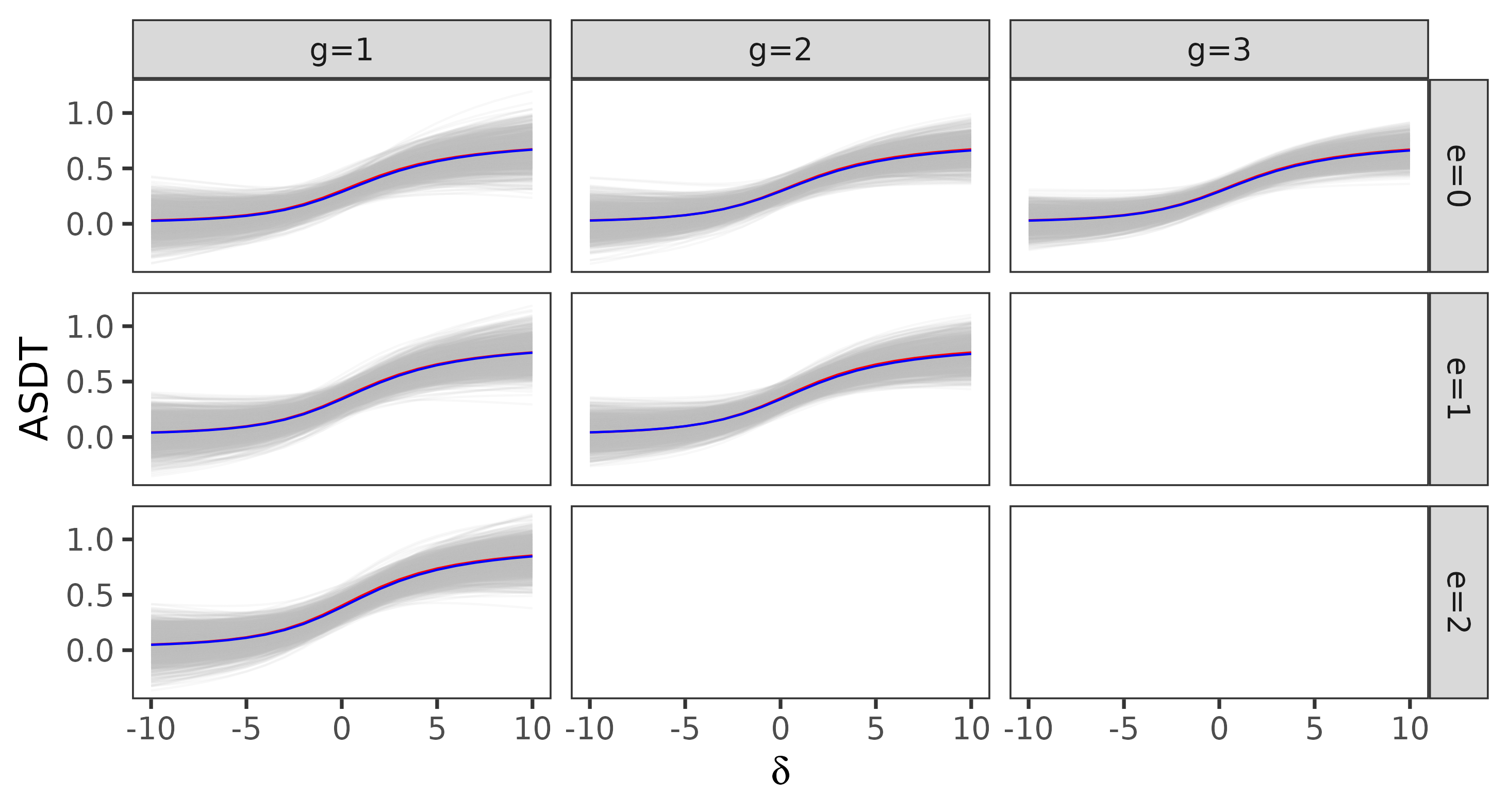}
        \caption{Oracle}
    \end{subfigure}
    \caption{Estimated $\ASDT(\delta, g, t)$ under the exponential tilt stochastic policy with varying increments $\delta$ from 1000 simulations generated under Scenario 1. Nuisance functions were estimated using (a) BART, or (b) oracle models. Grey lines denote estimates from a single simulation; the blue line is the average of the grey lines; and the red line is the estimand.}
    \label{fig:sc1_disagg-point}
\end{figure}

\begin{figure}[p!]
    \centering
    \begin{subfigure}{\textwidth}
        \centering
        \includegraphics[width=0.95\textwidth]{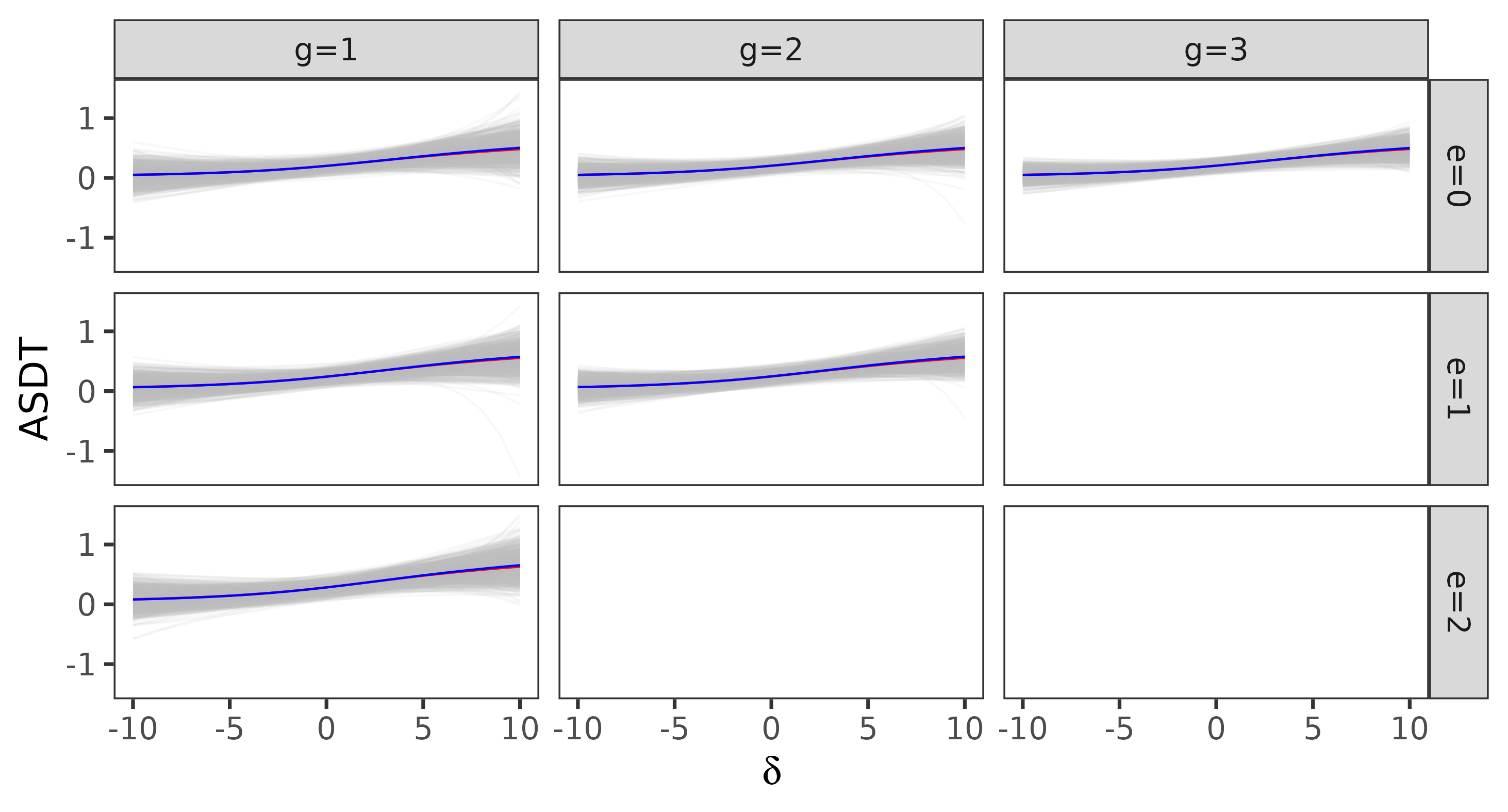}
        \caption{BART}
    \end{subfigure}
    \par \vspace{0.7cm}
    \begin{subfigure}{\textwidth}
        \centering
        \includegraphics[width=0.95\textwidth]{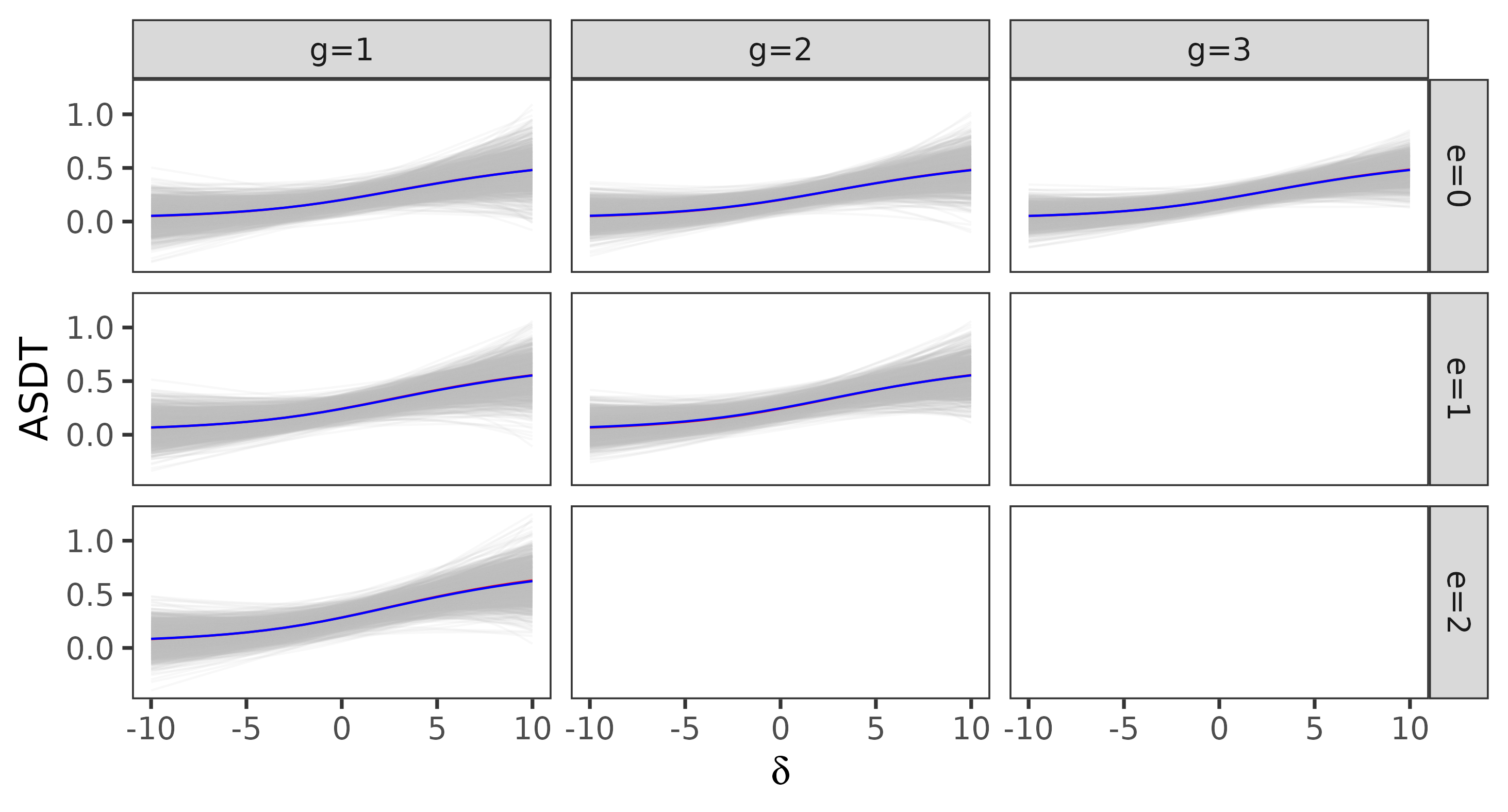}
        \caption{Oracle}
    \end{subfigure}
    \caption{Estimated $\ASDT(\delta, g, t)$ under the exponential tilt stochastic policy with varying increments $\delta$ from 1000 simulations generated under Scenario 2. Nuisance functions were estimated using (a) BART, or (b) oracle models. Grey lines denote estimates from a single simulation; the blue line is the average of the grey lines; and the red line is the estimand.}
    \label{fig:sc2_disagg-point}
\end{figure}

\begin{figure}[p!]
    \centering
    \begin{subfigure}{\textwidth}
        \centering
        \includegraphics[width=0.95\textwidth]{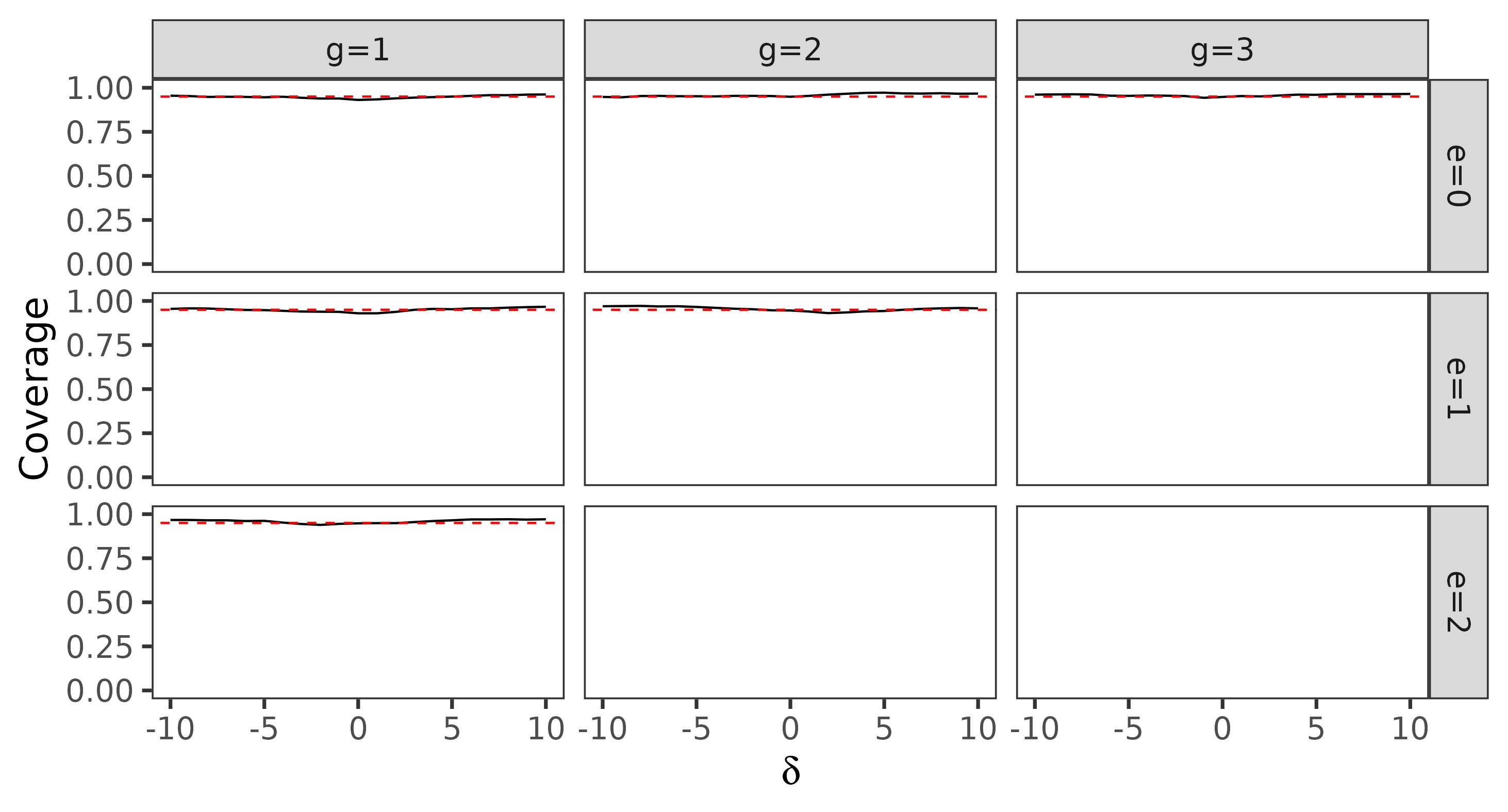}
        \caption{BART}
    \end{subfigure}
    \par \vspace{0.7cm}
    \begin{subfigure}{\textwidth}
        \centering
        \includegraphics[width=0.95\textwidth]{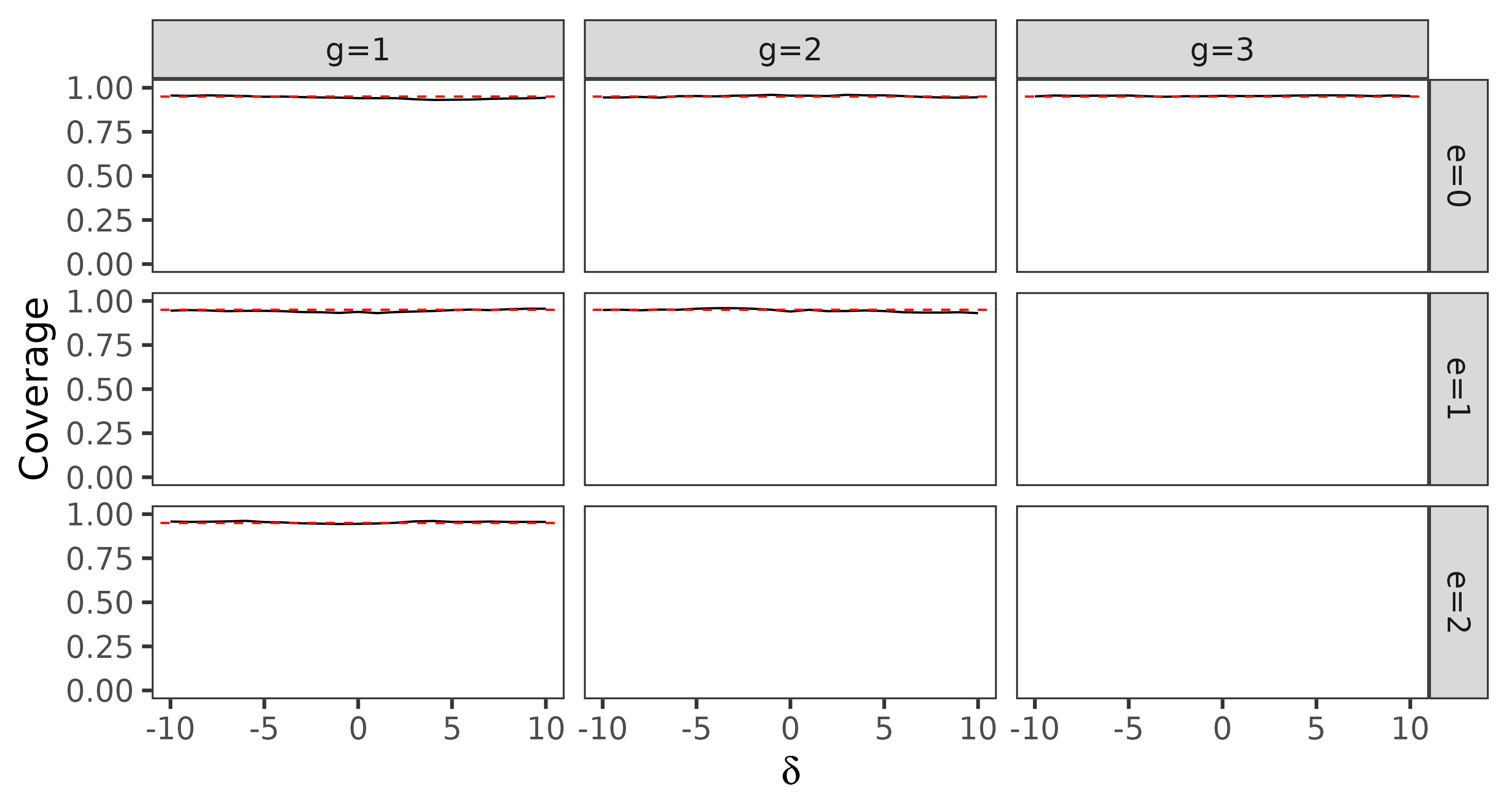}
        \caption{Oracle}
    \end{subfigure}
    \caption{Empirical coverage rate of 95\% confidence intervals from 1000 simulations generated under Scenario 1. Nuisance functions were estimated using (a) BART, or (b) oracle models. The black line shows the proportion of confidence intervals that contained the true estimand for each $\delta$ and the red dashed line is set at $0.95$.}
    \label{fig:sc1_disagg-cov}
\end{figure}

\begin{figure}[p!]
    \centering
    \begin{subfigure}{\textwidth}
        \centering
        \includegraphics[width=0.95\textwidth]{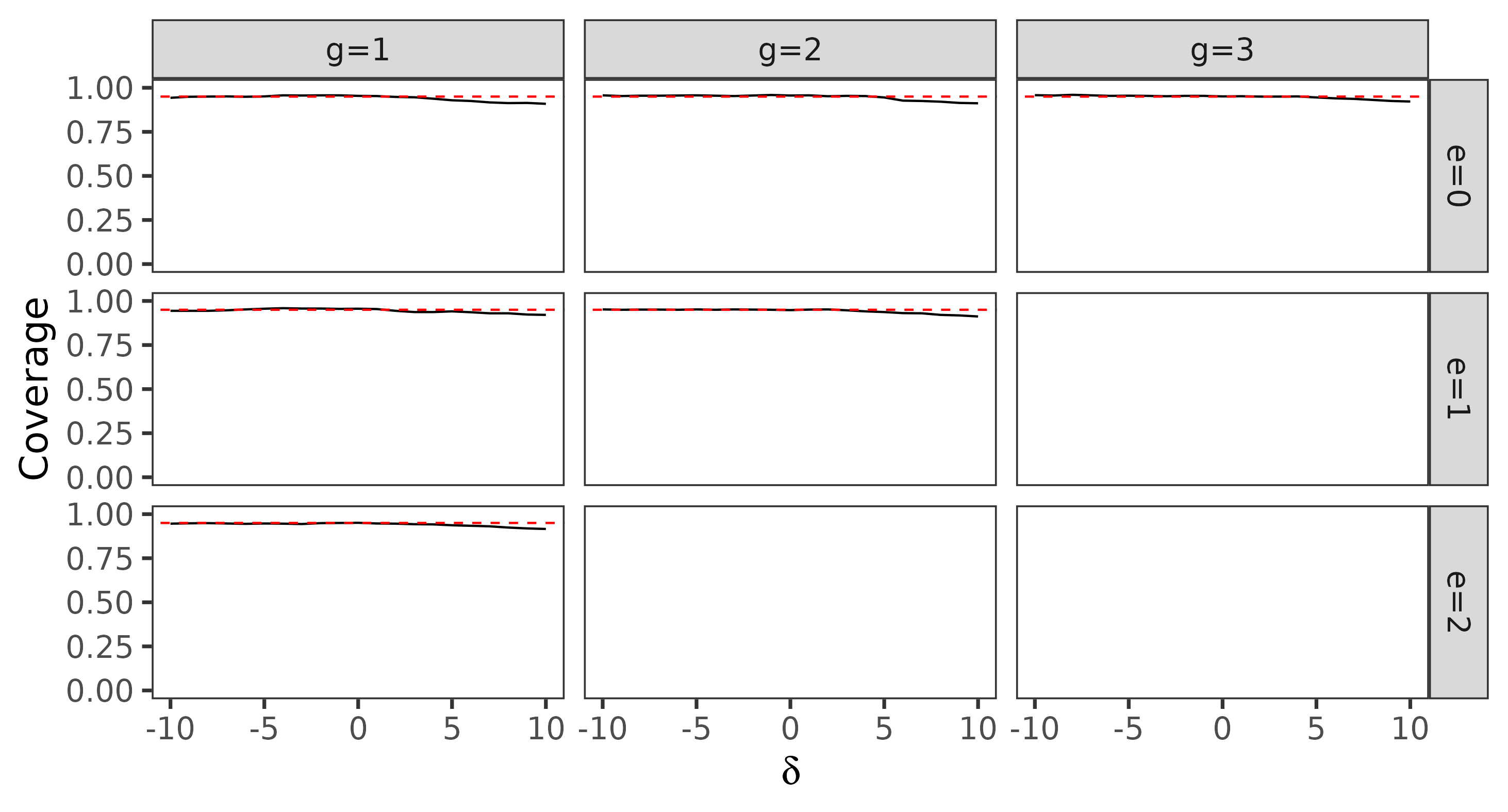}
        \caption{BART}
    \end{subfigure}
    \par \vspace{0.7cm}
    \begin{subfigure}{\textwidth}
        \centering
        \includegraphics[width=0.95\textwidth]{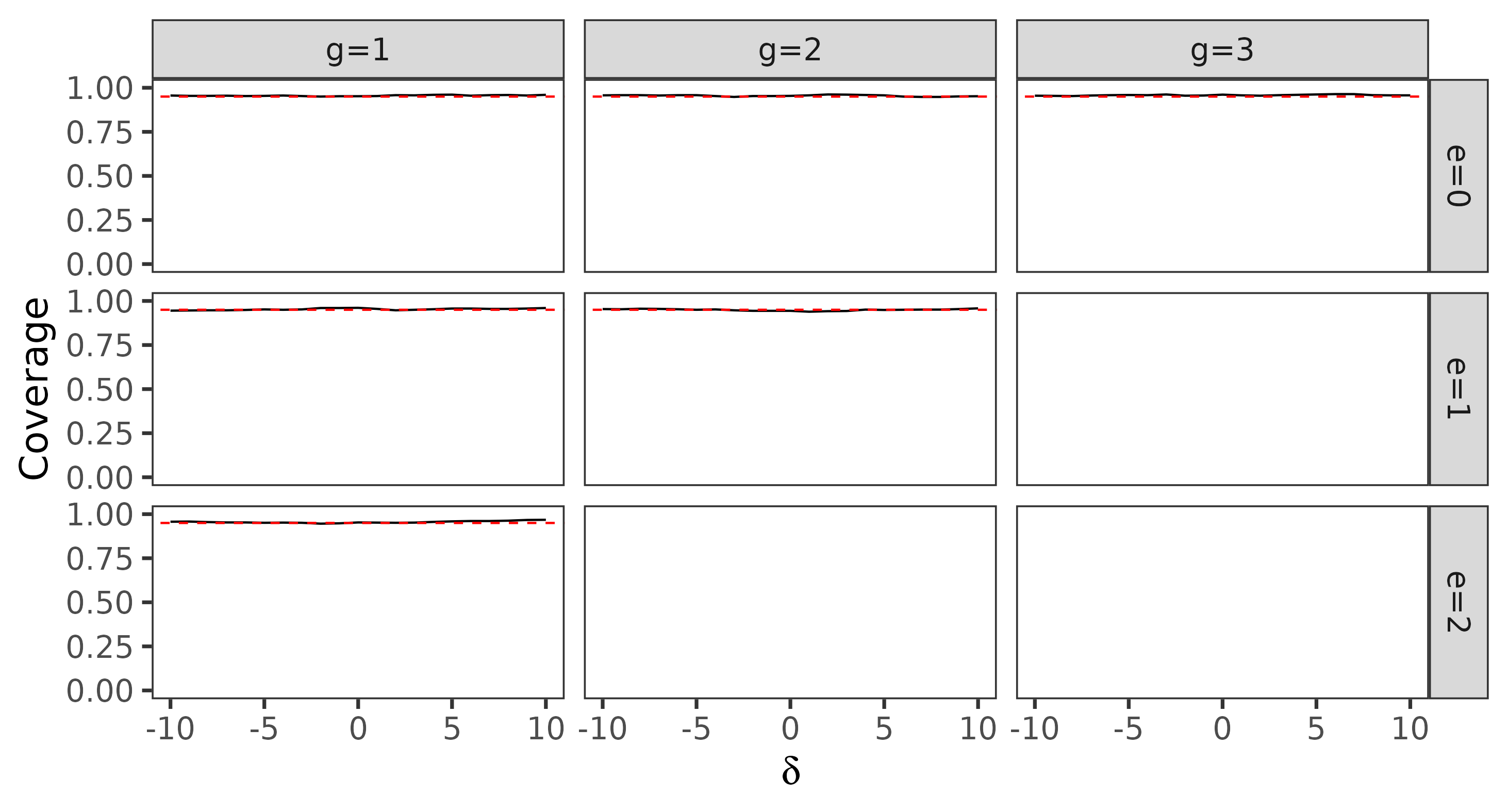}
        \caption{Oracle}
    \end{subfigure}
    \caption{Empirical coverage rate of 95\% confidence intervals from 1000 simulations generated under Scenario 2. Nuisance functions were estimated using (a) BART, or (b) oracle models. The black line shows the proportion of confidence intervals that contained the true estimand for each $\delta$ and the red dashed line is set at $0.95$.}
    \label{fig:sc2_disagg-cov}
\end{figure}

\begin{figure}[p!]
    \centering
    \begin{subfigure}{\textwidth}
        \centering
        \includegraphics[width=0.95\textwidth]{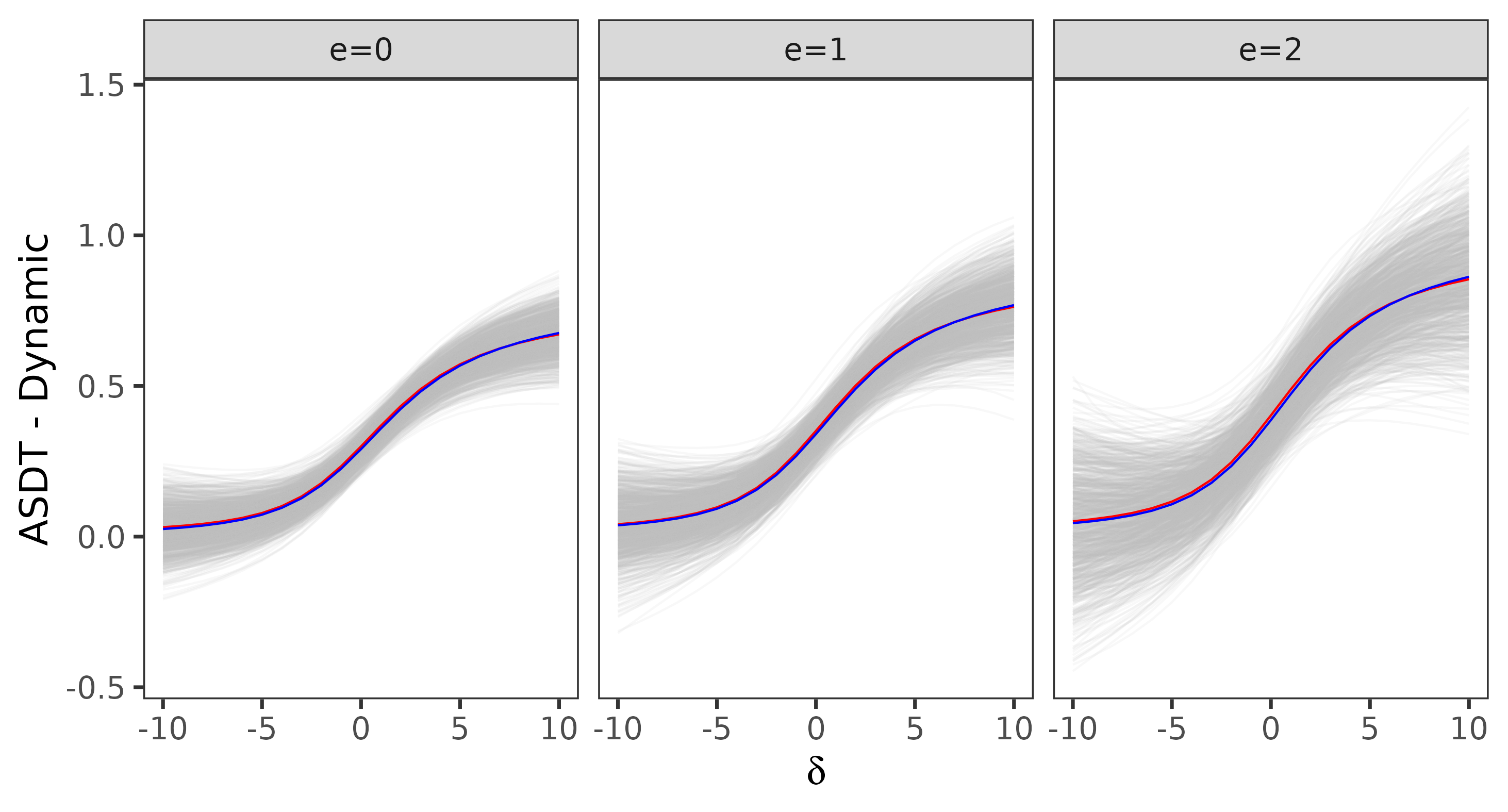}
        \caption{BART}
    \end{subfigure}
    \par \vspace{0.7cm}
    \begin{subfigure}{\textwidth}
        \centering
        \includegraphics[width=0.95\textwidth]{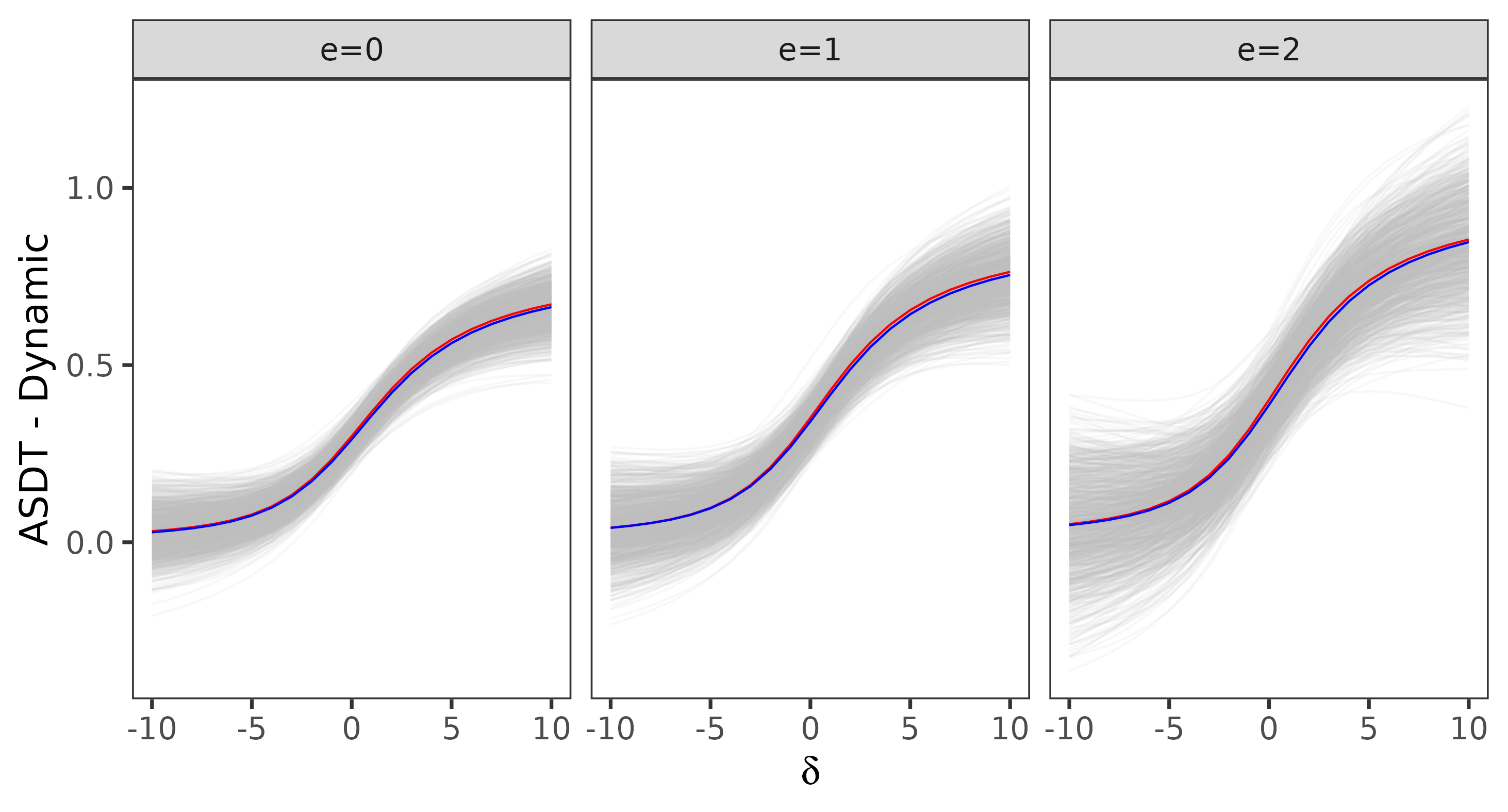}
        \caption{Oracle}
    \end{subfigure}
    \caption{Estimated $\ASDT^{\mathrm{es}}(\delta, e)$ under the exponential tilt stochastic policy with varying increments $\delta$ from 1000 simulations generated under Scenario 1. Nuisance functions were estimated using (a) BART, or (b) oracle models. Grey lines denote estimates from a single simulation; the blue line is the average of the grey lines; and the red line is the estimand.}
    \label{fig:sc1_agg_dynamic-point}
\end{figure}

\begin{figure}[p!]
    \centering
    \begin{subfigure}{\textwidth}
        \centering
        \includegraphics[width=0.95\textwidth]{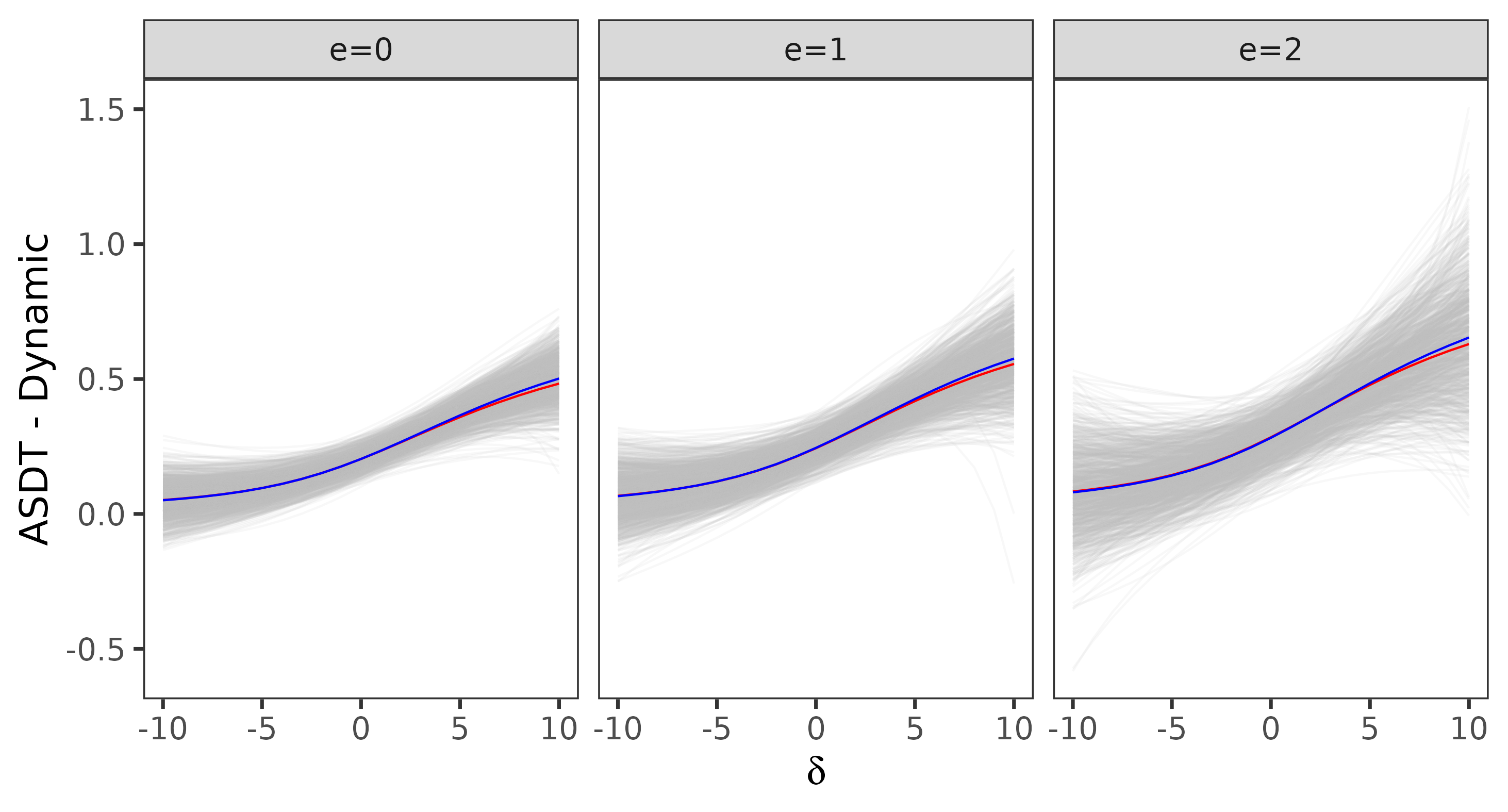}
        \caption{BART}
    \end{subfigure}
    \par \vspace{0.7cm}
    \begin{subfigure}{\textwidth}
        \centering
        \includegraphics[width=0.95\textwidth]{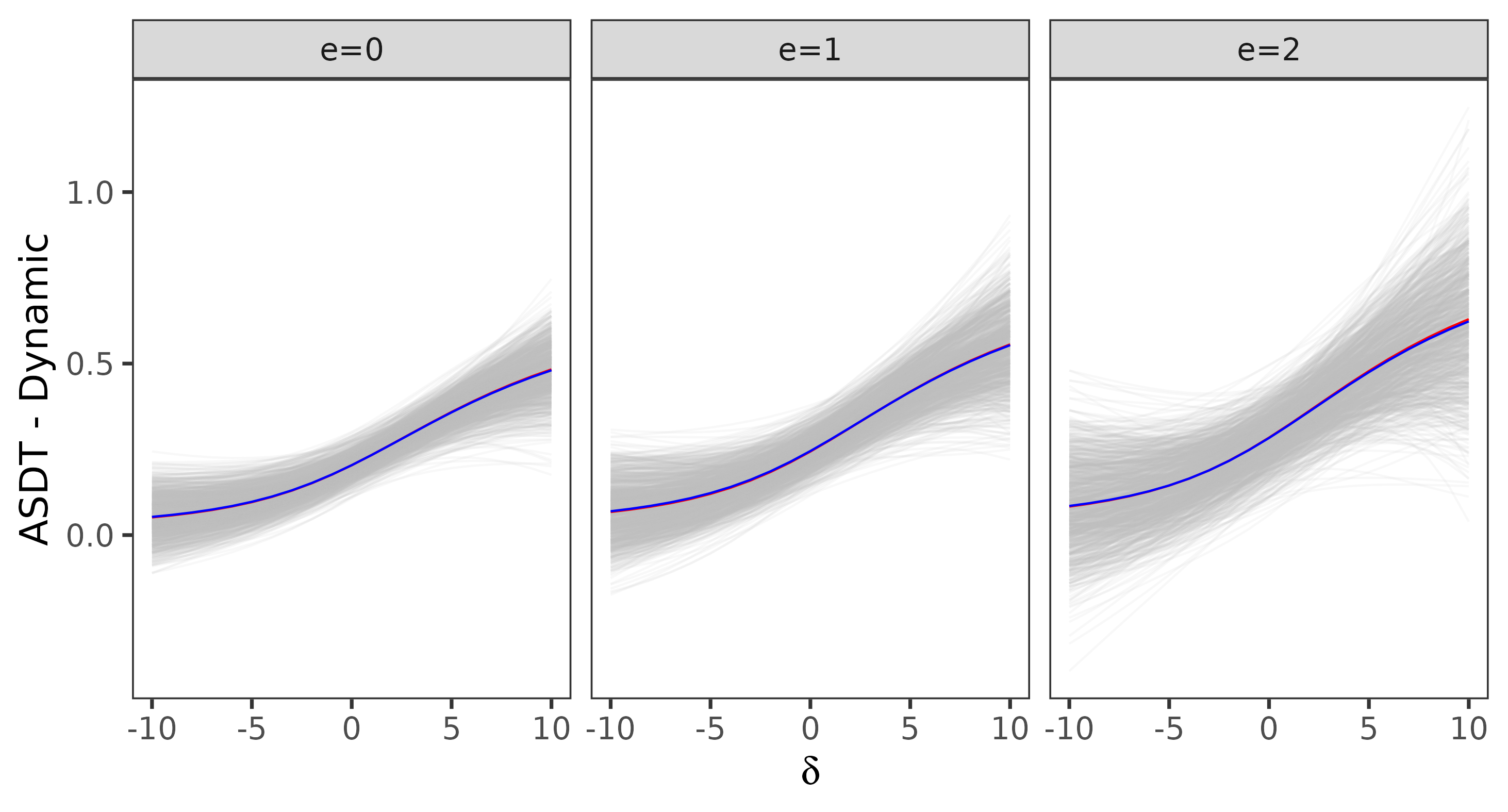}
        \caption{Oracle}
    \end{subfigure}
    \caption{Estimated $\ASDT^{\mathrm{es}}(\delta, e)$ under the exponential tilt stochastic policy with varying increments $\delta$ from 1000 simulations generated under Scenario 2. Nuisance functions were estimated using (a) BART, or (b) oracle models. Grey lines denote estimates from a single simulation; the blue line is the average of the grey lines; and the red line is the estimand.}
    \label{fig:sc2_agg_dynamic-point}
\end{figure}

\begin{figure}[p!]
    \centering
    \begin{subfigure}{\textwidth}
        \centering
        \includegraphics[width=0.95\textwidth]{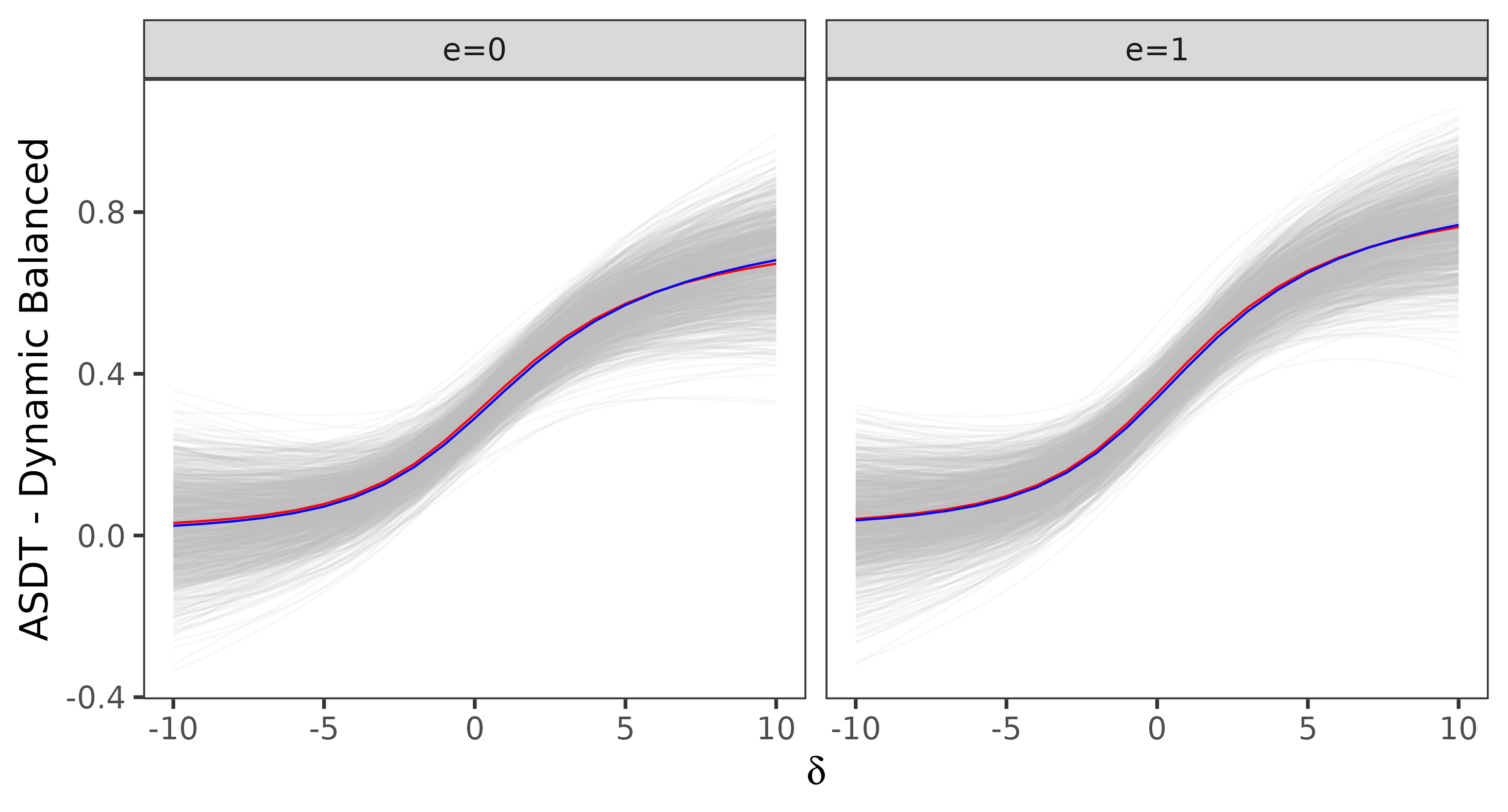}
        \caption{BART}
    \end{subfigure}
    \par \vspace{0.7cm}
    \begin{subfigure}{\textwidth}
        \centering
        \includegraphics[width=0.95\textwidth]{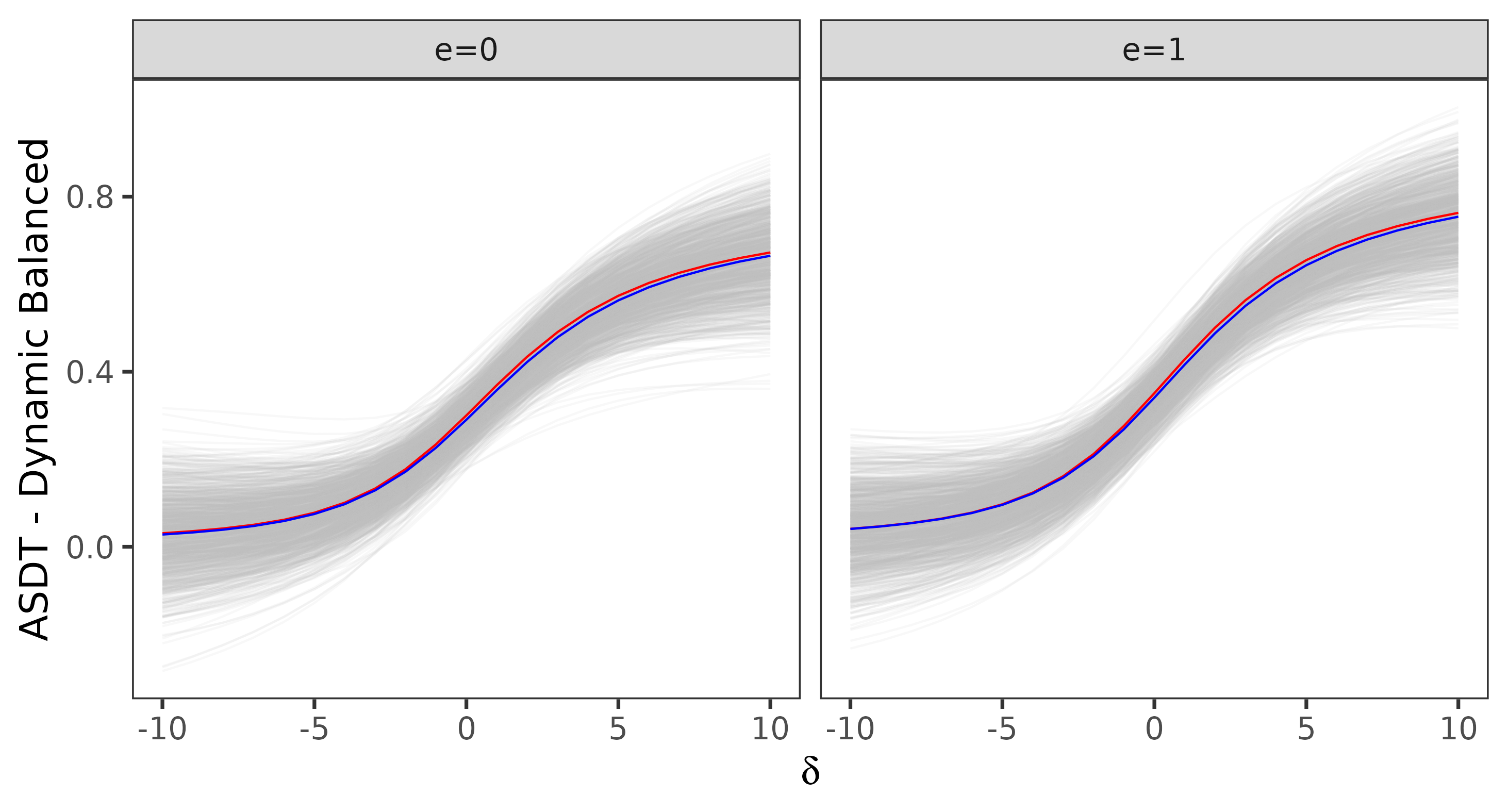}
        \caption{Oracle}
    \end{subfigure}
    \caption{Estimated $\ASDT^{\mathrm{es, bal}}(\delta, e; 1)$ under the exponential tilt stochastic policy with varying increments $\delta$ from 1000 simulations generated under Scenario 1. Nuisance functions were estimated using (a) BART, or (b) oracle models. Grey lines denote estimates from a single simulation; the blue line is the average of the grey lines; and the red line is the estimand.}
    \label{fig:sc1_agg_balanced-point}
\end{figure}

\begin{figure}[p!]
    \centering
    \begin{subfigure}{\textwidth}
        \centering
        \includegraphics[width=0.95\textwidth]{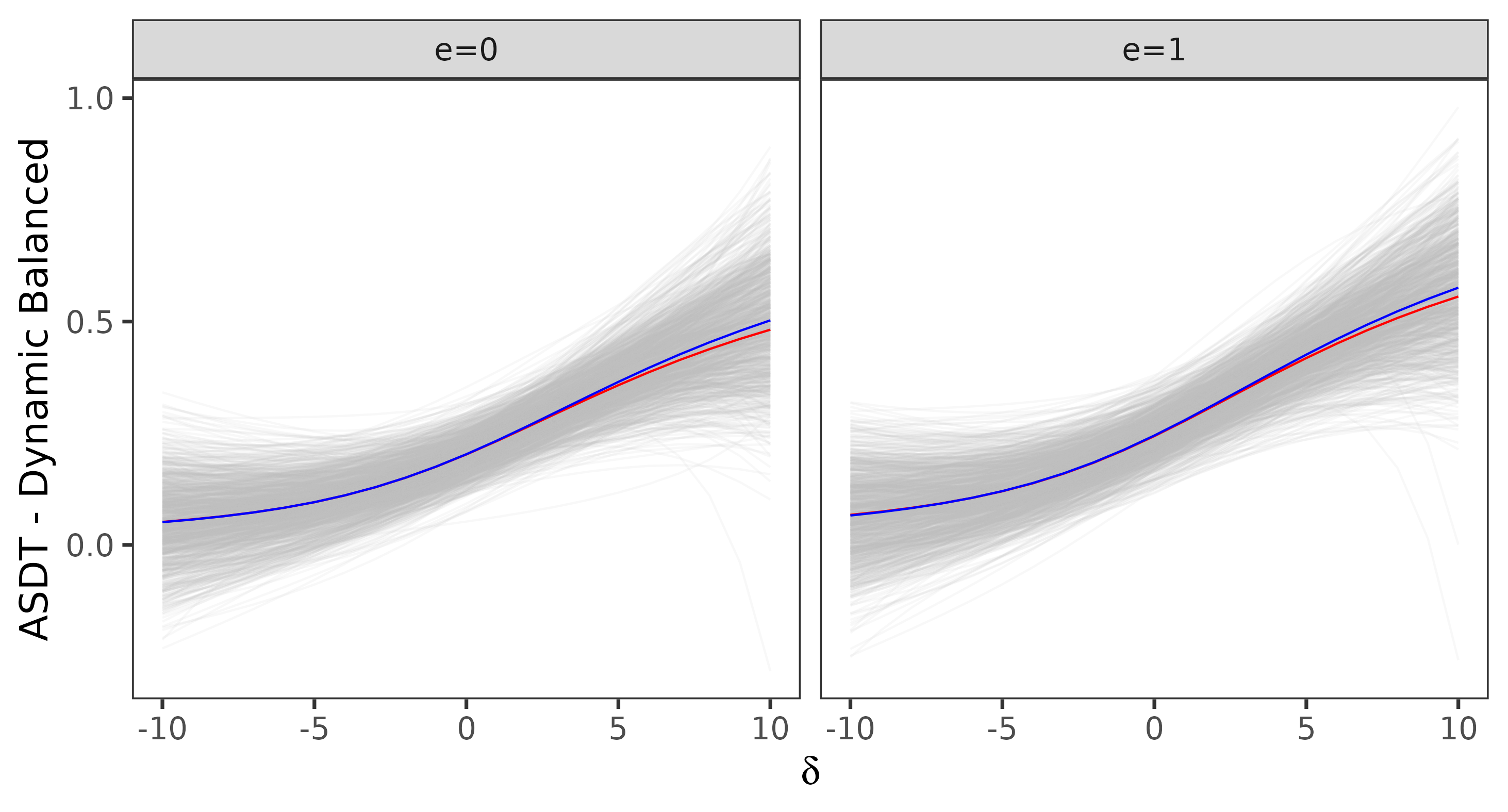}
        \caption{BART}
    \end{subfigure}
    \par \vspace{0.7cm}
    \begin{subfigure}{\textwidth}
        \centering
        \includegraphics[width=0.95\textwidth]{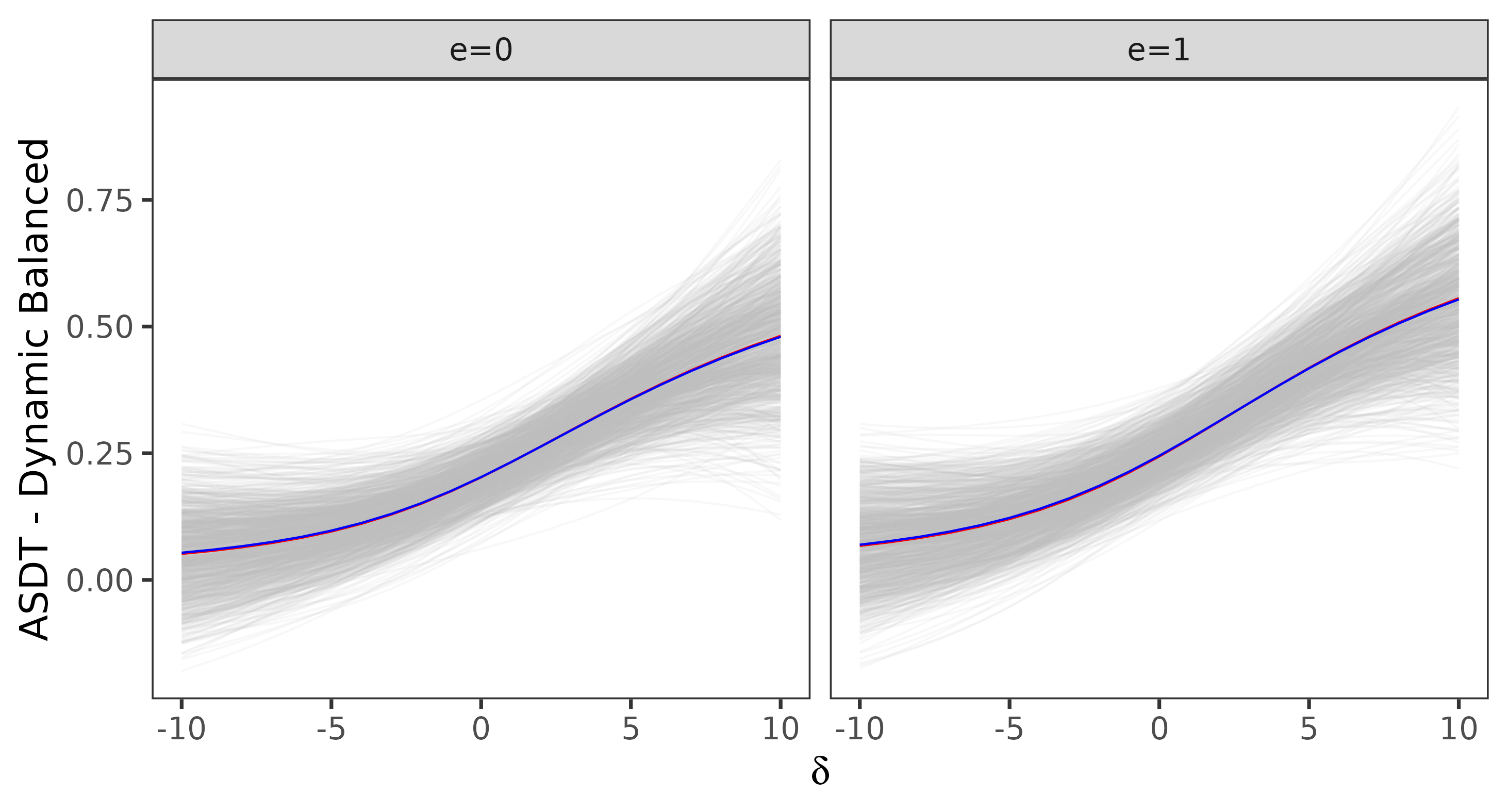}
        \caption{Oracle}
    \end{subfigure}
    \caption{Estimated $\ASDT^{\mathrm{es, bal}}(\delta, e; 1)$ under the exponential tilt stochastic policy with varying increments $\delta$ from 1000 simulations generated under Scenario 2. Nuisance functions were estimated using (a) BART, or (b) oracle models. Grey lines denote estimates from a single simulation; the blue line is the average of the grey lines; and the red line is the estimand.}
    \label{fig:sc2_agg_balanced-point}
\end{figure}

\begin{figure}[p!]
    \centering
    \begin{subfigure}{\textwidth}
        \centering
        \includegraphics[width=0.95\textwidth]{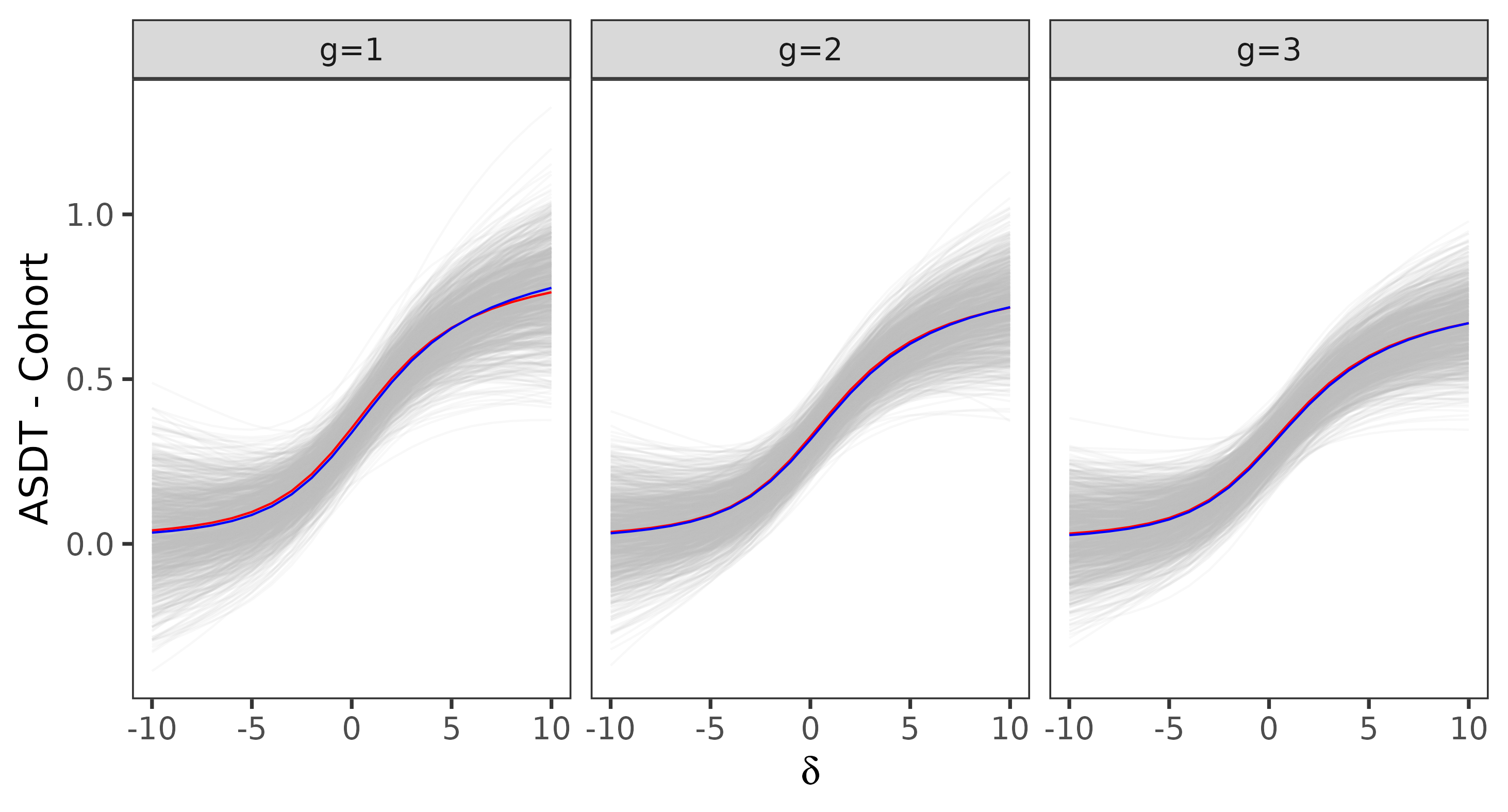}
        \caption{BART}
    \end{subfigure}
    \par \vspace{0.7cm}
    \begin{subfigure}{\textwidth}
        \centering
        \includegraphics[width=0.95\textwidth]{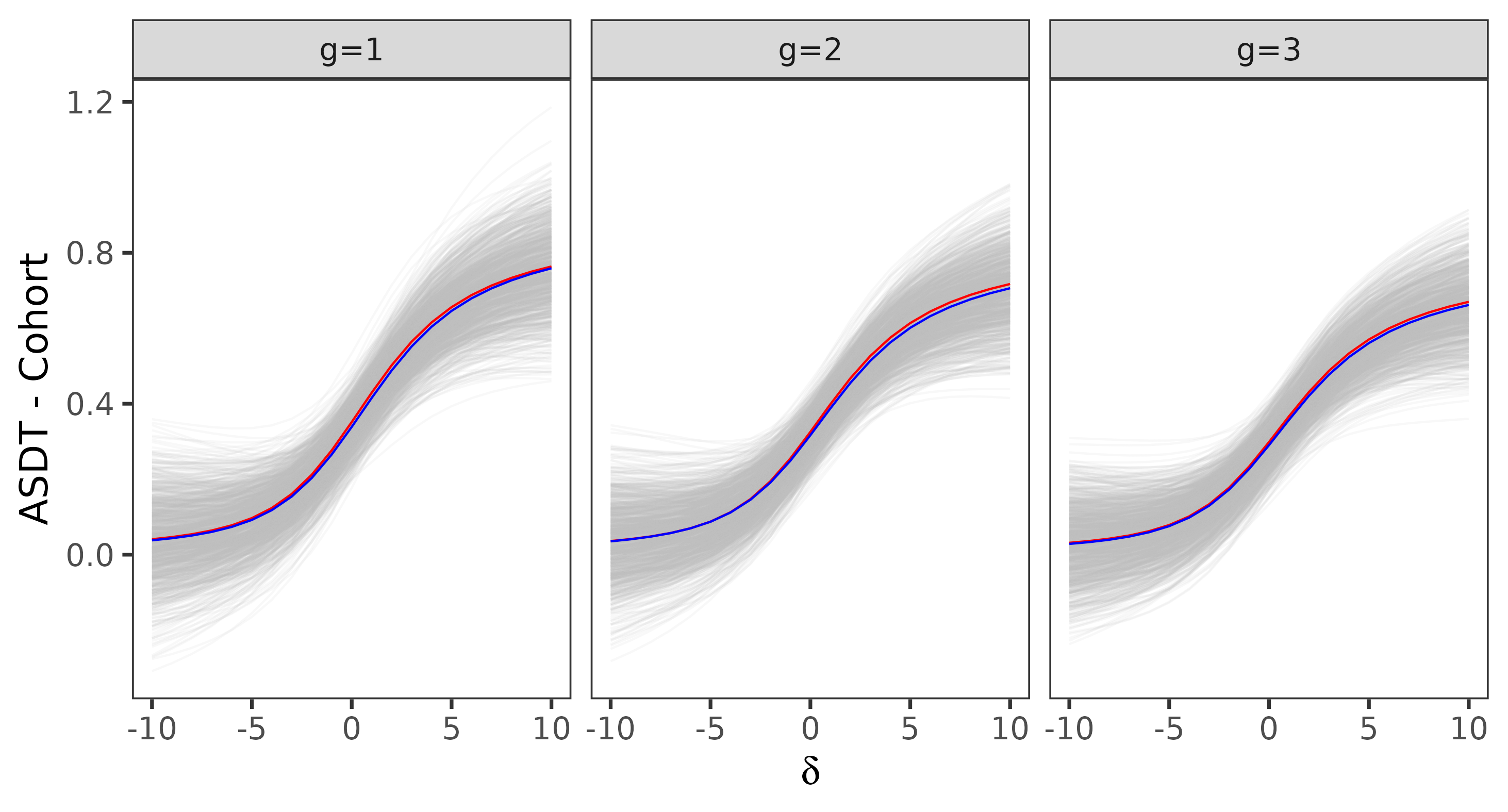}
        \caption{Oracle}
    \end{subfigure}
    \caption{Estimated $\ASDT^{\mathrm{cohort}}(\delta, g; 0, \mathcal{T}-g))$ under the exponential tilt stochastic policy with varying increments $\delta$ from 1000 simulations generated under Scenario 1. Nuisance functions were estimated using (a) BART, or (b) oracle models. Grey lines denote estimates from a single simulation; the blue line is the average of the grey lines; and the red line is the estimand.}
    \label{fig:sc1_agg_group-point}
\end{figure}

\begin{figure}[p!]
    \centering
    \begin{subfigure}{\textwidth}
        \centering
        \includegraphics[width=0.95\textwidth]{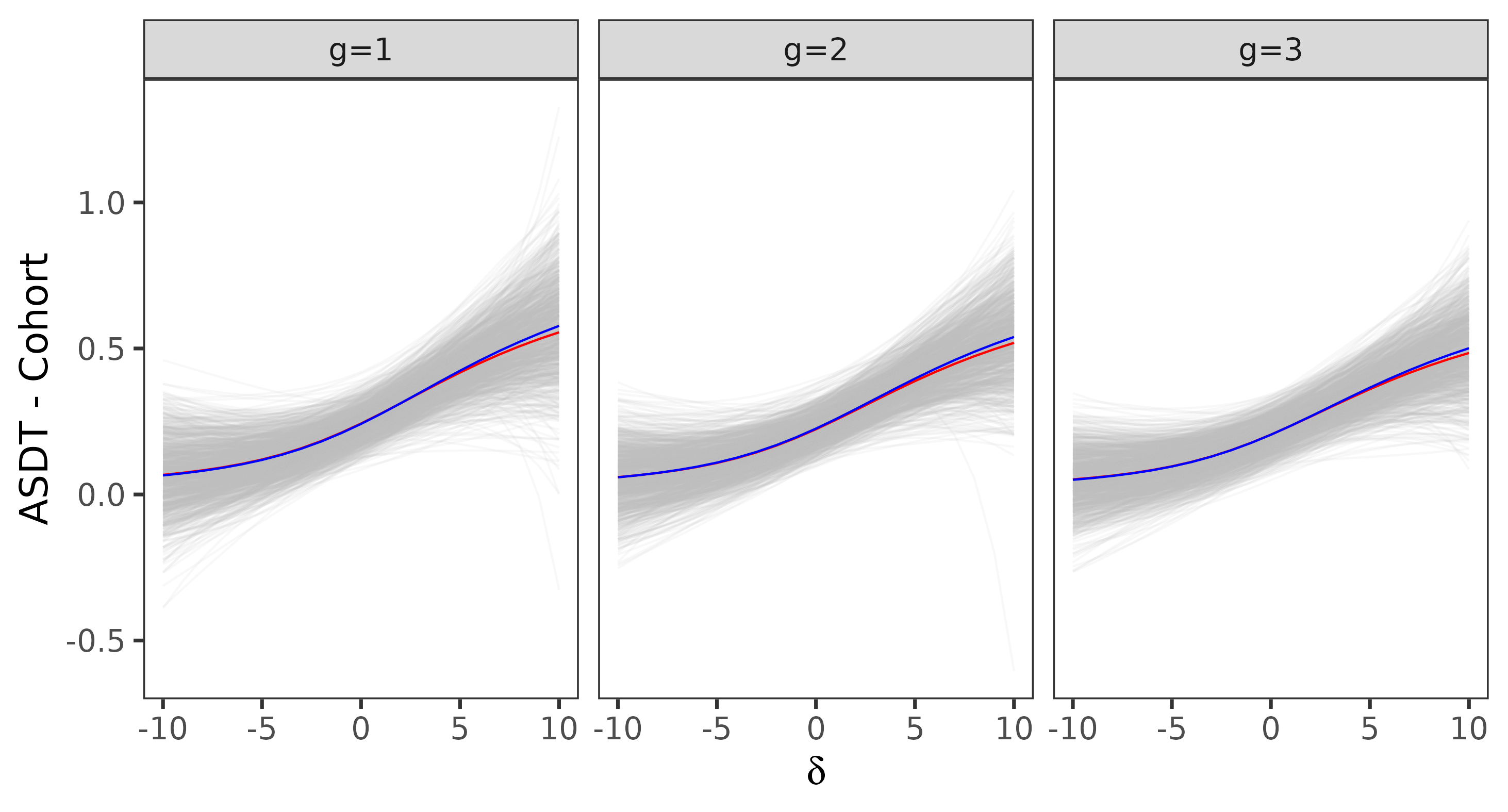}
        \caption{BART}
    \end{subfigure}
    \par \vspace{0.7cm}
    \begin{subfigure}{\textwidth}
        \centering
        \includegraphics[width=0.95\textwidth]{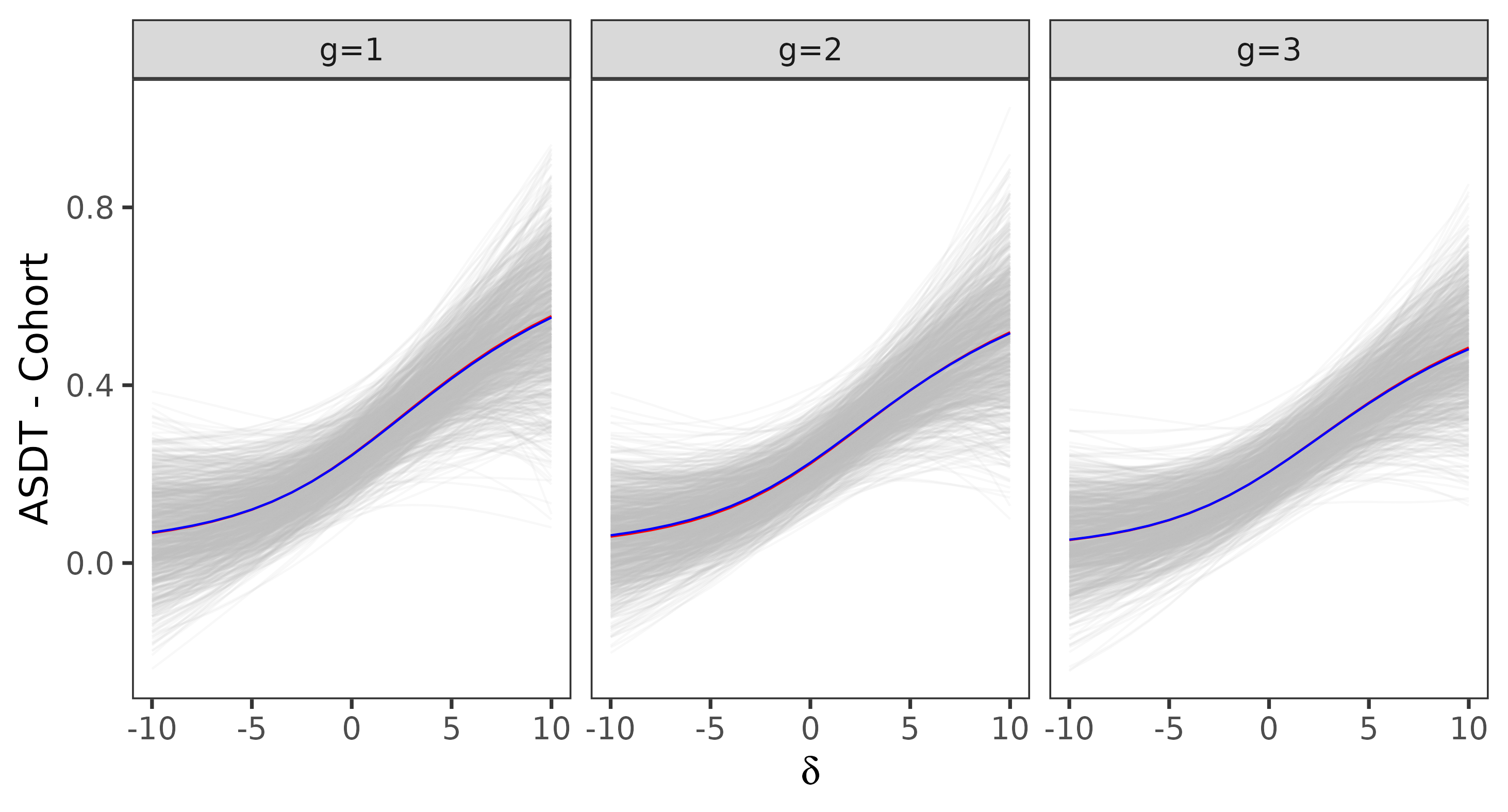}
        \caption{Oracle}
    \end{subfigure}
    \caption{Estimated $\ASDT^{\mathrm{cohort}}(\delta, g; 0, \mathcal{T}-g)$ under the exponential tilt stochastic policy with varying increments $\delta$ from 1000 simulations generated under Scenario 2. Nuisance functions were estimated using (a) BART, or (b) oracle models. Grey lines denote estimates from a single simulation; the blue line is the average of the grey lines; and the red line is the estimand.}
    \label{fig:sc2_agg_group-point}
\end{figure}

\begin{figure}[p!]
    \centering
    \begin{subfigure}{\textwidth}
        \centering
        \includegraphics[width=0.95\textwidth]{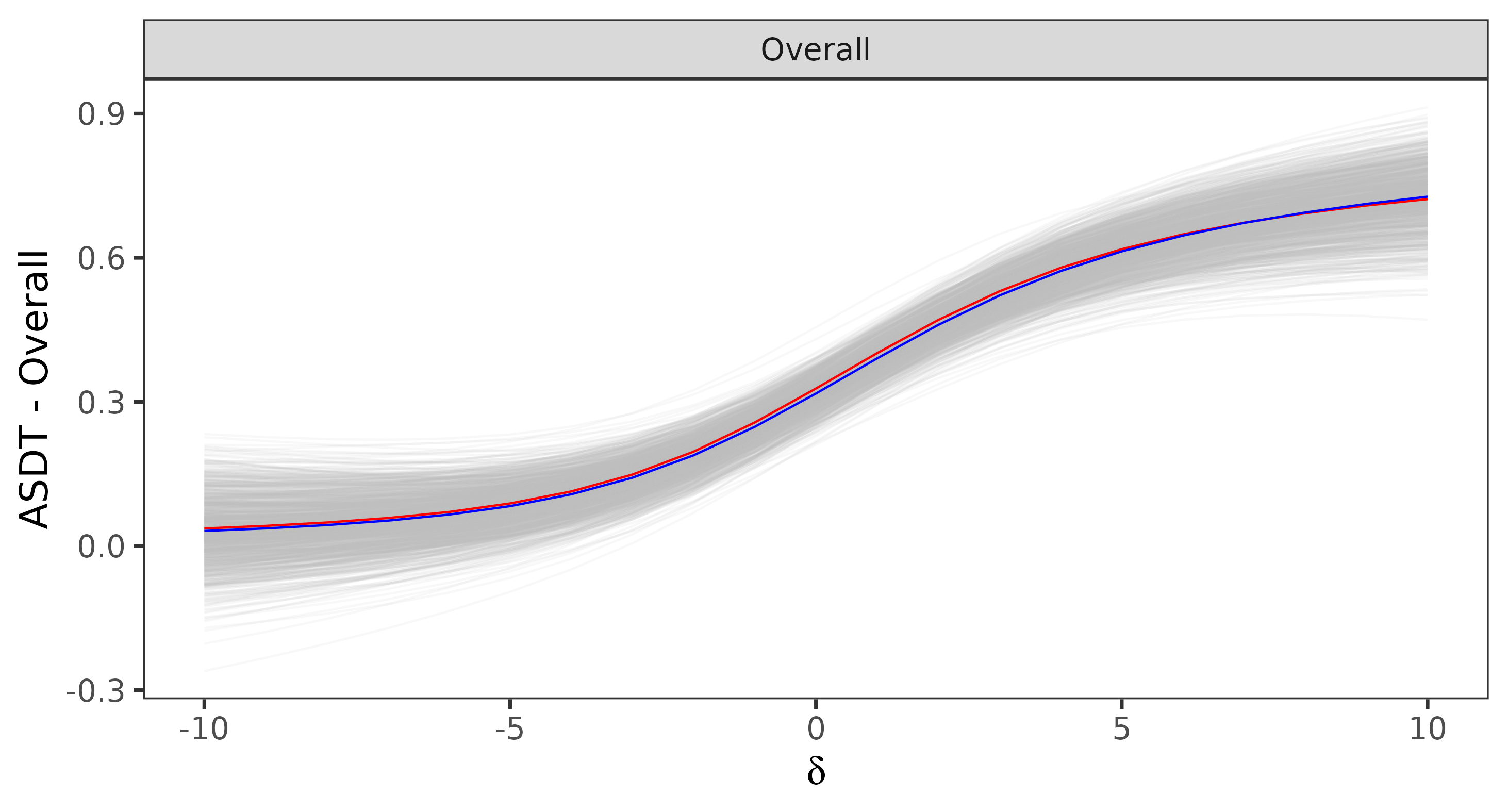}
        \caption{BART}
    \end{subfigure}
    \par \vspace{0.7cm}
    \begin{subfigure}{\textwidth}
        \centering
        \includegraphics[width=0.95\textwidth]{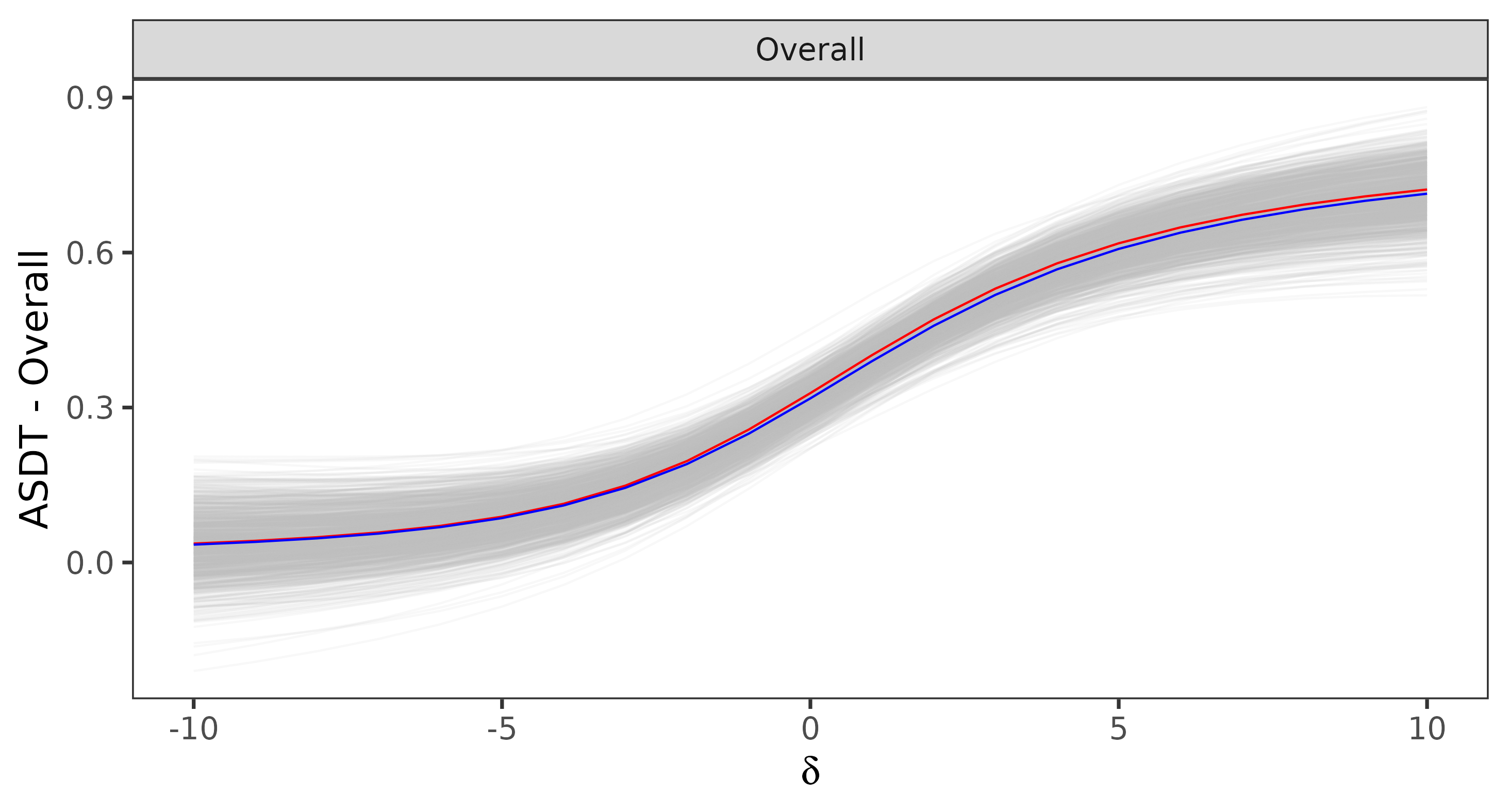}
        \caption{Oracle}
    \end{subfigure}
    \caption{Estimated $\ASDT^{\mathrm{overall}}(\delta)$ under the exponential tilt stochastic policy with varying increments $\delta$ from 1000 simulations generated under Scenario 1. Nuisance functions were estimated using (a) BART, or (b) oracle models. Grey lines denote estimates from a single simulation; the blue line is the average of the grey lines; and the red line is the estimand.}
    \label{fig:sc1_agg_overall-point}
\end{figure}

\begin{figure}[p!]
    \centering
    \begin{subfigure}{\textwidth}
        \centering
        \includegraphics[width=0.95\textwidth]{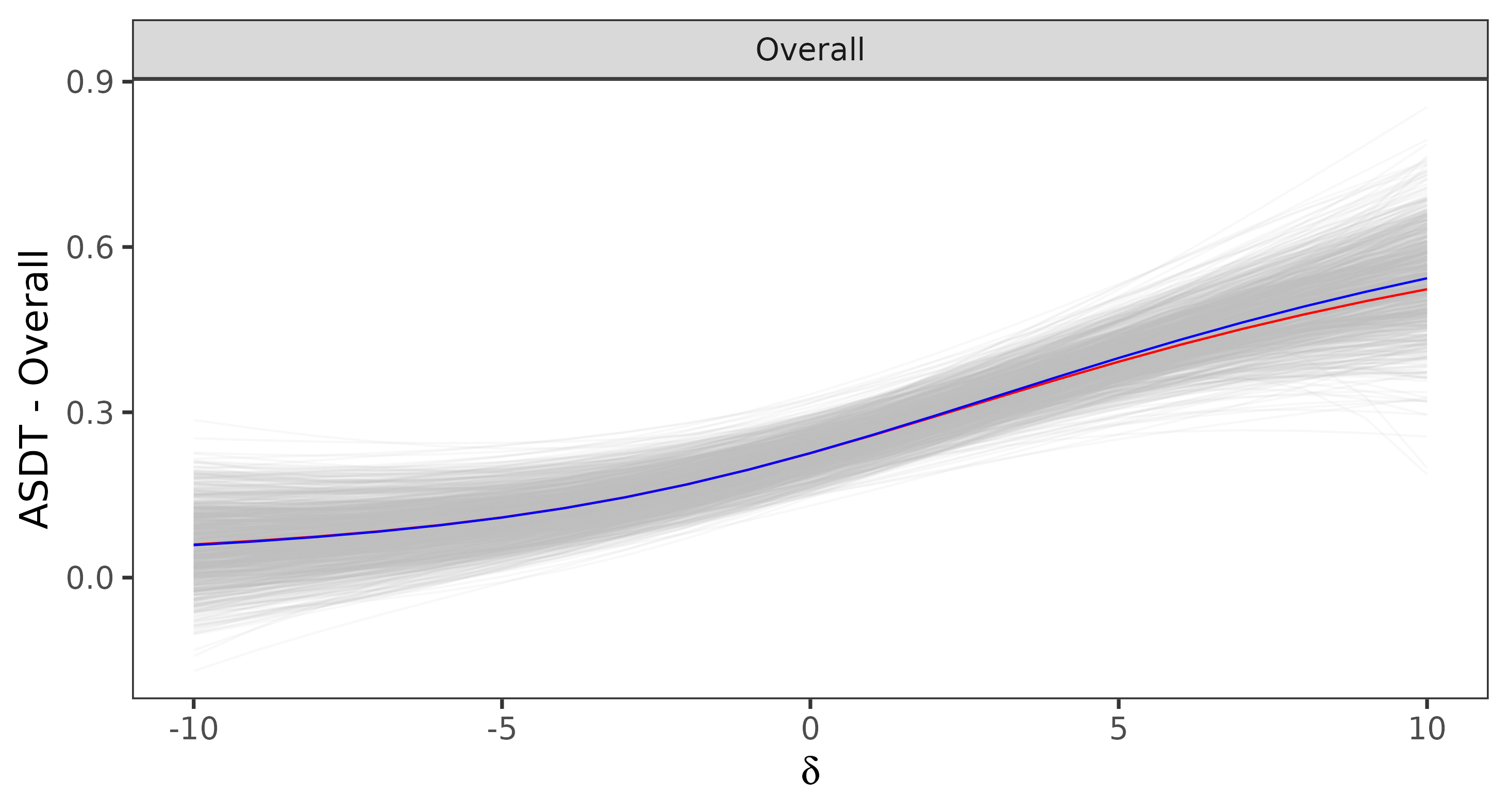}
        \caption{BART}
    \end{subfigure}
    \par \vspace{0.7cm}
    \begin{subfigure}{\textwidth}
        \centering
        \includegraphics[width=0.95\textwidth]{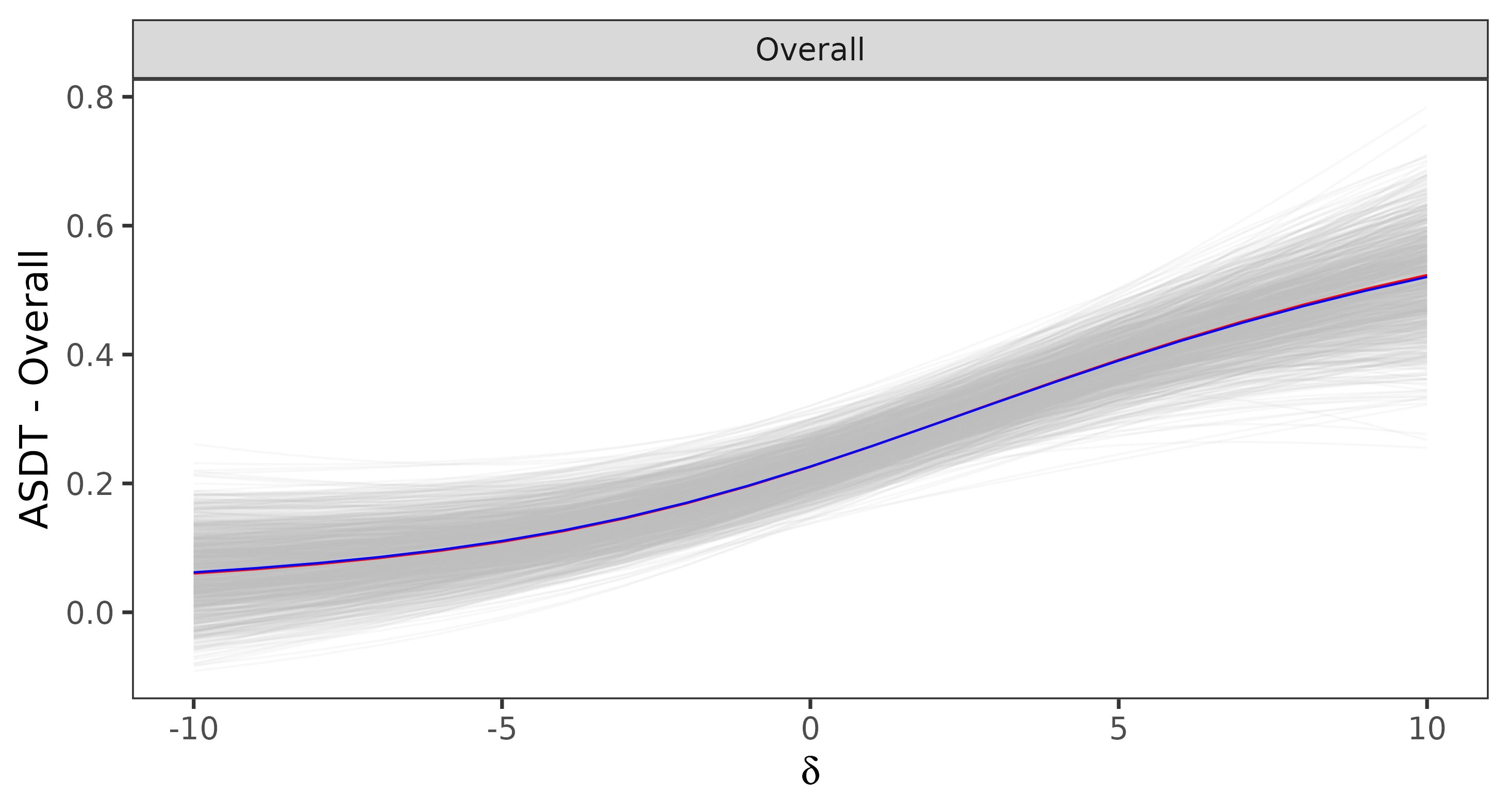}
        \caption{Oracle}
    \end{subfigure}
    \caption{Estimated $\ASDT^{\mathrm{overall}}(\delta)$ under the exponential tilt stochastic policy with varying increments $\delta$ from 1000 simulations generated under Scenario 2. Nuisance functions were estimated using (a) BART, or (b) oracle models. Grey lines denote estimates from a single simulation; the blue line is the average of the grey lines; and the red line is the estimand.}
    \label{fig:sc2_agg_overall-point}
\end{figure}

\begin{figure}[p!]
    \centering
    \begin{subfigure}{\textwidth}
        \centering
        \includegraphics[width=0.95\textwidth]{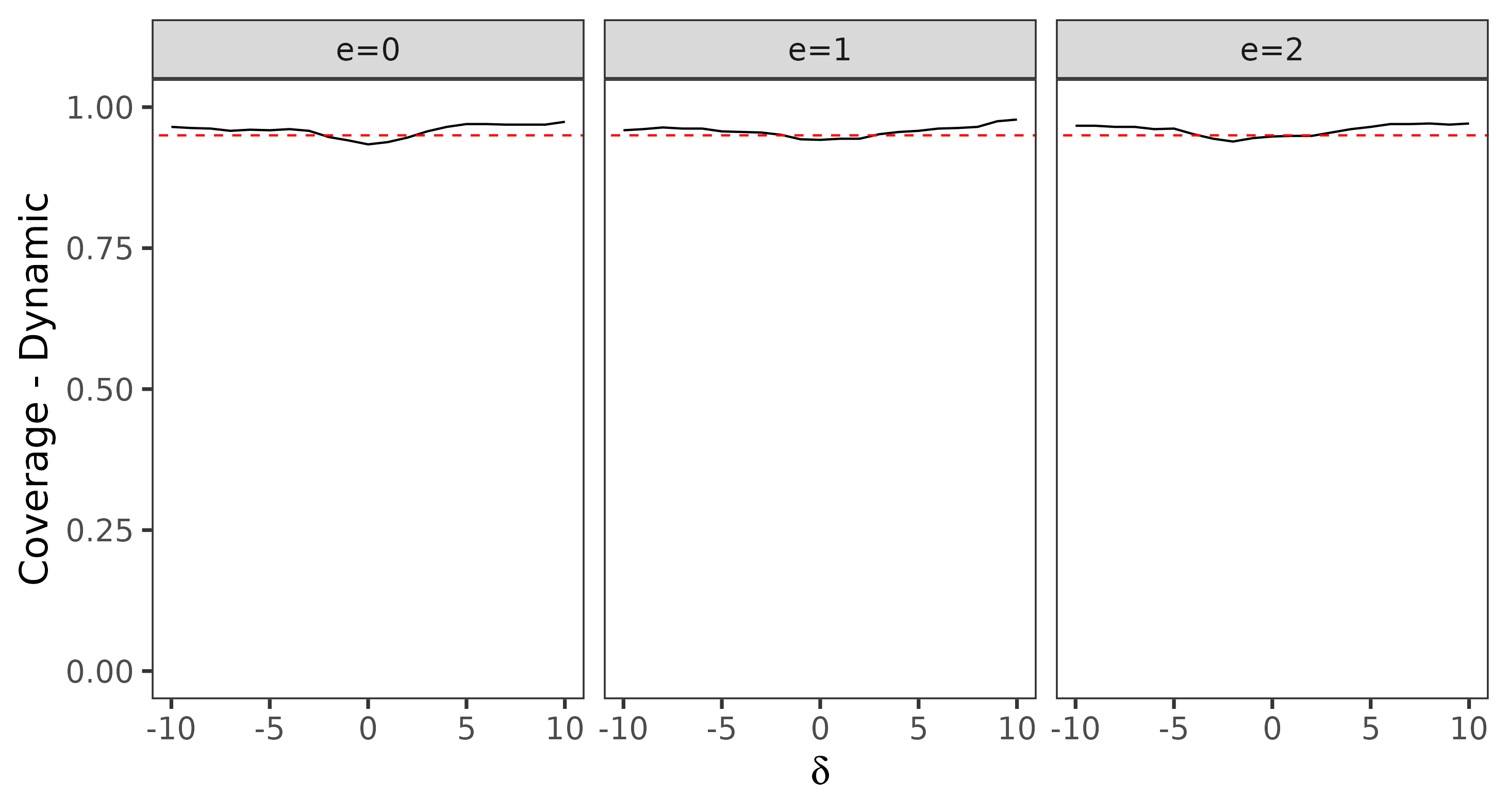}
        \caption{BART}
    \end{subfigure}
    \par \vspace{0.7cm}
    \begin{subfigure}{\textwidth}
        \centering
        \includegraphics[width=0.95\textwidth]{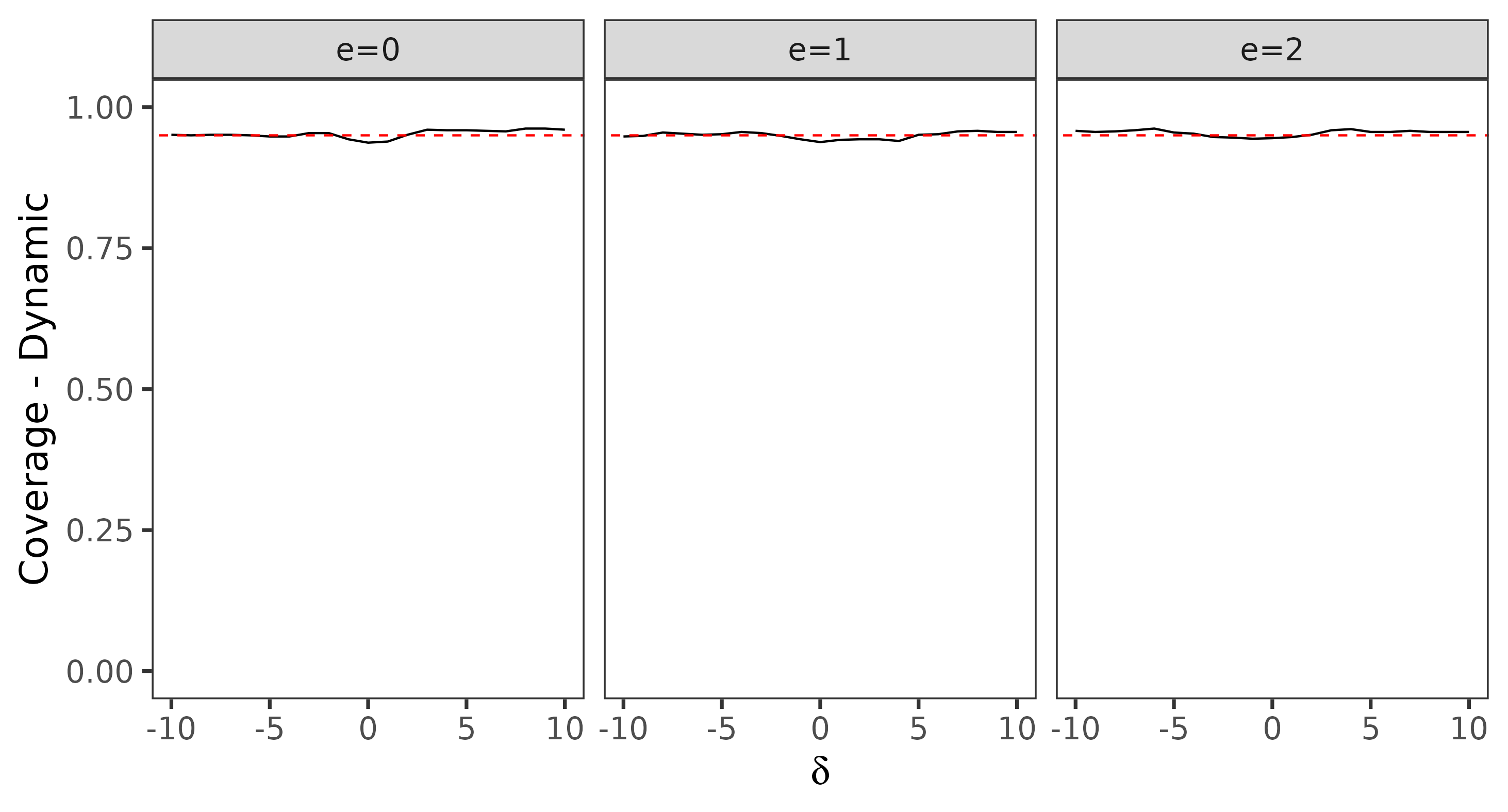}
        \caption{Oracle}
    \end{subfigure}
    \caption{Empirical coverage rate of 95\% confidence intervals for estimated $\ASDT^{\mathrm{es}}(\delta, e)$ under the exponential tilt stochastic policy with varying increments $\delta$ from 1000 simulations generated under Scenario 1. Nuisance functions were estimated using (a) BART, or (b) oracle models. The black line shows the proportion of confidence intervals that contained the true estimand for each $\delta$ and the red dashed line is set at $0.95$.}
    \label{fig:sc1_agg_dynamic-cov}
\end{figure}

\begin{figure}[p!]
    \centering
    \begin{subfigure}{\textwidth}
        \centering
        \includegraphics[width=0.95\textwidth]{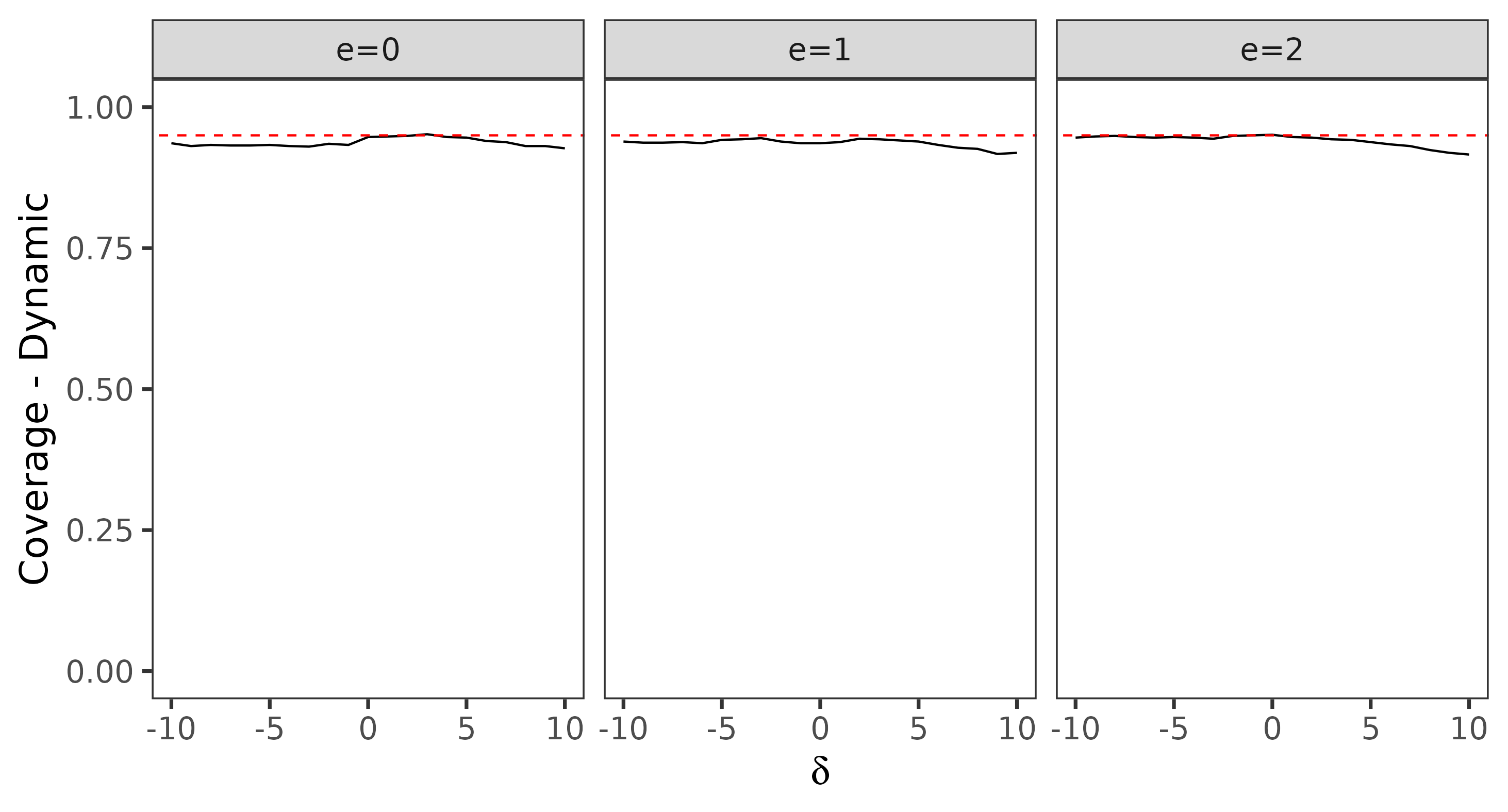}
        \caption{BART}
    \end{subfigure}
    \par \vspace{0.7cm}
    \begin{subfigure}{\textwidth}
        \centering
        \includegraphics[width=0.95\textwidth]{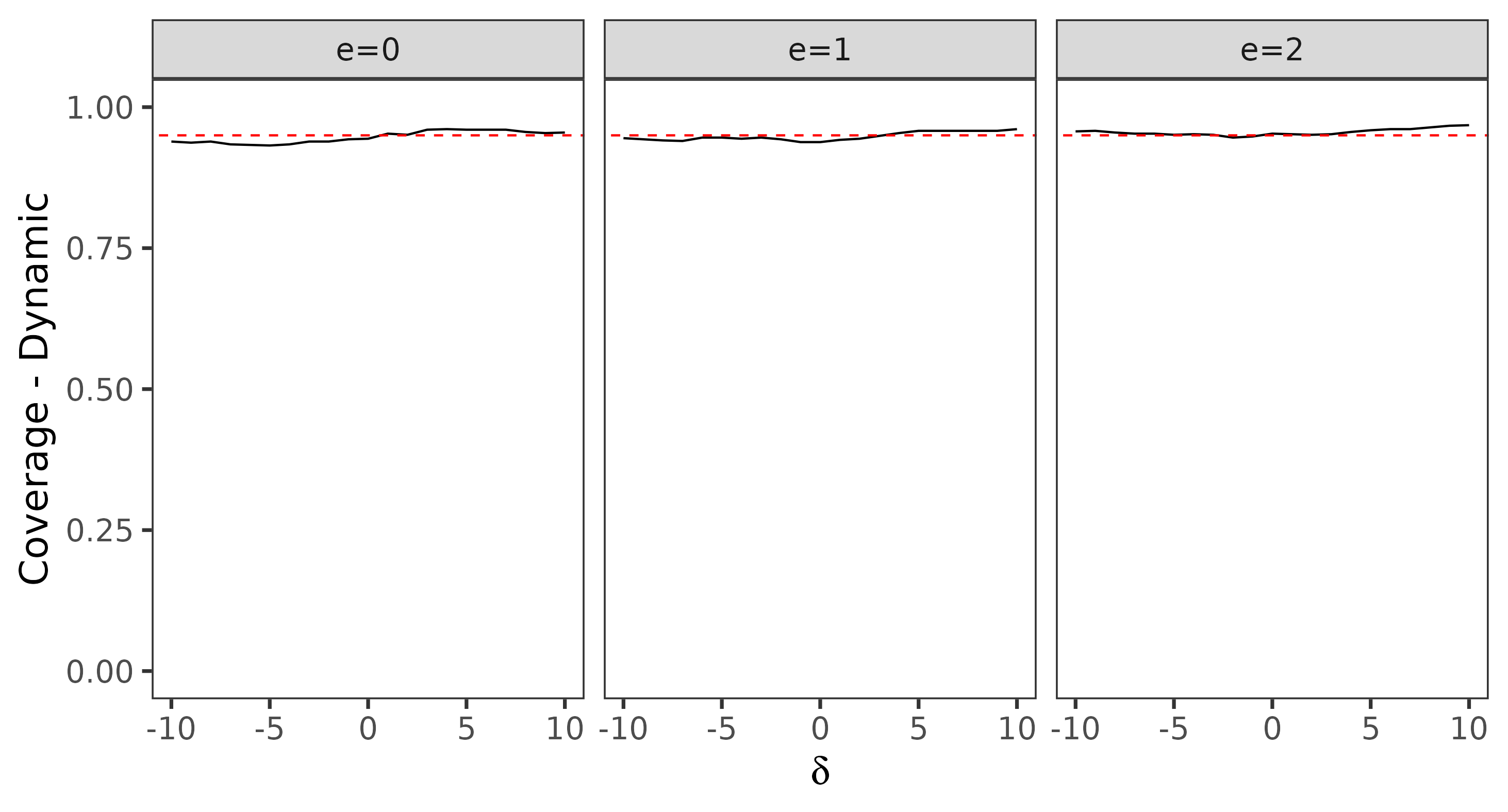}
        \caption{Oracle}
    \end{subfigure}
    \caption{Empirical coverage rate of 95\% confidence intervals for estimated $\ASDT^{\mathrm{es}}(\delta, e)$ under the exponential tilt stochastic policy with varying increments $\delta$ from 1000 simulations generated under Scenario 2. Nuisance functions were estimated using (a) BART, or (b) oracle models. The black line shows the proportion of confidence intervals that contained the true estimand for each $\delta$ and the red dashed line is set at $0.95$.}
    \label{fig:sc2_agg_dynamic-cov}
\end{figure}

\begin{figure}[p!]
    \centering
    \begin{subfigure}{\textwidth}
        \centering
        \includegraphics[width=0.95\textwidth]{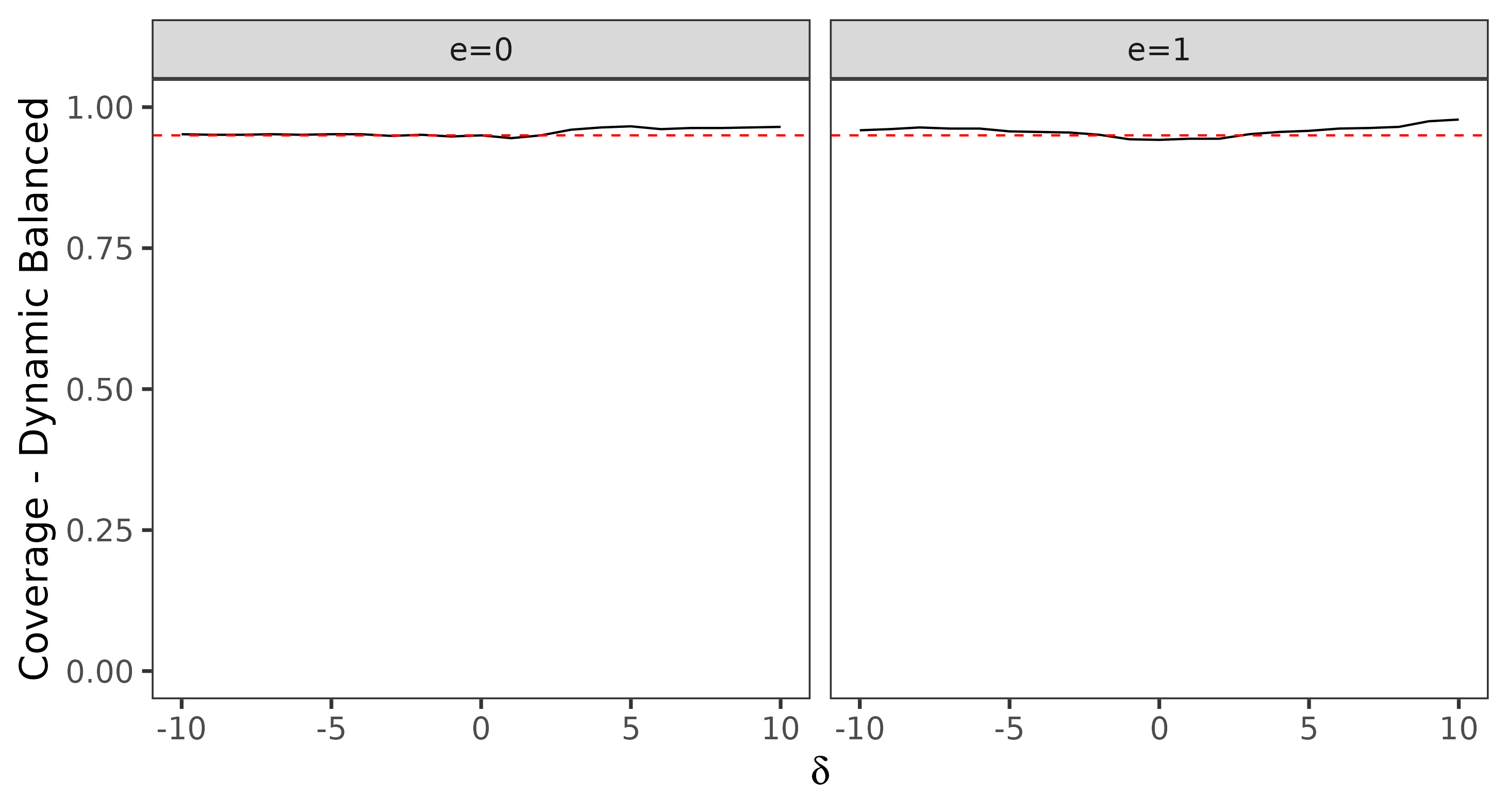}
        \caption{BART}
    \end{subfigure}
    \par \vspace{0.7cm}
    \begin{subfigure}{\textwidth}
        \centering
        \includegraphics[width=0.95\textwidth]{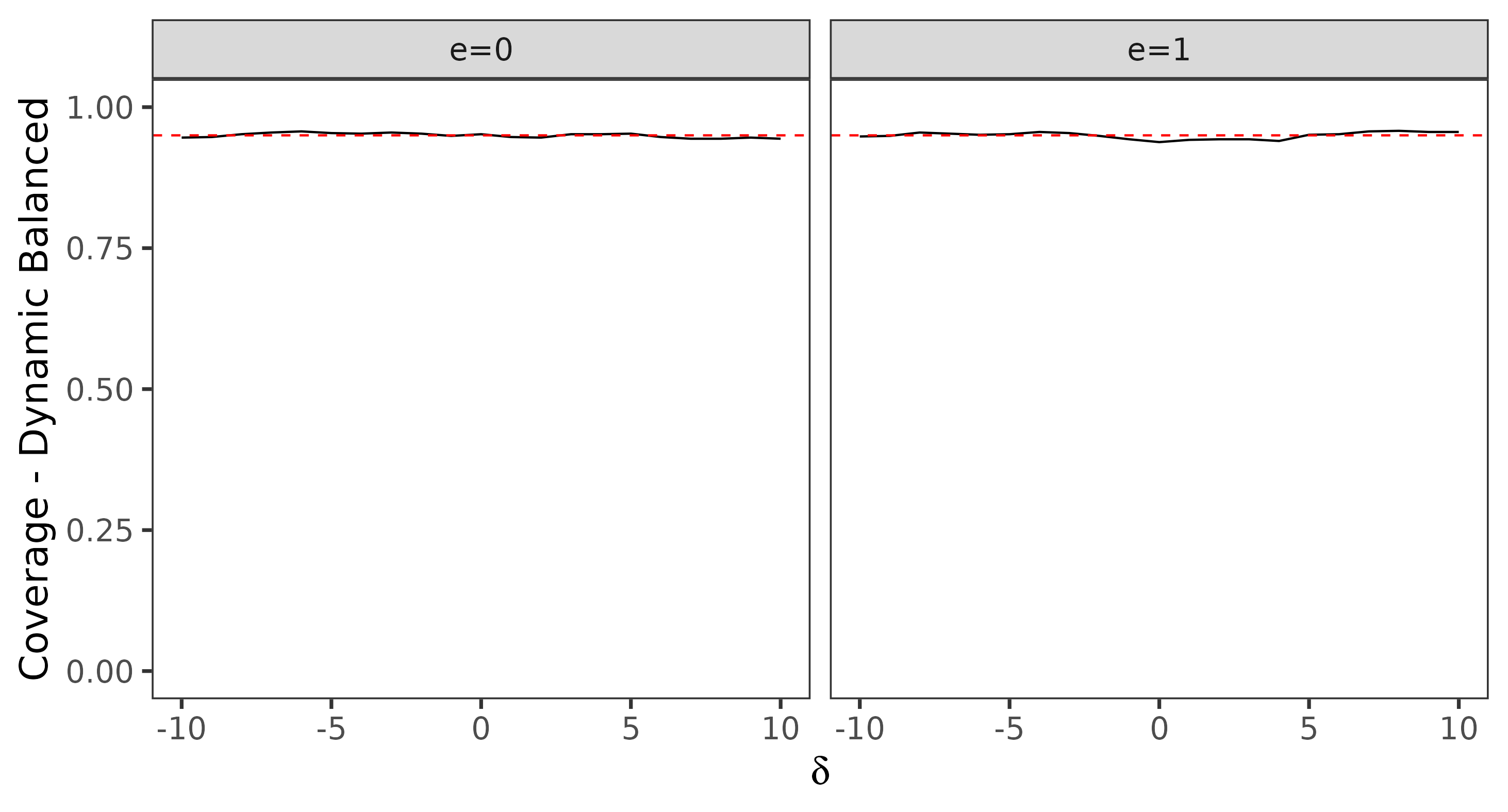}
        \caption{Oracle}
    \end{subfigure}
    \caption{Empirical coverage rate of 95\% confidence intervals for estimated $\ASDT^{\mathrm{es, bal}}(\delta, e; 1)$ under the exponential tilt stochastic policy with varying increments $\delta$ from 1000 simulations generated under Scenario 1. Nuisance functions were estimated using (a) BART, or (b) oracle models. The black line shows the proportion of confidence intervals that contained the true estimand for each $\delta$ and the red dashed line is set at $0.95$.}
    \label{fig:sc1_agg_balanced-cov}
\end{figure}

\begin{figure}[p!]
    \centering
    \begin{subfigure}{\textwidth}
        \centering
        \includegraphics[width=0.95\textwidth]{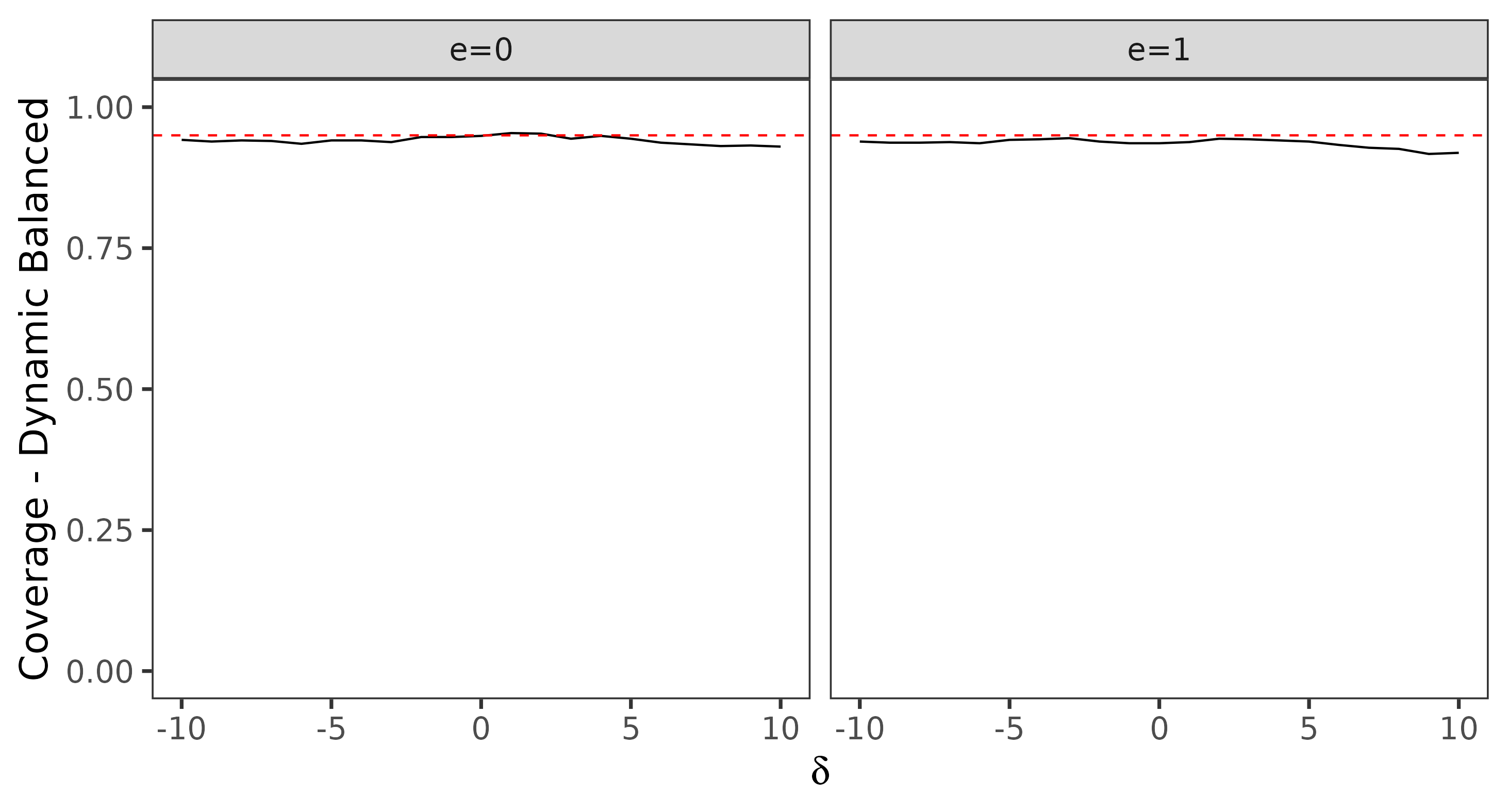}
        \caption{BART}
    \end{subfigure}
    \par \vspace{0.7cm}
    \begin{subfigure}{\textwidth}
        \centering
        \includegraphics[width=0.95\textwidth]{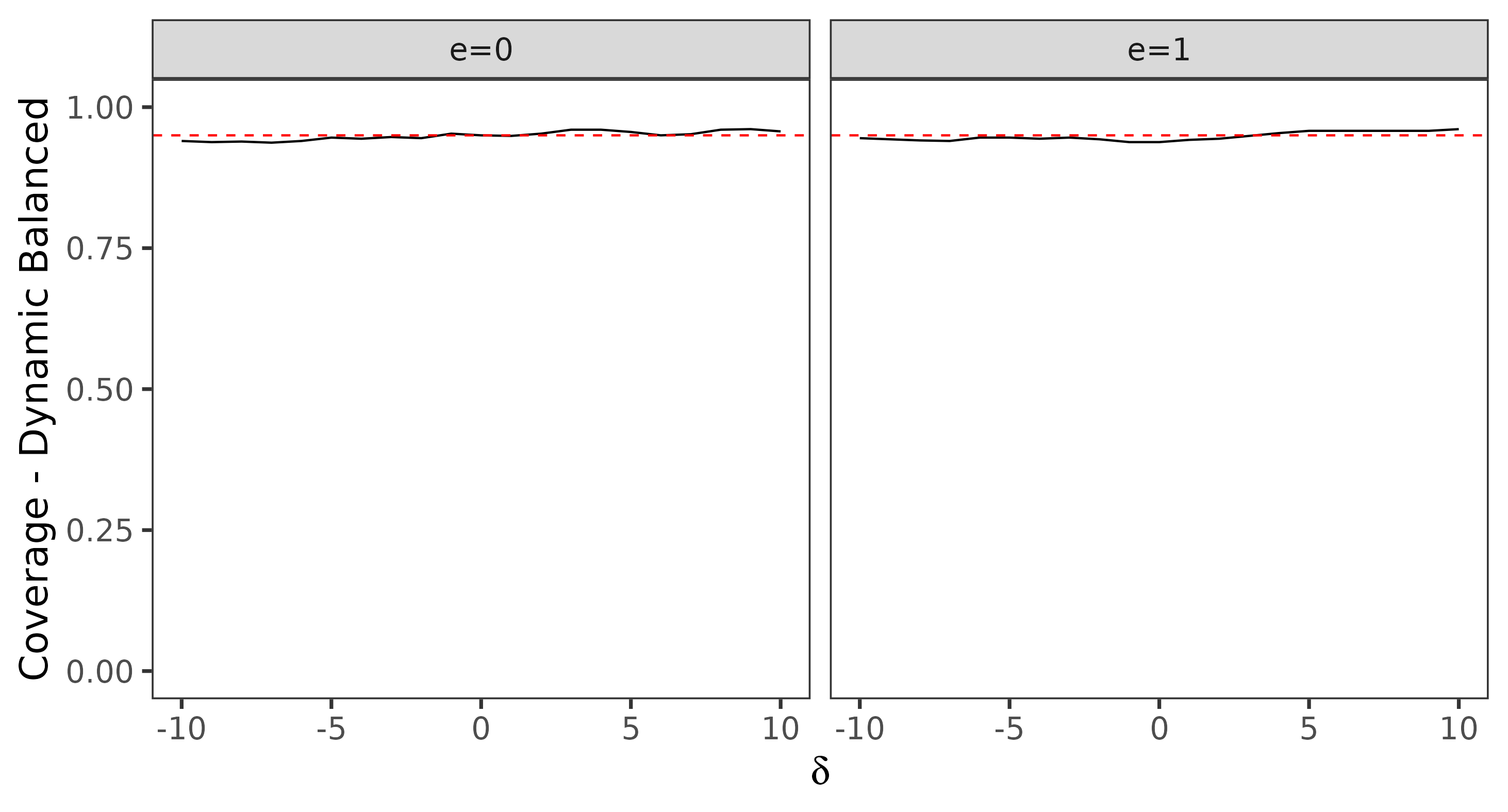}
        \caption{Oracle}
    \end{subfigure}
    \caption{Empirical coverage rate of 95\% confidence intervals for estimated $\ASDT^{\mathrm{es, bal}}(\delta, e; 1)$ under the exponential tilt stochastic policy with varying increments $\delta$ from 1000 simulations generated under Scenario 2. Nuisance functions were estimated using (a) BART, or (b) oracle models. The black line shows the proportion of confidence intervals that contained the true estimand for each $\delta$ and the red dashed line is set at $0.95$.}
    \label{fig:sc2_agg_balanced-cov}
\end{figure}

\begin{figure}[p!]
    \centering
    \begin{subfigure}{\textwidth}
        \centering
        \includegraphics[width=0.95\textwidth]{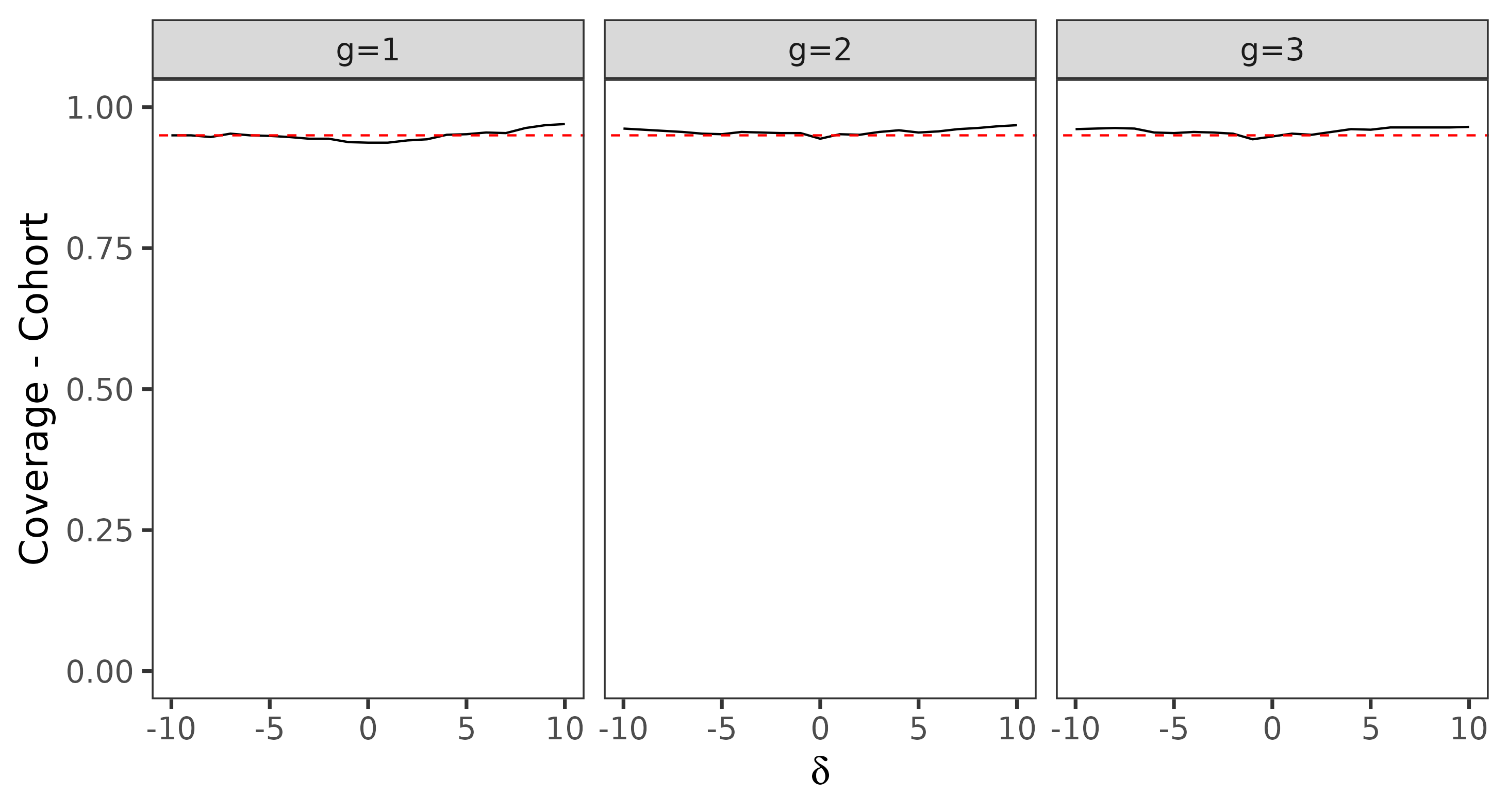}
        \caption{BART}
    \end{subfigure}
    \par \vspace{0.7cm}
    \begin{subfigure}{\textwidth}
        \centering
        \includegraphics[width=0.95\textwidth]{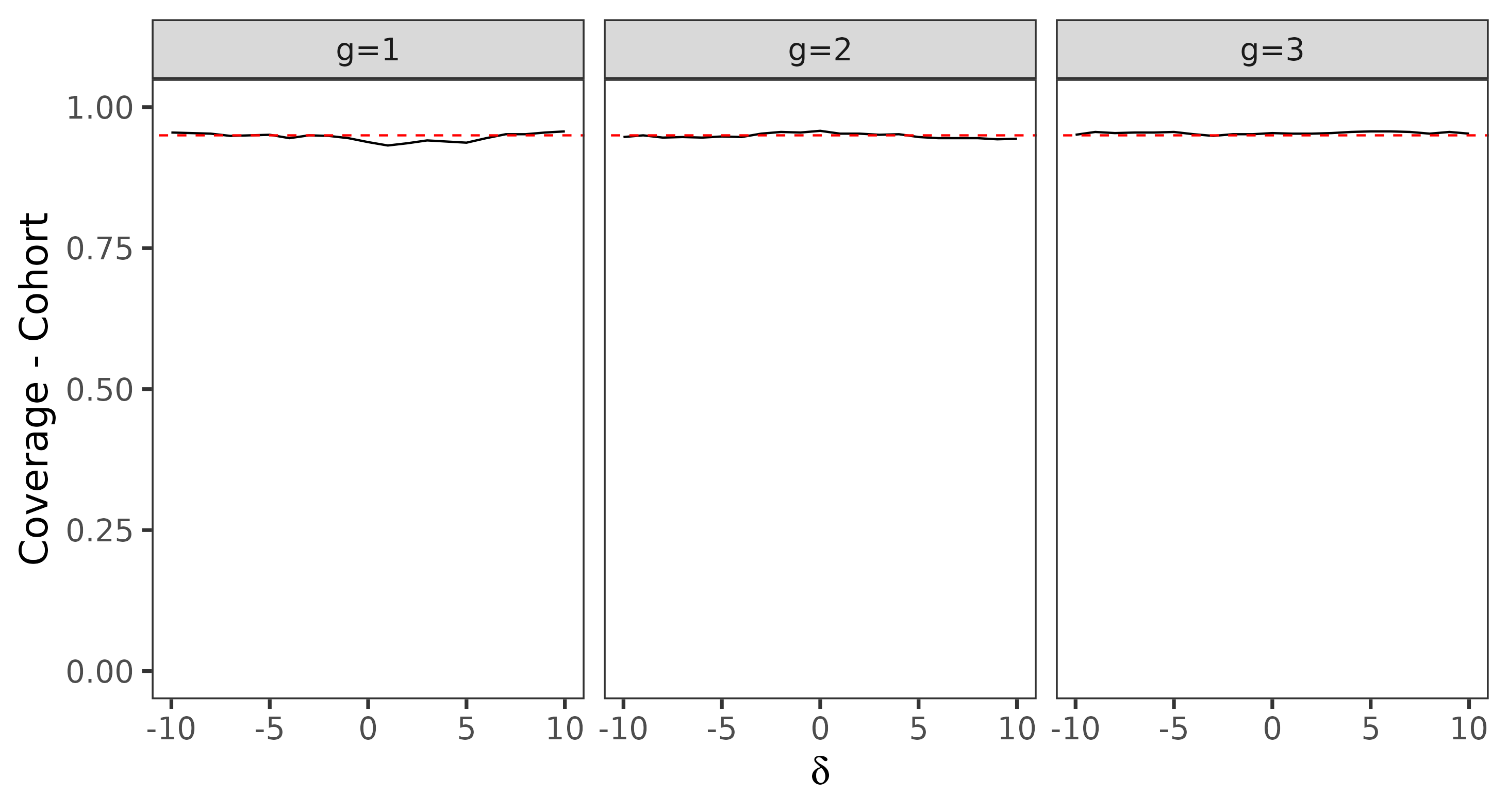}
        \caption{Oracle}
    \end{subfigure}
    \caption{Empirical coverage rate of 95\% confidence intervals for estimated $\ASDT^{\mathrm{cohort}}(\delta, g; 0, \mathcal{T}-g)$ under the exponential tilt stochastic policy with varying increments $\delta$ from 1000 simulations generated under Scenario 1. Nuisance functions were estimated using (a) BART, or (b) oracle models. The black line shows the proportion of confidence intervals that contained the true estimand for each $\delta$ and the red dashed line is set at $0.95$.}
    \label{fig:sc1_agg_group-cov}
\end{figure}

\begin{figure}[p!]
    \centering
    \begin{subfigure}{\textwidth}
        \centering
        \includegraphics[width=0.95\textwidth]{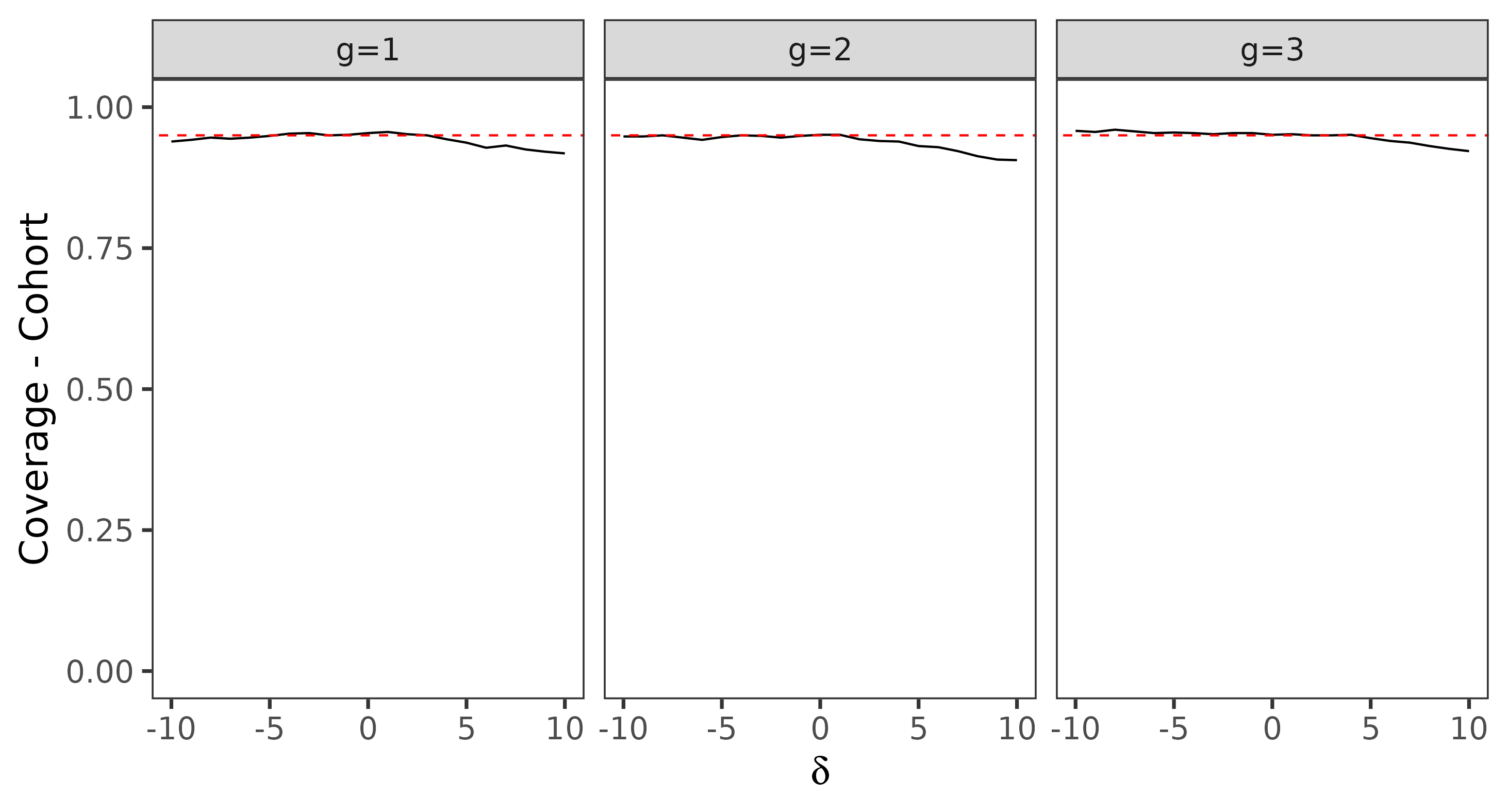}
        \caption{BART}
    \end{subfigure}
    \par \vspace{0.7cm}
    \begin{subfigure}{\textwidth}
        \centering
        \includegraphics[width=0.95\textwidth]{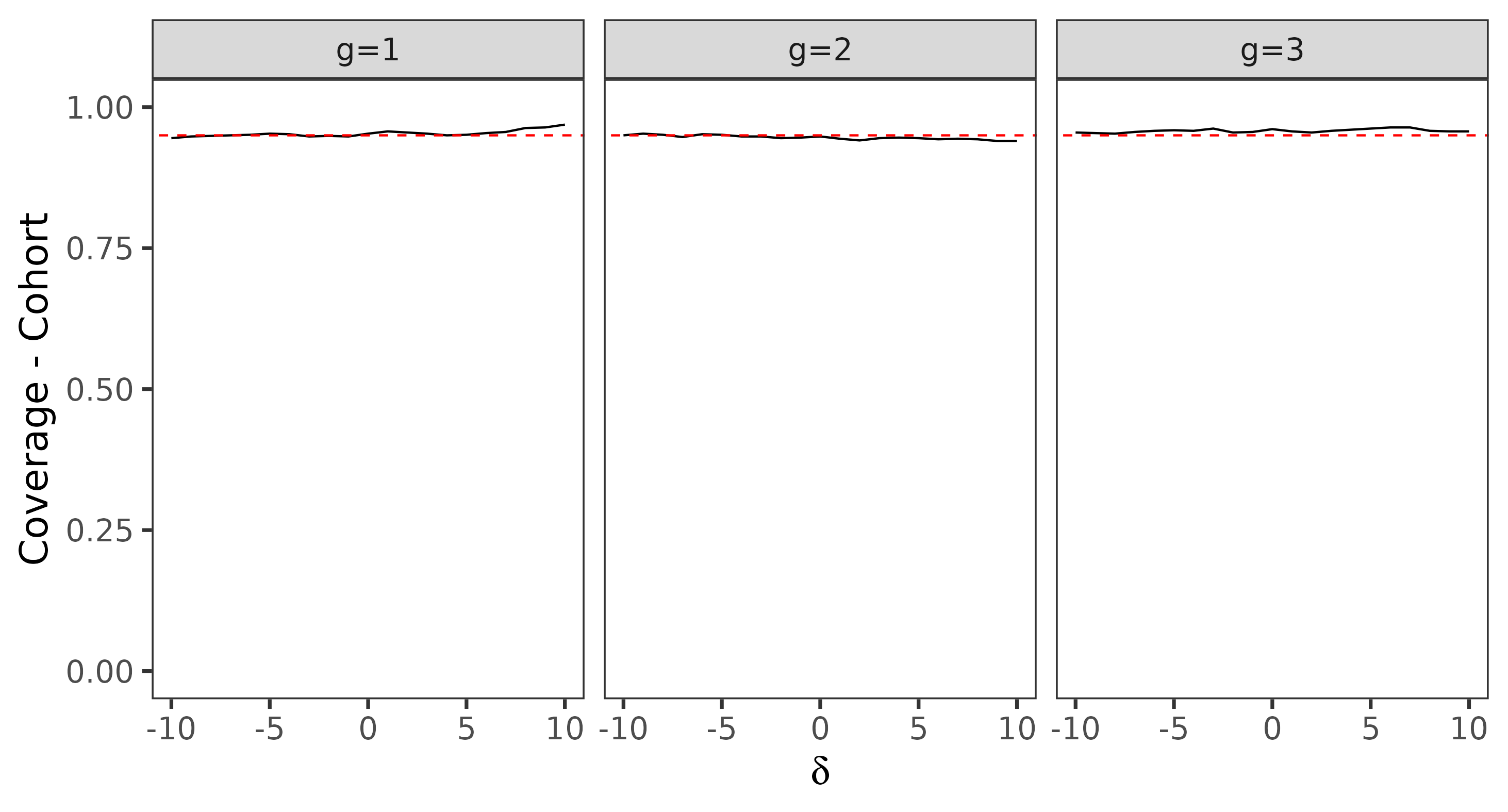}
        \caption{Oracle}
    \end{subfigure}
    \caption{Empirical coverage rate of 95\% confidence intervals for estimated $\ASDT^{\mathrm{cohort}}(\delta, g; 0, \mathcal{T}-g)$ under the exponential tilt stochastic policy with varying increments $\delta$ from 1000 simulations generated under Scenario 2. Nuisance functions were estimated using (a) BART, or (b) oracle models. The black line shows the proportion of confidence intervals that contained the true estimand for each $\delta$ and the red dashed line is set at $0.95$.}
    \label{fig:sc2_agg_group-cov}
\end{figure}

\begin{figure}[p!]
    \centering
    \begin{subfigure}{\textwidth}
        \centering
        \includegraphics[width=0.95\textwidth]{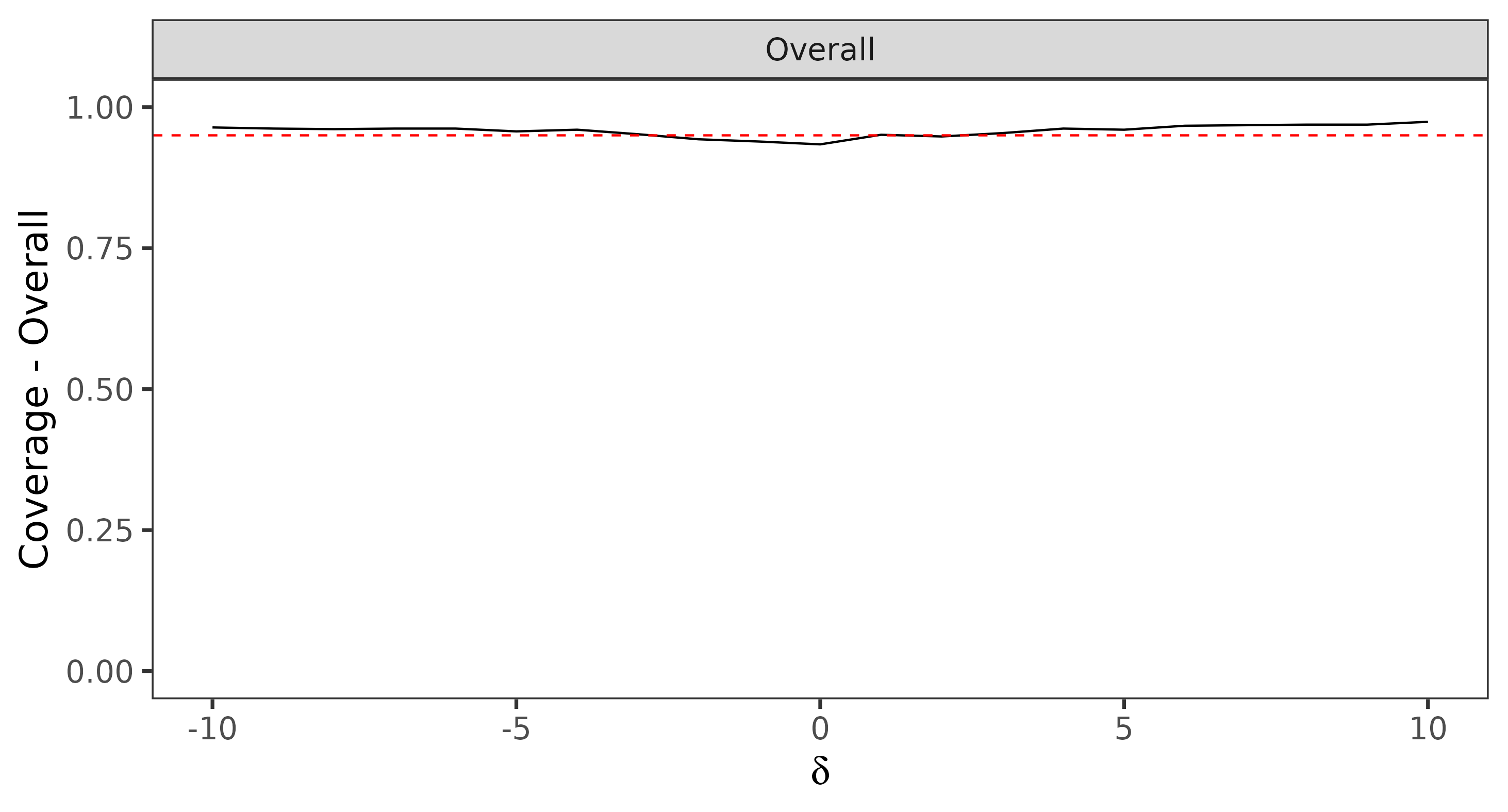}
        \caption{BART}
    \end{subfigure}
    \par \vspace{0.7cm}
    \begin{subfigure}{\textwidth}
        \centering
        \includegraphics[width=0.95\textwidth]{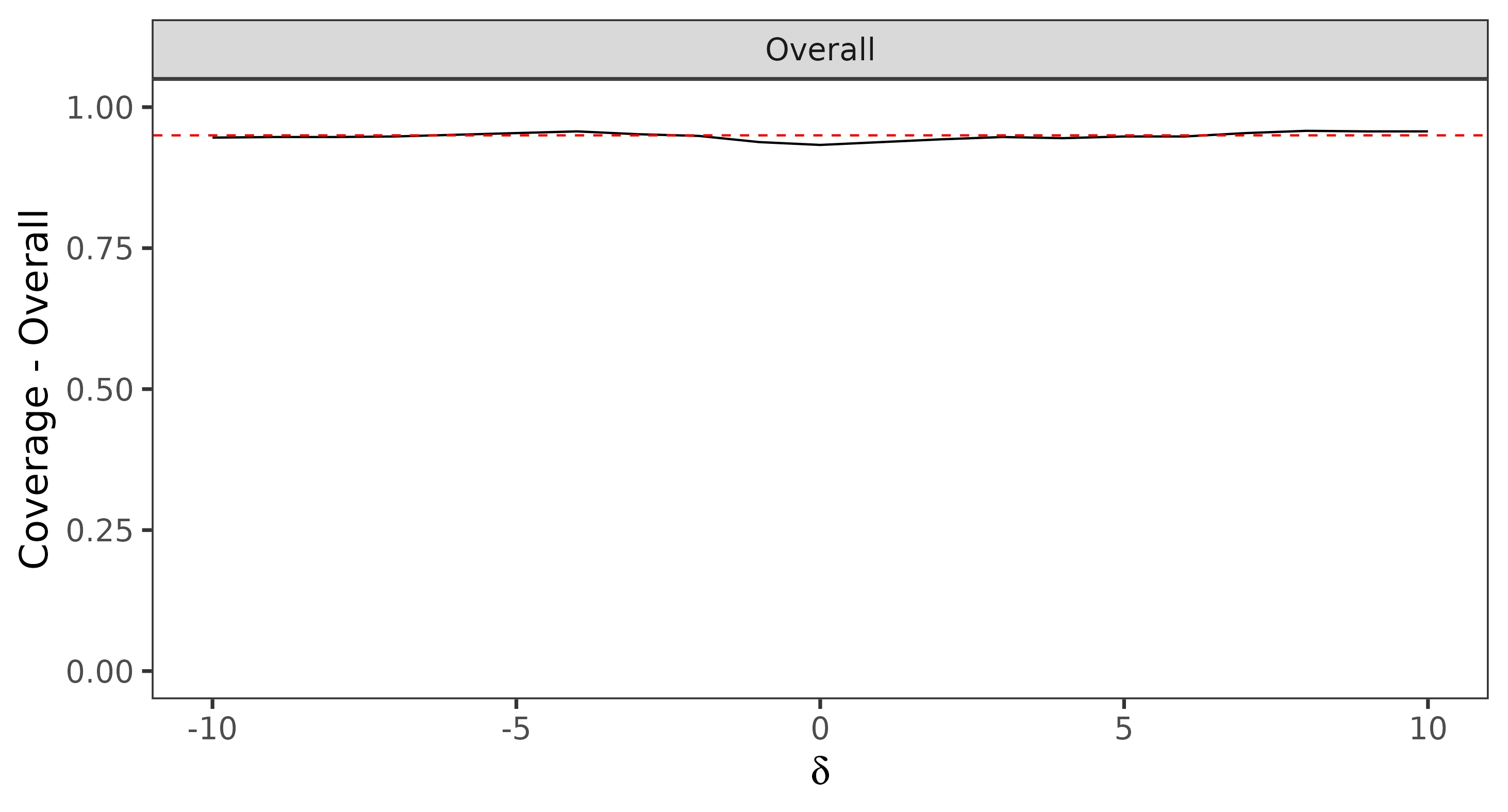}
        \caption{Oracle}
    \end{subfigure}
    \caption{Empirical coverage rate of 95\% confidence intervals for estimated $\ASDT^{\mathrm{overall}}(\delta)$ under the exponential tilt stochastic policy with varying increments $\delta$ from 1000 simulations generated under Scenario 1. Nuisance functions were estimated using (a) BART, or (b) oracle models. The black line shows the proportion of confidence intervals that contained the true estimand for each $\delta$ and the red dashed line is set at $0.95$.}
    \label{fig:sc1_agg_overall-cov}
\end{figure}

\begin{figure}[p!]
    \centering
    \begin{subfigure}{\textwidth}
        \centering
        \includegraphics[width=0.95\textwidth]{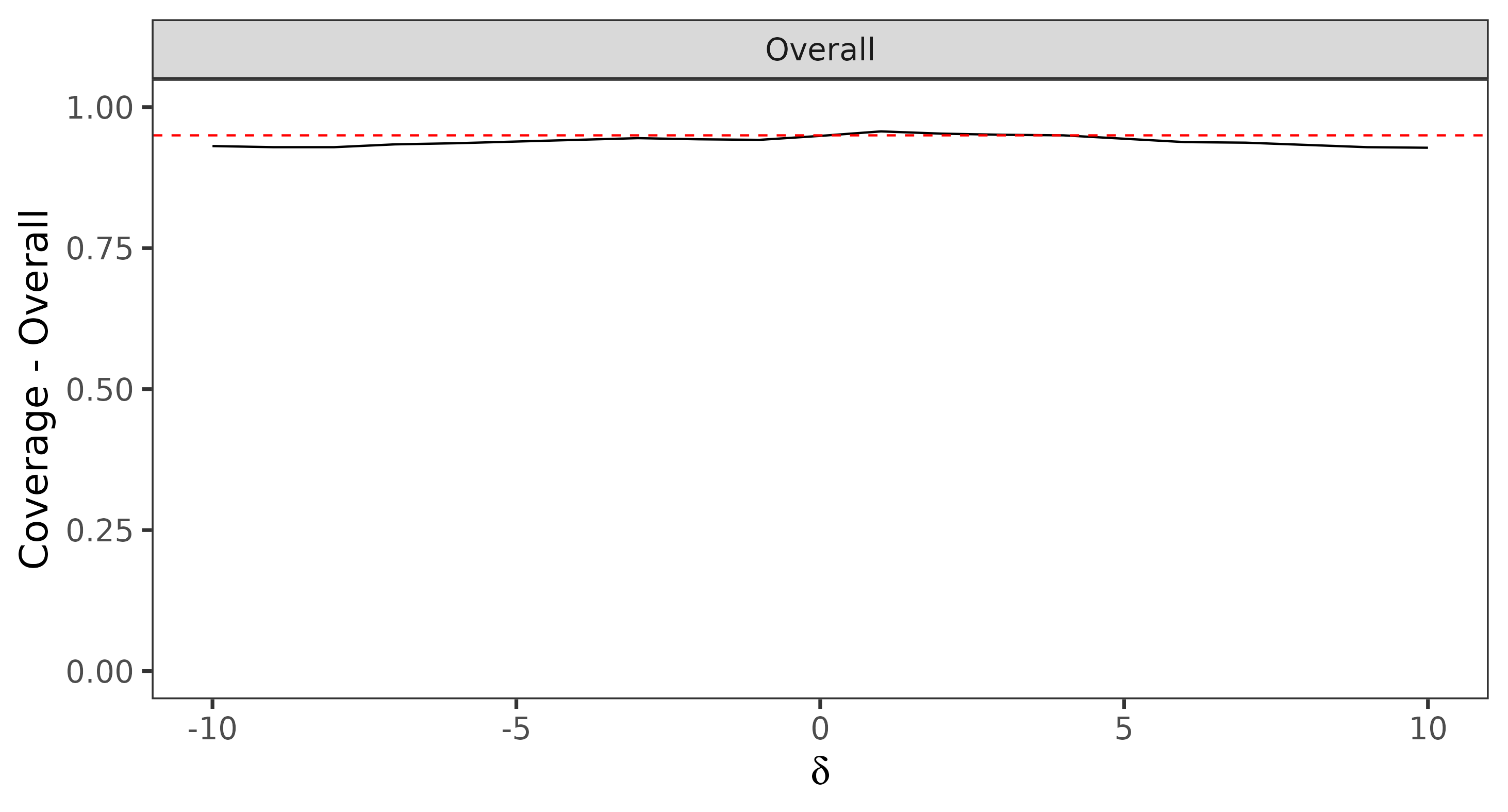}
        \caption{BART}
    \end{subfigure}
    \par \vspace{0.7cm}
    \begin{subfigure}{\textwidth}
        \centering
        \includegraphics[width=0.95\textwidth]{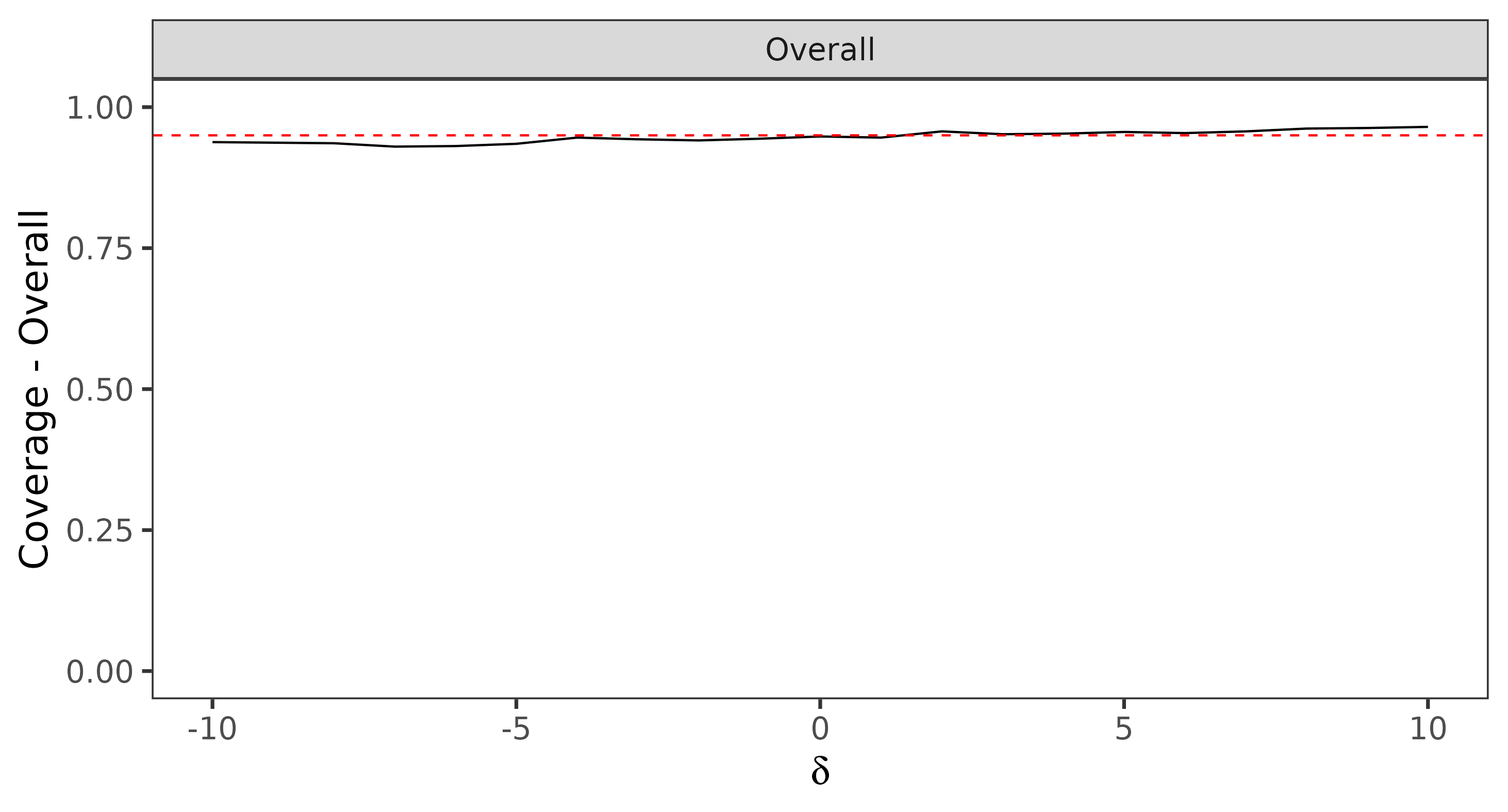}
        \caption{Oracle}
    \end{subfigure}
    \caption{Empirical coverage rate of 95\% confidence intervals for estimated $\ASDT^{\mathrm{overall}}(\delta)$ under the exponential tilt stochastic policy with varying increments $\delta$ from 1000 simulations generated under Scenario 2. Nuisance functions were estimated using (a) BART, or (b) oracle models. The black line shows the proportion of confidence intervals that contained the true estimand for each $\delta$ and the red dashed line is set at $0.95$.}
    \label{fig:sc2_agg_overall-cov}
\end{figure}

The results for estimating $\psi^{\mathrm{MTP, CF}}(\delta,g,t) = \ASDT(\delta, g, t)$ are presented in Figures \ref{fig:sc1_disagg-point} -- \ref{fig:sc2_disagg-cov}, where Figures \ref{fig:sc1_disagg-point} and \ref{fig:sc2_disagg-point} display the point estimates and true estimand for both scenarios, and Figures \ref{fig:sc1_disagg-cov} and \ref{fig:sc2_disagg-cov} show the empirical coverage of 95\% confidence intervals. There is negligible average bias for all values of $\delta$ when BART is used to estimate nuisance functions, and performance matches that of the oracle estimator. Similarly, estimating nuisance functions with BART results in similar coverage as when using oracle nuisance function estimators. The point estimates for the aggregated parameters are presented in Figures \ref{fig:sc1_agg_dynamic-point} -- \ref{fig:sc2_agg_overall-point}, and empirical coverage results are shown in Figures \ref{fig:sc1_agg_dynamic-cov} -- \ref{fig:sc2_agg_overall-cov}. For all aggregated parameters, using BART to estimate nuisance functions leads to low bias and similar coverage rates as when using oracle nuisance function estimators. When the nonlinear nuisance functions were mis-specified with GLMs, performance was poor for all parameters as expected; results are omitted for brevity.


\newpage

\subsection{Additional details on the data application}

The time-varying sample sizes for the initiated treatment and not-yet-treated groups are shown in Table \ref{tab:sample_size}. Treated counties ``drop out" of the study in subsequent years in the sense that in each treatment cohort, only counties that initiated treatment or were not-yet-treated are analyzed. Only years 2008 and 2009 contained sample sizes for the treated group that were large enough to distinguish between different dose levels; thus, other years were not analyzed. 

\begin{table}[!h]
\caption{Number of counties that initiated treatment (fracking) or were not-yet-treated over time.}
\label{tab:sample_size}
\centering
\begin{tabular}{lcc}
\hline
\textbf{Year} & \textbf{Not-yet-treated} & \textbf{Treated} \\ \hline
2008 & 166 & 156 \\
2009 & 81 & 85 \\
2010 & 60 & 21 \\
2011 & 60 & 0 \\
2012 & 38 & 22 \\
2013 & 38 & 0 \\
2014 & 38 & 0 \\ \hline
\end{tabular}
\end{table}

Tables \ref{tab:cov-summ-2008} and \ref{tab:cov-summ-2009} show covariate summary statistics for the 2008 and 2009 cohorts, respectively. In both years, there is a moderate amount of differences in the treated and untreated groups. Thus, it is possible that these covariates affect outcome trends in such a way that unconditional parallel trends does not hold but conditional parallel trends does hold. 

\begin{table}[!h]
\caption{Covariate summary statistics by untreated and any treatment groups, 2008.}
\label{tab:cov-summ-2008}
\centering
\begin{tabular}{lll}
\hline
\multirow{2}{*}{\textbf{Covariate}} & \multicolumn{2}{c}{\textbf{Mean (SD)}} \\
& \textbf{Untreated} & \textbf{Treated} \\ \hline
Percent White & 88.5 (11.0) & 91.7 (10.4) \\
Percent Black & 3.8 (6.1) & 5.3 (9.1) \\
Percent Hispanic & 11.6 (20.1) & 1.7 (2.5) \\
Percent female & 50.2 (1.8) & 51.0 (1.3) \\
Age (years) & 37.5 (4.3) & 38.4 (2.9) \\
Population (log) & 10.1 (1.6) & 10.6 (1.1) \\ \hline
\end{tabular} \par
\smallskip
SD: standard deviation.
\end{table}

\begin{table}[!h]
\caption{Covariate summary statistics by untreated and any treatment groups, 2009.}
\label{tab:cov-summ-2009}
\centering
\begin{tabular}{lll}
\hline
\multirow{2}{*}{\textbf{Covariate}} & \multicolumn{2}{c}{\textbf{Mean (SD)}} \\
& \textbf{Untreated} & \textbf{Treated} \\ \hline
Percent White & 89.3 (12.1) & 87.7 (9.9) \\
Percent Black & 4.3 (7.1) & 3.3 (4.9) \\
Percent Hispanic & 4.7 (6.7) & 18.1 (25.7) \\
Percent female & 50.3 (1.3) & 50.1 (2.2) \\
Age & 37 (3.6) & 38 (4.8) \\
Population (log) & 10.8 (1.5) & 9.4 (1.2) \\ \hline
\end{tabular} \par
\smallskip
SD: standard deviation.
\end{table}

\begin{figure}[!h]
    \centering
    \includegraphics[width=1\linewidth]{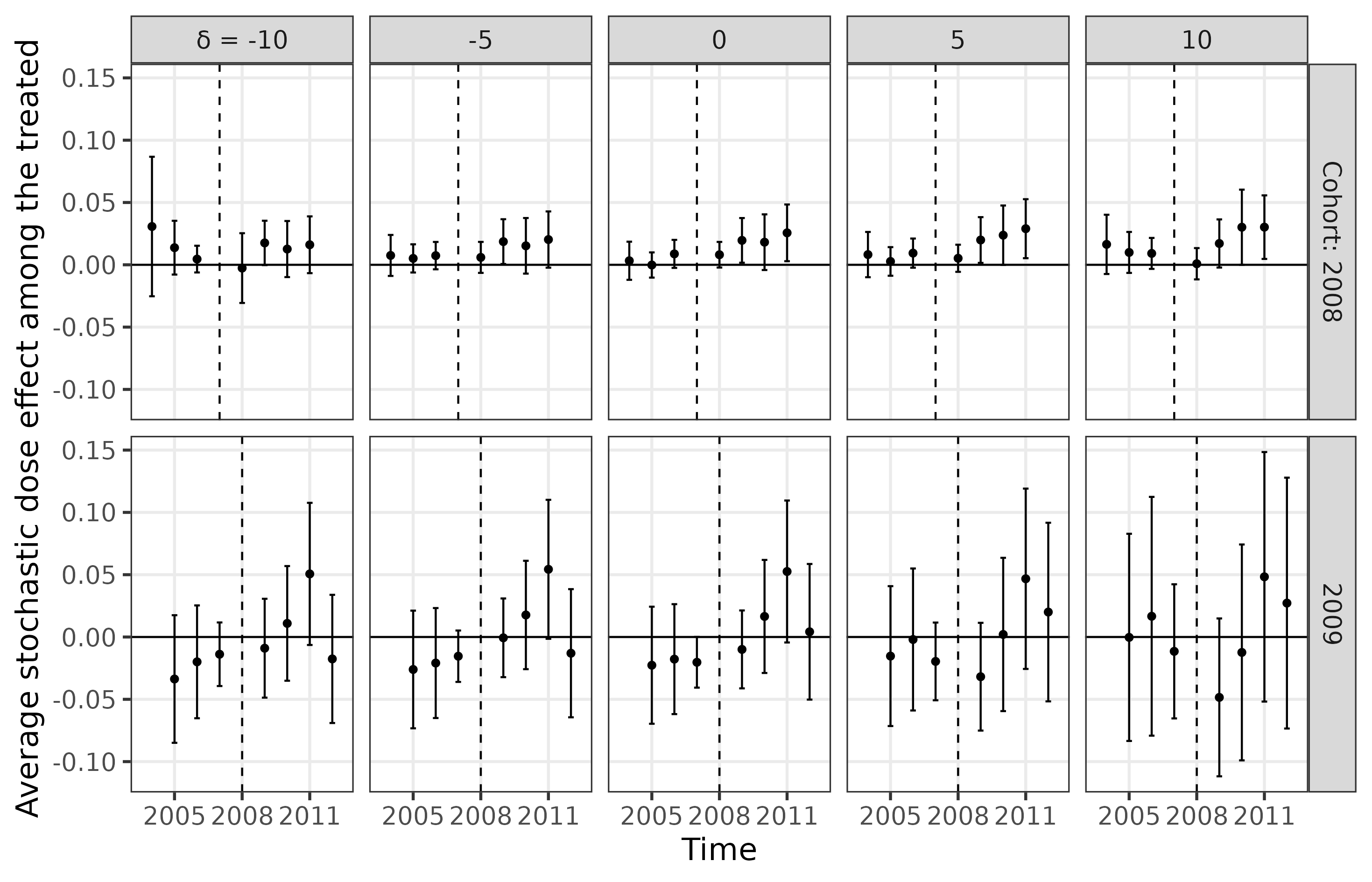}
    \caption{Income effects due to shifts in probability distribution of fracking potential in 2008 and 2009. A larger $\delta$ denotes a shift of the probability distribution towards the maximal fracking potential value. The dashed line denotes the reference year, circles represent point estimates, and vertical bars represent pointwise 95\% confidence intervals. Estimates to the left of the dashed lines are placebo tests.}
    \label{fig:results-inc}
\end{figure}

Figure \ref{fig:results-inc} displays the estimated ASDT across treatment cohorts and calendar time for the outcome of (log) total annual income, in thousands of dollars. The results are similar to the employment results where there appears to be increasing effects up to 2011 and increasing effects due to incrementing the dose distribution to towards the maximum.  

\begin{figure}[!h]
    \centering
    \includegraphics[width=1\linewidth]{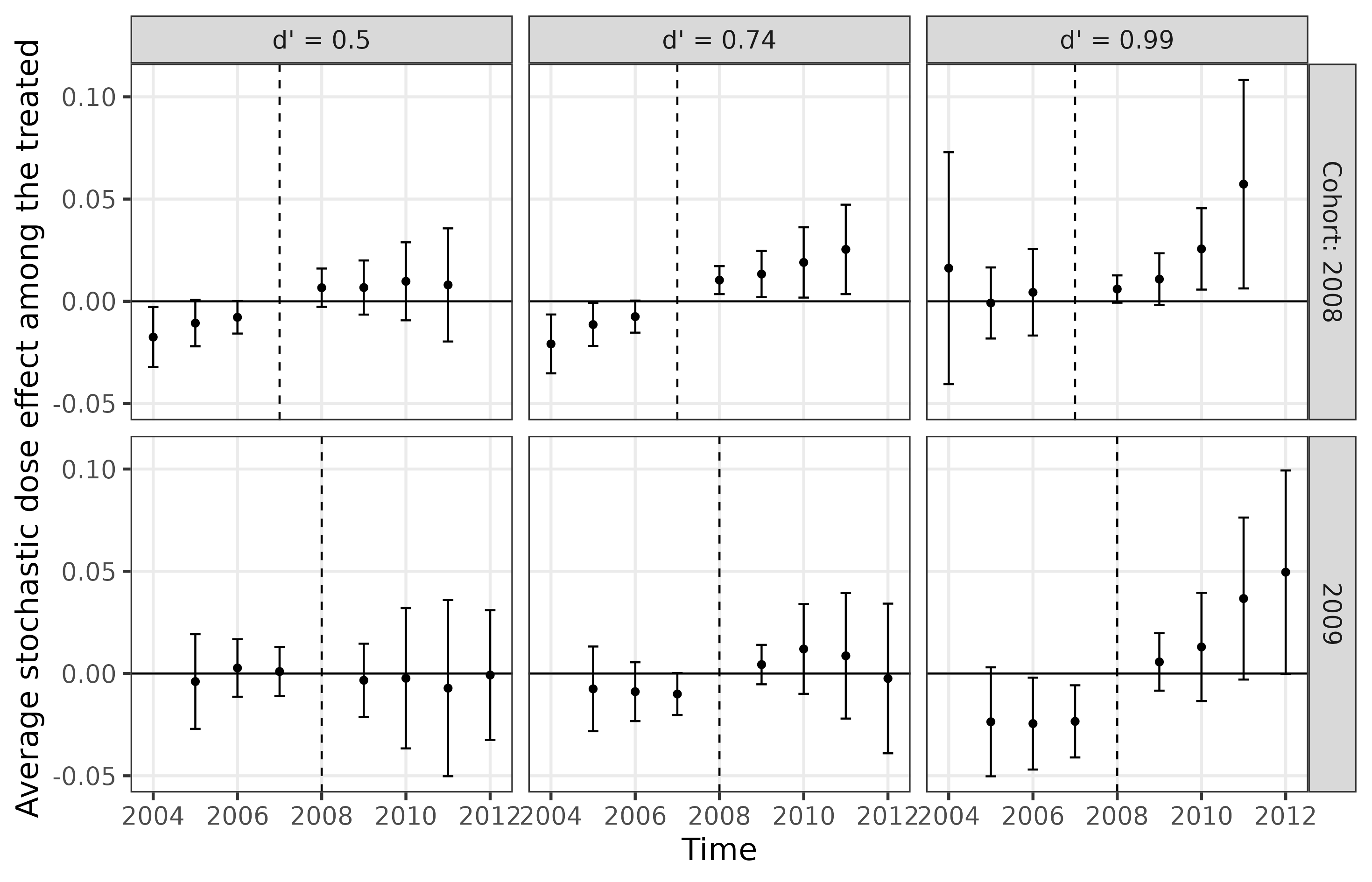}
    \caption{Employment effects due to shifts in probability distribution of fracking potential in 2008 and 2009. The fracking potential probability distribution was shifted according to the Gaussian concentration policy with length parameter $l=0.1$ and concentration points $d' \in \{0.5, 0.74, 0.99\}$. The dashed line denotes the reference outcome year, circles represent point estimates, and vertical bars represent pointwise 95\% confidence intervals. Estimates to the left of the dashed lines are placebo tests.}
    \label{fig:results-emp-gauss}
\end{figure}

Figure \ref{fig:results-emp-gauss} show the results for the ASDT under the Gaussian concentration stochastic policy with length parameter $l=0.1$ and concentration points $d' \in \{0.5, 0.74, 0.99\}$. By fixing the length parameter but varying the concentration points, Figure \ref{fig:results-emp-gauss} can be read like a dose-effect function when fixing cohort $g$ and calendar time $t$, i.e., one can examine how the ASDT varies by dose level. However, in contrast to typical dose-effect function estimation, strong positivity assumptions are not needed with the Gaussian concentration policy. In these results, the ASDT effect on total employment tends to be the same or grow larger with $d'$. 

\end{document}